\documentclass[aps,preprintnumbers,amsmath,amssymb,floatfix,10pt,prd,twocolumn,
superscriptaddress,nofootinbib]{revtex4-2}
\usepackage{silence}
\usepackage[bottom=2.5cm, right=1.5cm, left=1.5cm, top=2.5cm]{geometry}
\usepackage{bm}
\usepackage{amsfonts}
\usepackage{graphicx}
\usepackage{amsmath}
\usepackage[utf8]{inputenc}
\usepackage{textcomp}
\usepackage{booktabs}
\usepackage{dcolumn}
\usepackage{ragged2e}
\usepackage{xcolor}
\usepackage{microtype}
\usepackage{subcaption}
\usepackage{orcidlink}

\allowdisplaybreaks[1]

\newcommand{\csch}{\operatorname{csch}}
\newcommand{\sech}{\operatorname{sech}}
\begin{document}
\title{Optical and Dynamical Signatures of Pad\'{e} Approximated Wormholes in \texorpdfstring{$f(T)$}{f(T)} Gravity}
\author{Sneha Pradhan\orcidlink{0000-0002-3223-4085}}
\email{snehapradhan2211@gmail.com}
\affiliation{Chennai Mathematical Institute,
H1 SIPCOT IT Park, Siruseri 603103, India}
\affiliation{Department of Mathematics, Birla Institute of Technology and Science,
Pilani, Hyderabad Campus, Jawahar Nagar, Kapra Mandal,
Medchal District, Telangana 500078, India}
\author{Paras Balani\orcidlink{0009-0009-1806-5869}}
\email{f20230738@hyderabad.bits-pilani.ac.in}
\affiliation{Department of Mathematics and Department of Computer Science,
Birla Institute of Technology and Science, Pilani, Hyderabad Campus,
Jawahar Nagar, Kapra Mandal, Medchal District, Telangana 500078, India}

\author{P.K. Sahoo\orcidlink{0000-0003-2130-8832}}
\email{pksahoo@hyderabad.bits-pilani.ac.in}
\affiliation{Department of Mathematics, Birla Institute of Technology and Science,
Pilani, Hyderabad Campus, Jawahar Nagar, Kapra Mandal,
Medchal District, Telangana 500078, India}
\begin{abstract}
We construct traversable wormhole solutions in the linear teleparallel model $f(T) = \alpha T + \beta$, adopting Pad\'e approximants of order [1/0] and [0/1] as shape functions together with the redshift function $\Phi(r) = -M/r$, and derive the corresponding source terms from the modified field equations. We assess the physical viability of both configurations through the energy conditions, Tolman--Oppenheimer--Volkoff equilibrium, the volume integral quantifier, thin-shell junction matching to an exterior Schwarzschild geometry, embedding diagrams, proper radial distance, the gravitational energy and the active mass. We then examine the observational signatures of each geometry by analysing equatorial null geodesics and the photon-sphere structure, computing shadow maps and radial intensity profiles, ray-tracing relativistic images of a geometrically thin accretion disc, and integrating timelike rosette orbits. Finally, we study the dynamical stability of both wormholes through a third-order WKB quasinormal-mode analysis across scalar, electromagnetic, and axial gravitational perturbations, cross-checked against time-domain evolution and the eikonal photon-sphere correspondence, and examine the parameter dependence of the mode spectrum together with the possibility of trapped-mode echoes.
\end{abstract}

\maketitle
\section{Introduction}
\label{sec:intro}
The quest to extend Einstein's general theory of relativity beyond its classical domain
has driven substantial progress in gravitational physics over the past several decades.
Among the most well-studied extensions is teleparallel gravity and its generalizations,
wherein the gravitational field is described not through spacetime curvature but through
torsion arising from the Weitzenb\"{o}ck connection~\cite{Nojiri2007,Starobinsky1980,
Copeland2006,Capozziello2011}. The $f(T)$ gravity framework, which replaces the torsion
scalar $T$ in the teleparallel equivalent of general relativity with an arbitrary function
$f(T)$, has attracted considerable attention as a geometrically distinct, yet physically
rich alternative to both general relativity and curvature-based modified theories such as
$f(R)$ gravity~\cite{Brans1961,Bertolami2007,Houndjo2012}. The modified field equations
in $f(T)$ gravity yield richer phenomenology in strong-field regimes, including
corrections to black hole thermodynamics, modified gravitational wave propagation, and new
classes of compact object solutions. These features make $f(T)$ gravity a compelling arena for investigating exotic spacetime geometries, particularly traversable wormholes, whose construction in general relativity typically requires violations of the null energy condition attributed to exotic matter. However, 
in $f(T)$ gravity, such configurations can be supported more naturally by torsion-induced geometric corrections~\cite{Naz2023,Nazavari2023,Solanki2023,Tayde2023,Saleem2023torsion}.

Traversable wormholes, first placed on rigorous footing by Morris and
Thorne~\cite{Morris1988}, represent topological bridges connecting distinct regions of
spacetime and have evolved from purely theoretical constructs~\cite{Flamm1916,
Einstein1935} into objects of active observational interest. In modified gravity
frameworks, including $f(R)$, $f(R,T)$, and teleparallel formulations, the effective
stress-energy contributions arising from geometric modifications can partially or fully
replace the exotic matter otherwise required to sustain the wormhole
throat~\cite{Saiedi2011,Najafi2015,Rahaman2016,Samanta2019a,Godani2019a,Samanta2019b,
Godani2019b,Zubair2016frt,Elizalde2019,Sharif2019ann,Azizi2013,Mishra2020,Ahmed2022,
Naseer2023,Yousaf2017}.

Recent investigations have further explored traversable wormhole configurations in a variety of modified gravity theories, considering different geometric conditions, matter sources, and additional physical fields. In particular, charged wormhole solutions have been investigated in scalar-tensor and curvature-based modified gravity frameworks, where the effects of electromagnetic fields, scalar potentials, and embedding conditions on the geometry and energy conditions have been examined~\cite{Malik2026Charge,Malik2026ScalarPotential,Naz2024Karmarkar,Fayyaz2024Charged}. Wormhole configurations involving global monopole charge have also been studied in the context of $f(G)$ gravity, highlighting the influence of topological charge on the energy conditions and stability properties~\cite{Yousaf2025Monopole}. Furthermore, evolving and embedded wormhole geometries in $f(R,G)$ gravity have been investigated to understand the role of higher-curvature contributions in supporting traversable configurations~\cite{Naz2023Evolving}.

The astrophysical significance of traversable
wormholes has grown in parallel with the maturation of gravitational-wave astronomy and
high-resolution very-long-baseline interferometry, both of which open windows onto
compact object signatures that may distinguish wormholes from black
holes~\cite{Dai2019,Jafferis2022,Simonetti2021,Guerrero2022,Olmo2015}. In particular,
the formation of photon spheres, the divergence of the deflection angle in the strong-field
limit, and the structure of shadow images cast by compact objects under strong
gravitational lensing provide powerful observational probes of the underlying spacetime
geometry~\cite{Schneider1992,Frittelli2000,Virbhadra2000,Bozza2001,Bozza2002,
Bhadra2003,Bozza2005,Vasquez2004}. Gravitational lensing in wormhole spacetimes has been
extensively studied, including deflection angle calculations~\cite{Nandi2006,Rahaman2007,
Dey2008,Bhattacharya2010,Abe2010,Nakajima2012,Sharif2015,Tsukamoto2016,Tsukamoto2017,
Nandi2017a,Nandi2017b}, analyses of negative-mass wormholes~\cite{Cramer1995,
Safonova2002}, convergent lens behavior~\cite{Tejeiro2005}, and the characterization of
photon and antiphoton sphere structures~\cite{Shaikh2019jcap,Bronnikov2019,Shaikh2019plb,
DeFalco2020,DeFalco2021,Alhamzawi2016,Jusufi2017,Shaikh2017}. The Event Horizon
Telescope's imaging of compact object shadows has further sharpened the need for detailed
theoretical predictions of shadow morphology in alternative spacetime backgrounds,
including those generated by wormhole geometries in modified gravity~\cite{Jusufi2018}.

A central technical challenge in modeling wormhole observables is the construction of
physically admissible shape functions that simultaneously satisfy throat regularity, the
flare-out condition, and asymptotic flatness across the full radial domain~\cite{Visser1995,
Mishra2020,Ahmed2022,Naseer2023}. Shape functions derived from truncated Taylor series
often fail to reproduce physically consistent behavior at large radial distances or in the
strong-field region near the throat. Pad\'{e} approximants, rational functions constructed
as ratios of polynomials matched to a given power series~\cite{Baker1961}, provide a
systematic and analytically tractable improvement over Taylor truncations by encoding pole
structure and extending the convergence domain~\cite{Capozziello2020,Gruber2014}. Their
application to wormhole modeling has been demonstrated to transform inadequate or
analytically intractable shape functions into viable candidates that satisfy all required
geometric constraints, while remaining amenable to closed-form analysis and numerical
integration of geodesic equations~\cite{Capozziello2021,Sultan2025,Cataldo2017}. The
stability and convergence behavior of Pad\'{e} approximants, including the role of
spurious poles and the limitations of higher-order approximations, have been analyzed in
the mathematical literature and carry direct implications for their use in wormhole
modeling~\cite{Stahl1998,Beardon1968,Suetin2002}. The utility of Pad\'{e} approximants
in cosmography~\cite{Capozziello2020,Gruber2014} and in $f(R)$ and $f(R,T)$ wormhole
frameworks~\cite{Capozziello2021,Sultan2025} further motivates their systematic
deployment in torsion-based gravitational theories.

Previous investigations have examined wormhole solutions in various modified gravity settings
and have separately studied Pad\'{e} approximants as tools for shape function construction
or cosmographic modeling. Shadow formation and photon sphere structure in wormhole
spacetimes have been studied within general relativity and selected curvature-based
modified gravity frameworks~\cite{Shaikh2019jcap,Guerrero2022,DeFalco2021,Bronnikov2019}.
Pad\'{e}-based shape functions have been applied in $f(R)$ and $f(R,T)$ gravity to
improve analytical control of wormhole geometries~\cite{Capozziello2021,Sultan2025}.
Charged wormhole geometries with photon sphere analysis and energy condition investigations
have been conducted within the $f(R,T)$ framework~\cite{Godani2019a,Samanta2019b}, and
wormhole solutions in teleparallel-inspired models have been reported in the
literature~\cite{Nazavari2023,Saleem2023torsion}.

More recently, wormhole solutions supported by quantum-inspired matter sources have also received attention within modified gravity theories. In particular, Yukawa-corrected Casimir wormholes have been investigated in $f(R)$ gravity, while generalized uncertainty principle (GUP)-corrected Casimir wormholes have been explored in extended teleparallel gravity with matter coupling~\cite{Venkatesha2024Casimir,Chalavadi2024GUP}. These studies demonstrate the continued interest in exploring alternative matter sources and quantum corrections in the construction of physically viable wormhole geometries.
Nevertheless, no prior study has
integrated the $f(T)$ gravity framework with Pad\'{e}-approximated metric functions to
perform a combined analysis that encompasses null geodesic structure, determination of the photon sphere
 by ray tracing, shadow imaging, and emission of accretion disks with intensity
mapping. This gap in the literature leaves the observational signatures of $f(T)$
wormholes quantitatively unexplored and the utility of Pad\'{e} approximation within
torsion-based gravity unaddressed.

The present work addresses these open questions by constructing traversable wormhole solutions within $f(T)$ gravity and applying Pad\'{e} approximants to the relevant metric functions to obtain physically consistent and analytically tractable forms. The null geodesic equations are derived and analyzed to determine the structure of the photon sphere, with the effective potential examined as a function of the Pad\'{e} approximant parameters and the $f(T)$ coupling. Ray tracing techniques are employed to map null trajectories in the wormhole spacetime and compute the resulting shadow images, and the bound photon trajectories are further investigated through the rosette orbit structure. The study further incorporates a relativistic accretion disk model around the wormhole throat, from which specific intensity maps are produced to characterize the luminosity profile as seen by a distant observer. We additionally examine the dynamical stability of the constructed wormhole configurations through a quasinormal mode (QNM) analysis, computing the characteristic ringdown frequencies under scalar perturbations using the WKB approximation~\cite{SchutzWill1985,IyerWill1987,Konoplya2003} and studying their dependence on the Pad\'{e} parameters and the $f(T)$ coupling. Quasinormal modes encode the response of the wormhole to external perturbations and serve as a robust probe of dynamical stability and a potential observational discriminator between wormholes and black holes~\cite{Kokkotas1999,Berti2009,Konoplya2011,Cardoso2016,Churilova2019,Bronnikov2021}. Together, these analyses provide a comprehensive theoretical framework for the observational phenomenology of $f(T)$ wormholes and establish Pad\'{e} approximation as a practical and powerful tool within torsion-based modified gravity.

The outline of the present paper is as follows:
 Sec.~\ref{sec:gravity} sets up the wormhole
construction in teleparallel $f(T)$ gravity, and Sec.~\ref{sec:pade} derives the
$[1/0]$ and $[0/1]$ Pad\'{e} shape functions and their source terms.
Sec.~\ref{sec:ec} assesses the physical viability of both configurations, while
Sec.~\ref{sec:obsanalysis} examines their observational signatures. The
quasinormal modes and linear stability are studied in Sec.~\ref{sec:qnm}, and we
conclude in Sec.~\ref{sec:conclusion}.

\section{Wormhole Construction in Teleparallel \texorpdfstring{$f(T)$}{f(T)} Gravity}
\label{sec:gravity}
In teleparallel gravity the dynamical variable is the tetrad field
$e^{A}{}_{\mu}$, related to the metric by
$g_{\mu\nu} = \eta_{AB}\,e^{A}{}_{\mu} e^{B}{}_{\nu}$, with
$\eta_{AB} = \mathrm{diag}(-1,1,1,1)$, inverse $e_{A}{}^{\mu}$ obeying
$e^{A}{}_{\mu} e_{A}{}^{\nu} = \delta^{\nu}_{\mu}$, and
$e = \det(e^{A}{}_{\mu}) = \sqrt{-g}$. Capital Latin indices label the tangent
frame, Greek indices the spacetime coordinates. The curvature-free
Weitzenb\"{o}ck connection ($\hat{\Gamma}^{\lambda}{}_{\nu\mu}$) defines the torsion tensor,
\begin{equation}\label{eq1}
T^{\lambda}{}_{\mu\nu} = \hat{\Gamma}^{\lambda}{}_{\nu\mu} - \hat{\Gamma}^{\lambda}{}_{\mu\nu} = e^{\lambda}_{A}\left(\partial_{\mu}e^{A}{}_{\nu} - \partial_{\nu}e^{A}{}_{\mu}\right),
\end{equation}
from which the torsion scalar is constructed as
\begin{equation}\label{eq2}
T = S_{\lambda}{}^{\mu\nu}T^{\lambda}{}_{\mu\nu},
\end{equation}
where $S_{\lambda}{}^{\mu\nu} = \frac{1}{2}\left(K^{\mu\nu}{}_{\lambda} + \delta^{\mu}_{\lambda}T^{\alpha\nu}{}_{\alpha} - \delta^{\nu}_{\lambda}T^{\alpha\mu}{}_{\alpha}\right)$ and $K^{\mu\nu}{}_{\lambda}$ is the contorsion tensor given by
\begin{equation}
    K^{\alpha}_{~~\mu\nu} = \frac{1}{2} T^{\alpha}_{~\mu\nu} + T_{(\mu\,~\nu)}^{~\,\alpha}
    = \frac{1}{2} g^{\alpha\lambda}\left( T_{\mu\lambda\nu} + T_{\nu\lambda\mu} + T_{\lambda\mu\nu} \right),
    \label{eq:contortion}
\end{equation}
The Einstein-Hilbert action for $f(T)$ gravity is
\begin{equation}
\mathcal{S} = \frac{1}{2}\int e\,f(T)\,d^{4}x + \mathcal{S}_{\mathrm{matter}},
\end{equation}
where $e = \det(e^{A}{}_{\mu}) = \sqrt{-g}$, and we use units with $8\pi G=c=1$ throughout. The variation with respect to the tetrad yields the field equations \cite{aldrovandi2013}
\begin{equation}
\begin{split}
&e^{-1}\partial_{\mu}\!\left(e\,e_{A}{}^{\rho}S_{\rho}{}^{\sigma\mu}\right)f_{T}
- e_{A}{}^{\lambda}T^{\rho}{}_{\mu\lambda}S_{\rho}{}^{\mu\sigma}f_{T}\\
&\quad + e_{A}{}^{\rho}S_{\rho}{}^{\sigma\mu}\,\partial_{\mu}(T)\,f_{TT}
+ \tfrac14\,e_{A}{}^{\sigma}f(T)
= \tfrac12\,e_{A}{}^{\rho}\,\mathcal{T}_{\rho}{}^{\sigma},
\end{split}
\label{eq:fT_field}
\end{equation}
where $f_{T} = df/dT$, $f_{TT} = d^{2}f/dT^{2}$, and
$\mathcal{T}_{\rho}{}^{\sigma}$ is the matter energy-momentum tensor, can be expressed as
\begin{equation}
\mathcal{T}_{\mu\nu}
=
-\frac{2}{\sqrt{-g}}
\frac{\delta \left(\sqrt{-g}\,\mathcal{L}_{m}\right)}
{\delta g^{\mu\nu}},
\end{equation}
The static, spherically symmetric Morris--Thorne metric is~\cite{Morris1988}
\begin{equation}
ds^{2} = -e^{2\Phi(r)}\,dt^{2} + \frac{dr^{2}}{1-b(r)/r} + r^{2}\left(d\theta^{2}+\sin^{2}\theta\,d\phi^{2}\right),
\label{eq:MTmetric}
\end{equation}
where $\Phi(r)$ is the redshift function and $b(r)$ is the shape function. The radial coordinate satisfies $r\in[r_{0},\infty)$, with $r_{0}$ the throat radius. For a traversable wormhole, $b(r)$ must satisfy:

\begin{itemize}
\item $b(r_{0})=r_{0}$ (throat condition),
\item $[b(r)-b'(r)r]/b^{2}>0$ (flare-out condition),
\item $b'(r_{0})\leq 1$ (derivative bound at throat),
\item $b(r)/r < 1$ for all $r>r_{0}$,
\item $b(r)/r\to 0$ as $r\to\infty$ (asymptotic flatness); this requirement is met by the $[0/1]$ configuration, whereas the $[1/0]$ configuration has a nonzero limit of $b(r)/r$ and is not asymptotically flat (see Sec.~\ref{sec:pade}).
\end{itemize}
In addition, $\Phi(r)$ must remain finite for all $r\geq r_{0}$ to prevent event horizon formation.

The diagonal tetrad compatible with Eq.~\eqref{eq:MTmetric} is
\begin{equation}
e^{A}{}_{\mu} = \mathrm{diag}\!\left(e^{\Phi},\;
\frac{1}{\sqrt{1-b/r}},\;r,\;r\sin\theta\right),
\label{eq:tetrad}
\end{equation}
for the above choice of tetrad, the non-vanishing torsion components~\eqref{eq1} are
\begin{equation}
T^{t}{}_{rt}=\Phi',\quad
T^{\theta}{}_{r\theta}=T^{\phi}{}_{r\phi}=\frac{1}{r},\quad
T^{\phi}{}_{\theta\phi}=\cot\theta,
\end{equation}
where, $'\equiv d/dr$. Therefore,
inserting Eq.~\eqref{eq:tetrad} into
Eq.~\eqref{eq1}--\eqref{eq2} gives
\begin{equation}
T(r) = -\frac{2}{r^{2}}\left(1-\frac{b}{r}\right)\bigl(1+2r\Phi'\bigr).
\label{eq:T_scalar}
\end{equation}

In this work, we adopt the linear $f(T)$ model,
\begin{equation}
f(T) = \alpha\,T + \beta,\qquad f_{T}=\alpha,\qquad f_{TT}=0,
\label{eq:linearmodel}
\end{equation}
where $\alpha$ and $\beta$ are the model parameters. With $f_{TT}=0$ the
derivative-coupling term in~\eqref{eq:fT_field} vanishes and, for the diagonal
tetrad~\eqref{eq:tetrad}, the off-diagonal field equation is satisfied
identically. The diagonal components reduce to the effective relations
\begin{equation}
\rho = \alpha\,G^{t}{}_{t} + \tfrac{\beta}{2},\quad
p_{r} = -\alpha\,G^{r}{}_{r} - \tfrac{\beta}{2},\quad
p_{t} = -\alpha\,G^{\theta}{}_{\theta} - \tfrac{\beta}{2},
\label{eq:eff_eqs}
\end{equation}
where $\mathcal{T}_{\mu}{}^{\nu}=\mathrm{diag}(-\rho,p_{r},p_{t},p_{t})$ is the
anisotropic source and $G^{\mu}{}_{\nu}$ is the mixed Einstein tensor of the
Levi-Civita connection. The sign of the geometric term
follows the teleparallel convention fixed by~\eqref{eq:T_scalar}. The nonzero
components of $G^{\mu}{}_{\nu}$ are
\begin{align}
G^{t}{}_{t} &= -\frac{b'}{r^{2}},
\label{eq:Gtt}\\[4pt]
G^{r}{}_{r} &= \frac{2}{r}\left(1-\frac{b}{r}\right)\Phi'
- \frac{b}{r^{3}},
\label{eq:Grr}\\[4pt]
G^{\theta}{}_{\theta} &=
\left(1-\frac{b}{r}\right)\!\left(\Phi''+\Phi'^{2}\right)
+ \frac{\Phi'}{r}\!\left[1-\tfrac12\!\left(b'+\frac{b}{r}\right)\right]
\nonumber\\
&\quad - \frac{rb'-b}{2r^{3}}.
\label{eq:Gthth}
\end{align}
Substituting~\eqref{eq:Gtt}--\eqref{eq:Gthth}
into~\eqref{eq:eff_eqs} yields the field equation for $f(T)$ gravity,
\begin{align}
\rho &= \frac{\beta}{2} - \frac{\alpha\,b'}{r^{2}},
\label{eq:rho_gen}\\[4pt]
p_{r} &= -\frac{\beta}{2}
- \alpha\!\left[\frac{2}{r}\!\left(1-\frac{b}{r}\right)\Phi'
- \frac{b}{r^{3}}\right],
\label{eq:pr_gen}\\[4pt]
p_{t} &= -\frac{\beta}{2}
- \alpha\Bigl[\left(1-\tfrac{b}{r}\right)\!\left(\Phi''+\Phi'^{2}\right)
\nonumber\\
&\quad + \frac{\Phi'}{r}\!\left(1-\tfrac12 b'-\frac{b}{2r}\right)
- \frac{rb'-b}{2r^{3}}\Bigr].
\label{eq:pt_gen}
\end{align}
The explicit forms follow once $b(r)$ is fixed in
(Sec.~\ref{sec:pade}). Throughout this study, we adopt the asymptotically flat redshift function $\Phi(r)=-M/r$, a horizon-free inverse-$r$ ansatz that has been employed in wormhole studies of photon trajectories, photon spheres, and shadow observables \cite{Vagnozzi2023,Errehymy2025}. This choice ensures that $\Phi(r)$ remains finite throughout the spacetime and vanishes asymptotically as $r\to\infty$, thereby preserving traversability and asymptotic flatness. For this form, $\Phi'(r)=M/r^2$ and $\Phi''(r)=-2M/r^3$, both of which decay at large distances.

\section{Pad\'{e} Approximation and Source}
\label{sec:pade}

The Pad\'{e} approximation is a technique for representing a function as the ratio of two polynomials. For a function $f(x)$ whose Taylor (Maclaurin) series expansion about $x=0$ is known, the corresponding Pad\'{e} approximant reproduces that series up to the highest possible order while expressing the function in rational form. A Pad\'{e} approximant of order $[N/M]$ is written as~\cite{Baker1961}
\begin{equation}
P^{N}_{M}(x) = \frac{\sum_{n=0}^{N} a_{n}\, x^{n}}{\sum_{n=0}^{M} b_{n}\, x^{n}}
= \frac{a_{0} + a x +  \cdots + a_{N} x^{N}}{b_{0} + b_{1} x + \cdots + b_{M} x^{M}},
\label{eq:pade_general}
\end{equation}
where $N$ and $M$ are the degrees of the numerator and denominator polynomials, respectively, and the coefficients $a_{n}$ and $b_{n}$ are real constants. Without loss of generality, one conventionally normalizes the leading denominator coefficient as $b_{0}=1$, which fixes the overall scale of the rational function and leaves $N+M+1$ independent coefficients to be determined.

These coefficients are obtained by demanding that the Taylor expansion of $P^{N}_{M}(x)$ agree with the Taylor series of the function $f(x)$ to as many orders as possible, i.e.
\begin{equation}
f(x) - P^{N}_{M}(x) = \mathcal{O}\!\left(x^{N+M+1}\right),
\label{eq:pade_matching}
\end{equation}

so that the first $N+M+1$ terms of the two expansions coincide.

In the present work, we employ the Pad\'{e} approximation to model the shape function $b(r)$ of the wormhole geometry. By expressing $b(r)$ in rational form, we obtain a flexible yet analytically tractable ansatz whose free coefficients can be fixed by imposing the standard physical requirements on the shape function---namely the throat condition, the flaring-out condition, and where the order of the approximant allows it, asymptotic flatness. In this paper, we specifically consider the two lowest-order approximants, the $[1/0]$ and $[0/1]$ orders, which represent the simplest non-trivial rational forms of $b(r)$ and are analyzed in detail in the following subsections.

\subsection{The \texorpdfstring{$[1/0]$}{[1/0]}-Order Pad\'{e} Approximant}
\label{subsec:pade10}

For the $[1/0]$ order, the numerator is linear and the denominator reduces to a
constant (with $b_0=1$), so that Eq.~\eqref{eq:pade_general} takes the simple
polynomial form
\begin{equation}
P_{1,0}(x) = a_0 + a_1 x .
\label{eq:pade10_def}
\end{equation}
This represents the lowest-order non-trivial approximant, i.e.\ a linear function
of the radial coordinate.

To construct the shape function in this order, we begin from the seed function \cite{Goswami2024}
\begin{equation}
b(r) = r_0 \left[ \log\!\left(\frac{r}{r_0}\right)
       + \coth(r_0)\tanh(r) \right]^{a}
     \equiv r_0\,\bigl[ F(r) \bigr]^{a},
\label{eq:seed}
\end{equation}
where $a$ is a free parameter and we have defined,
\begin{equation}
F(r) \equiv \log\!\left(\frac{r}{r_0}\right) + \coth(r_0)\tanh(r) .
\label{eq:Fdef}
\end{equation}
Since the $[1/0]$ approximant is linear in $r$, we expand the seed function to
the first order about the throat $r=r_0$. The Taylor expansion of $F(r)$ around
$r_0$ reads
\begin{equation}
F(r) = F(r_0) + F'(r_0)\,(r-r_0)
     + \tfrac{1}{2}F''(r_0)\,(r-r_0)^2 + \cdots .
\label{eq:Ftaylor}
\end{equation}

The seed function is adopted as a phenomenological ansatz rather than being derived from a specific matter model. Its choice is motivated by its smooth and analytically tractable form, with the logarithmic and hyperbolic-tangent terms providing a flexible radial dependence. In particular, the normalization \(F(r_0)=1\) ensures that the throat condition \(b(r_0)=r_0\) is satisfied automatically, while the parameter \(a\) controls the radial profile of the shape function. This makes the ansatz convenient for constructing low-order Padé approximants and investigating different wormhole geometries.Its main advantages are that the throat condition holds for any value of $a$, its Taylor expansion about $r_0$ is regular, which makes the Padé construction straightforward, and the throat slope $b'(r_0)=a\,[1+r_0\operatorname{cosech}(r_0)\operatorname{sech}(r_0)]$ is controlled by the single parameter $a$.

Next, we expand the power $\bigl[F(r)\bigr]^{a}$ about $F(r_0)=1$, retaining only
the linear term consistent with the $[1/0]$ order,
\begin{equation}
\bigl[F(r)\bigr]^{a} = 1 + a\bigl(F(r)-F(r_0)\bigr)
                     = 1 + a\bigl(F(r)-1\bigr) ,
\label{eq:Fpow}
\end{equation}
 Truncating Eq.~\eqref{eq:Ftaylor} at first
order gives $F(r)-1 = F'(r_0)(r-r_0)$, so that
\begin{equation}
\bigl[F(r)\bigr]^{a} = 1 + a\,F'(r_0)\,(r-r_0) .
\label{eq:Fpow_lin}
\end{equation}
Substituting Eq.~\eqref{eq:Fpow_lin} into the seed function~\eqref{eq:seed}
produces the shape function at the $[1/0]$ order,
\begin{eqnarray}
&& \hspace{0cm} b(r) = r_0\Bigl[ 1 + a\,F'(r_0)\,(r-r_0) \Bigr] .
\\
&& \hspace{-0.6cm}  b(r) = r_0 - a\,(r_0-r)\Bigl[\,1 + r_0\,\operatorname{cosech}(r_0)\,
       \operatorname{sech}(r_0)\,\Bigr]~~~ .
\label{eq:b_pade10}
\end{eqnarray}
%


By construction this form satisfies the throat condition $b(r_0)=r_0$, as is
evident by setting $r=r_0$ in Eq.~\eqref{eq:b_pade10}. 

\begin{figure}[!htbp]
    \centering
    \includegraphics[width=0.7\columnwidth]{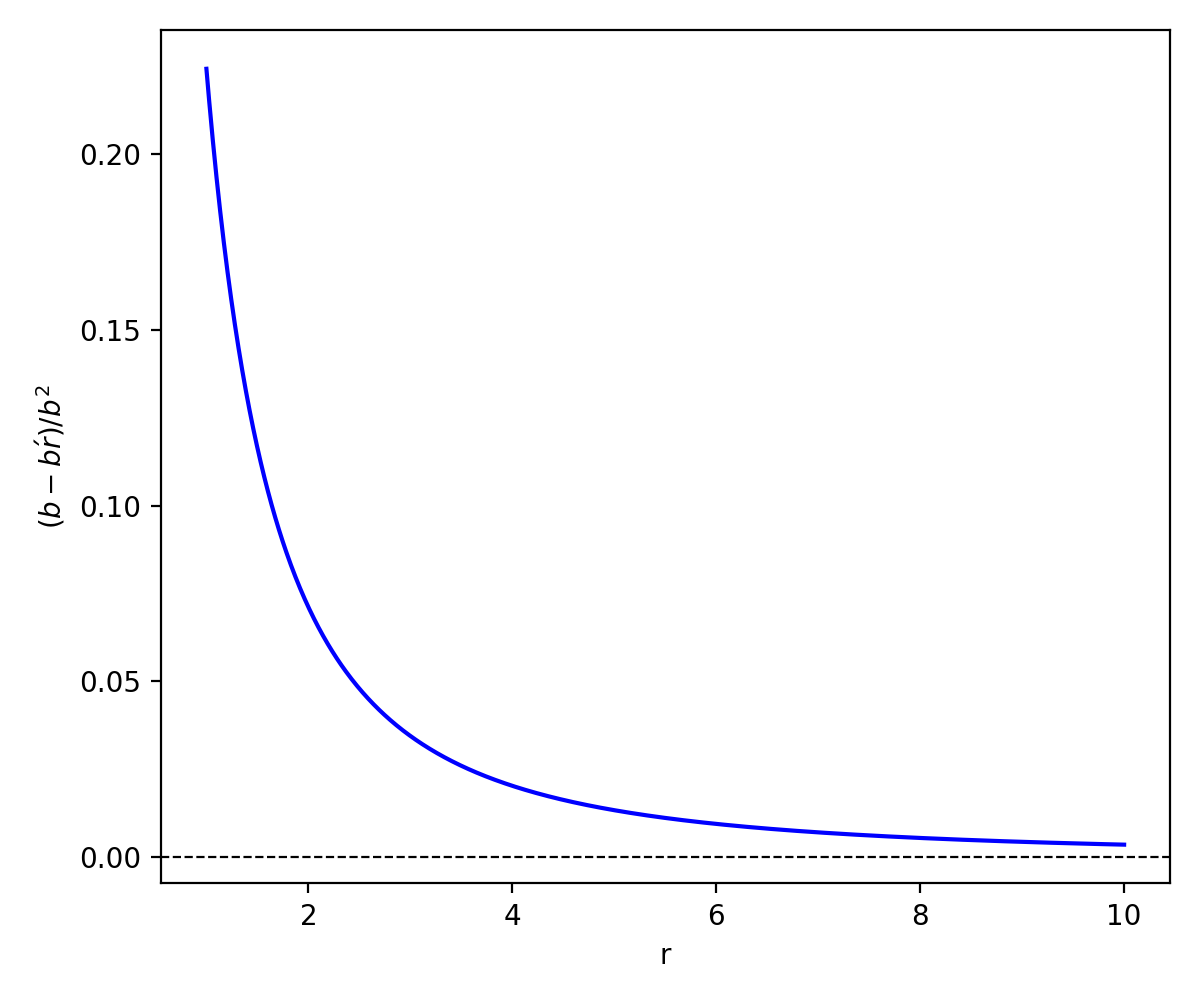}
    \caption{The flaring-out quantity $\bigl[b(r)-b'(r)\,r\bigr]/b^{2}$ for the
    $[1/0]$-order shape function.}
    \label{fig:flareOutPade10}
\end{figure}

 We now verify the geometric
requirements with the parameters $r_0=1$, $a=0.5$, $M=1.5$, and $\alpha=\beta=2$. The flaring-out
condition, expressed through the combination
$\bigl[b(r)-b'(r)\,r\bigr]/b^{2}>0$, ensures that the wormhole opens outward in
the vicinity of the throat. As shown in Fig.~\ref{fig:flareOutPade10}, this
quantity remains strictly positive over the entire radial domain, decreasing
monotonically from its maximal value at the throat and tending to zero at large
$r$, so that the flaring-out condition is satisfied. At the throat itself the
derivative of the shape function obeys the bound $b'(r_0)\leq 1$; for the
present case we find $b'(r_0)\simeq 0.776$, which lies safely below unity and is
consistent with the throat being a minimum of the embedded geometry.

\begin{figure}[!htbp]
    \centering
    \includegraphics[width=0.7\columnwidth]{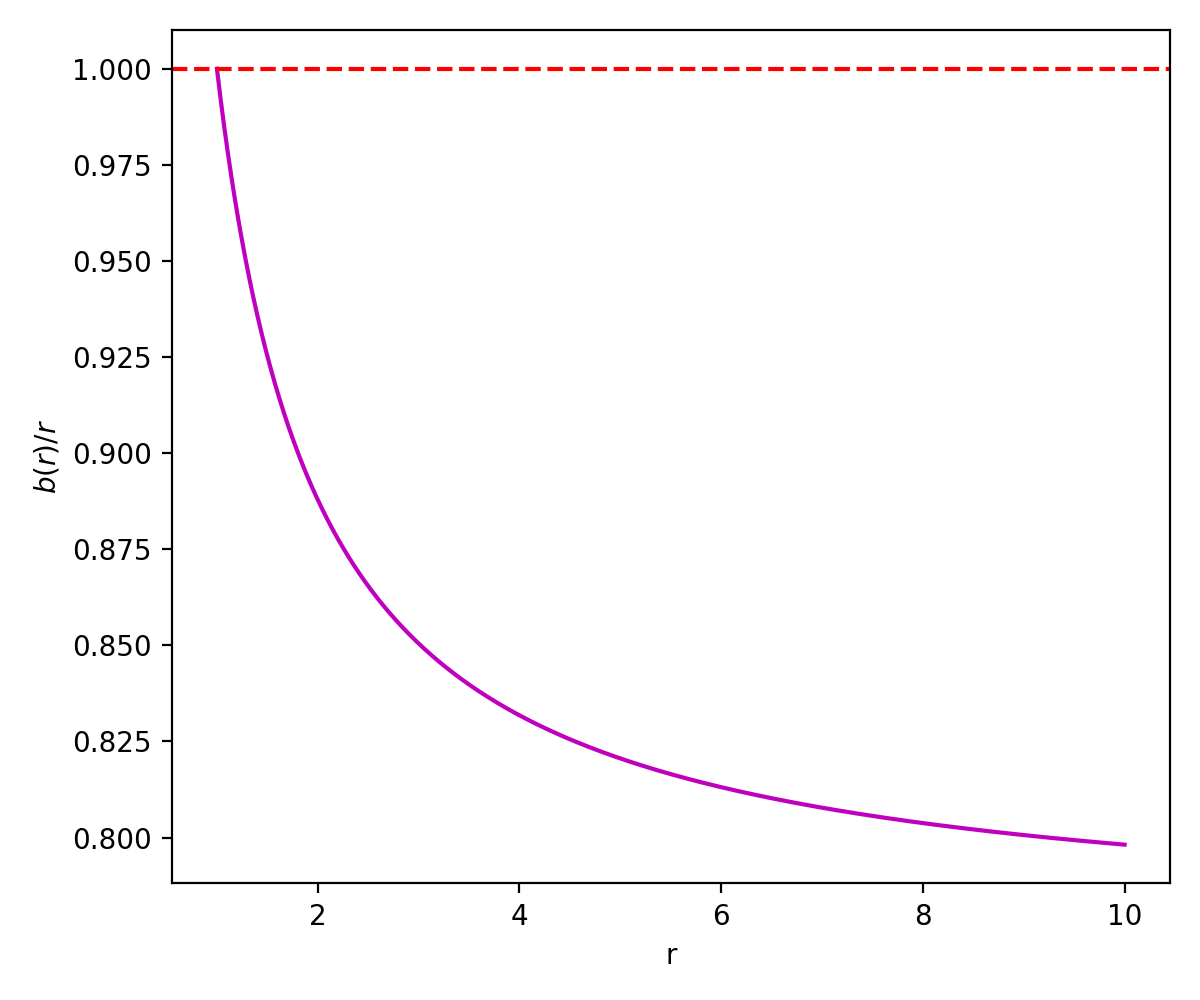}
    \caption{The ratio $b(r)/r$ for the $[1/0]$-order shape function.}
    \label{fig:bOverRPade10}
\end{figure}

The final constraint, $b(r)/r\to 0$ as $r\to\infty$, warrants more careful
analysis. As shown in Fig.~\ref{fig:bOverRPade10}, the ratio $b(r)/r$ starts
from unity at the throat and decreases monotonically thereafter, staying
strictly below $1$ for all $r>r_0$. However, since the $[1/0]$ shape function grows linearly in $r$, the ratio does not vanish at large $r$. Instead, it approaches the finite constant $\lim_{r\to\infty}b(r)/r=a\,[1+r_0\,\operatorname{cosech}(r_0)\,\operatorname{sech}(r_0)]\simeq 0.776$ for the chosen parameters, so the $[1/0]$ configuration is not asymptotically flat. This residual value does not signal a pathology: the linear growth of $b(r)$ endows the geometry with a solid-angle deficit at large scales, indicating that the wormhole behaves as a topological defect of the global-monopole type rather than as an asymptotically flat spacetime. Accordingly, the asymptotic-flatness requirement listed in Sec.~\ref{sec:gravity} is not imposed on this branch; throughout this work the $[1/0]$ solution is treated as a wormhole with a solid-angle deficit, and the term asymptotically flat is reserved for the $[0/1]$ configuration.

For this above $b(r)$ and the redshift function $\Phi(r)$ mentioned in section-\ref{sec:gravity}, the field Eqs. (\ref{eq:rho_gen}-\ref{eq:pt_gen}) reduces to,

\begin{eqnarray}
\rho &=& \frac{\beta}{2}
-\frac{a\alpha\left(1+r_{0}\csch(r_{0})\sech(r_{0})\right)}{r^{2}},
\label{eq:rho}
\\[2mm]
p_{r} &=& -\frac{\beta}{2}
+\frac{\alpha}{r^{4}}
\Bigl[
2(a-1)M(r-r_{0})
+r\left(ar+r_{0}-ar_{0}\right)
\nonumber\\
&&\qquad
+a(2M+r)(r-r_{0})r_{0}\csch(r_{0})\sech(r_{0})
\Bigr],
\label{eq:pr}
\\[2mm]
p_{t} &=& \frac{1}{2r^{5}}
\Bigl[\alpha(a-1)\mathcal{Q}-\beta r^{5}
+\alpha ar_{0}\mathcal{Q}\csch(r_{0})\sech(r_{0})\Bigr].
\label{eq:pt}
\end{eqnarray}
Where, $
\mathcal{Q}
=
2M^{2}(r-r_{0})
+r^{2}r_{0}
+Mr(3r_{0}-2r).$



\subsection{The \texorpdfstring{$[0/1]$}{[0/1]}-Order Pad\'{e} Approximant}
\label{subsec:pade01}

For the $[0/1]$ order, the situation is reversed with respect to the previous
case: the numerator reduces to a constant while the denominator is linear in the
radial coordinate. After normalizing the leading denominator coefficient,
Eq.~\eqref{eq:pade_general} takes the rational form
\begin{equation}
P_{0,1}(z) = \frac{a_0}{1 + b_0 + b_1 z}
           = \frac{a_0}{b_0' + b_1 z}
           \approx \frac{1}{1 + c_1 z} ,
\label{eq:pade01_def}
\end{equation}
where $c_1$ collects the single independent coefficient that survives after
normalization.

We proceed from the same seed function and Taylor expansion employed in the
previous subsection. Truncating Eq.~\eqref{eq:Ftaylor} at first order about the
throat gives
\begin{equation}
F(r) \approx 1 + F'(r_0)\,(r-r_0) ,
\label{eq:F_lin_again}
\end{equation}
In contrast to the $[1/0]$ case, where
the power $[F(r)]^a$ was represented as a linear polynomial, here we recast it in
rational form so as to match the $[0/1]$ structure of
Eq.~\eqref{eq:pade01_def}. To this end we write
\begin{equation}
\bigl[F(r)\bigr]^{a} = \frac{1}{1 - a\,(r-r_0)\,F'(r_0)} .
\label{eq:Fpow_rational}
\end{equation}
The equivalence of this representation with the linear one, to the order
considered, follows from the geometric-series expansion
\begin{equation}
\frac{1}{1-x} = (1-x)^{-1} = 1 + x + x^2 + \cdots \approx 1 + x ,
\label{eq:geom}
\end{equation}
so that, identifying $x = a\,(r-r_0)\,F'(r_0)$,
\begin{equation}
\frac{1}{1 - a\,(r-r_0)\,F'(r_0)}
   \approx 1 + a\,(r-r_0)\,F'(r_0) .
\label{eq:rational_check}
\end{equation}
Both the $[1/0]$ and $[0/1]$ approximants thus share the same first-order Taylor
expansion, as required, but differ in their behavior away from the throat.

Substituting Eq.~\eqref{eq:Fpow_rational} into the seed
function~\eqref{eq:seed}, $b(r)=r_0\,[F(r)]^a$, gives the closed form of $[0/1]$ order shape function as,
\begin{eqnarray}
&&\hspace{-0.5cm} b(r) = \frac{r_0}{\displaystyle 1 - a\,(r-r_0)
       \left[\dfrac{1}{r_0} + \operatorname{cosech}(r_0)\,
       \operatorname{sech}(r_0)\right]} ,~~~~
\label{eq:Fprime_r0_again}
\end{eqnarray}

The two Padé configurations exhibit different asymptotic behaviours. For $a<0$ the $[0/1]$ shape function decays as $b(r)\sim r_0/\{|a|\,[1/r_0+\operatorname{cosech}(r_0)\,\operatorname{sech}(r_0)]\,r\}$ at large $r$, so that $\lim_{r\to\infty}b(r)/r=0$ and this configuration is asymptotically flat. In contrast, the $[1/0]$ configuration has $\lim_{r\to\infty}b(r)/r=a\,[1+r_0\,\operatorname{cosech}(r_0)\,\operatorname{sech}(r_0)]\neq 0$, leading to a solid-angle deficit. Accordingly, only the $[0/1]$ branch is referred to as asymptotically flat in this work.


Since we retain only the lowest-order Padé approximants, their reliability is primarily assessed through their behaviour near the throat and over the radial domain in which the shape functions remain regular and satisfy the required wormhole conditions. Their behaviour at large distances is examined separately, since low-order approximants need not reproduce the asymptotic behaviour of the seed function exactly. For the parameters used here, both approximants agree with the seed to within $10\%$ for $r\lesssim1.5\,r_0$; beyond this near-throat region they should be regarded as distinct shape functions rather than as approximations to the seed.

For a given $a$, every higher-order approximant $[N/M]$ of the seed reproduces the same values $b(r_0)=r_0$ and $b'(r_0)$. Since Eqs.~\eqref{eq:rho_gen}--\eqref{eq:pt_gen} involve only $b$ and $b'$, the throat and flaring-out conditions at $r_0$ and the values of $\rho$, $p_r$ and $p_t$ at the throat, and hence the NEC violation there, are unchanged at any order. The photon sphere at $r=M$ and the critical impact parameter $b_c=eM$ depend only on $\Phi(r)$, and the TOV equilibrium is satisfied identically for any $b(r)$ in the linear model. Higher-order terms can, however, modify the geometry away from the throat: they may introduce spurious poles, which must lie outside $r\ge r_0$, and they fix the asymptotic behaviour, with $b(r)/r\to0$ only for $N\le M$. Global properties such as the VIQ, the lensed disc images and the QNM spectrum may therefore change quantitatively, whereas the throat-level conclusions are robust.

\begin{figure}[!htbp]
    \centering
    \includegraphics[width=0.7\linewidth]{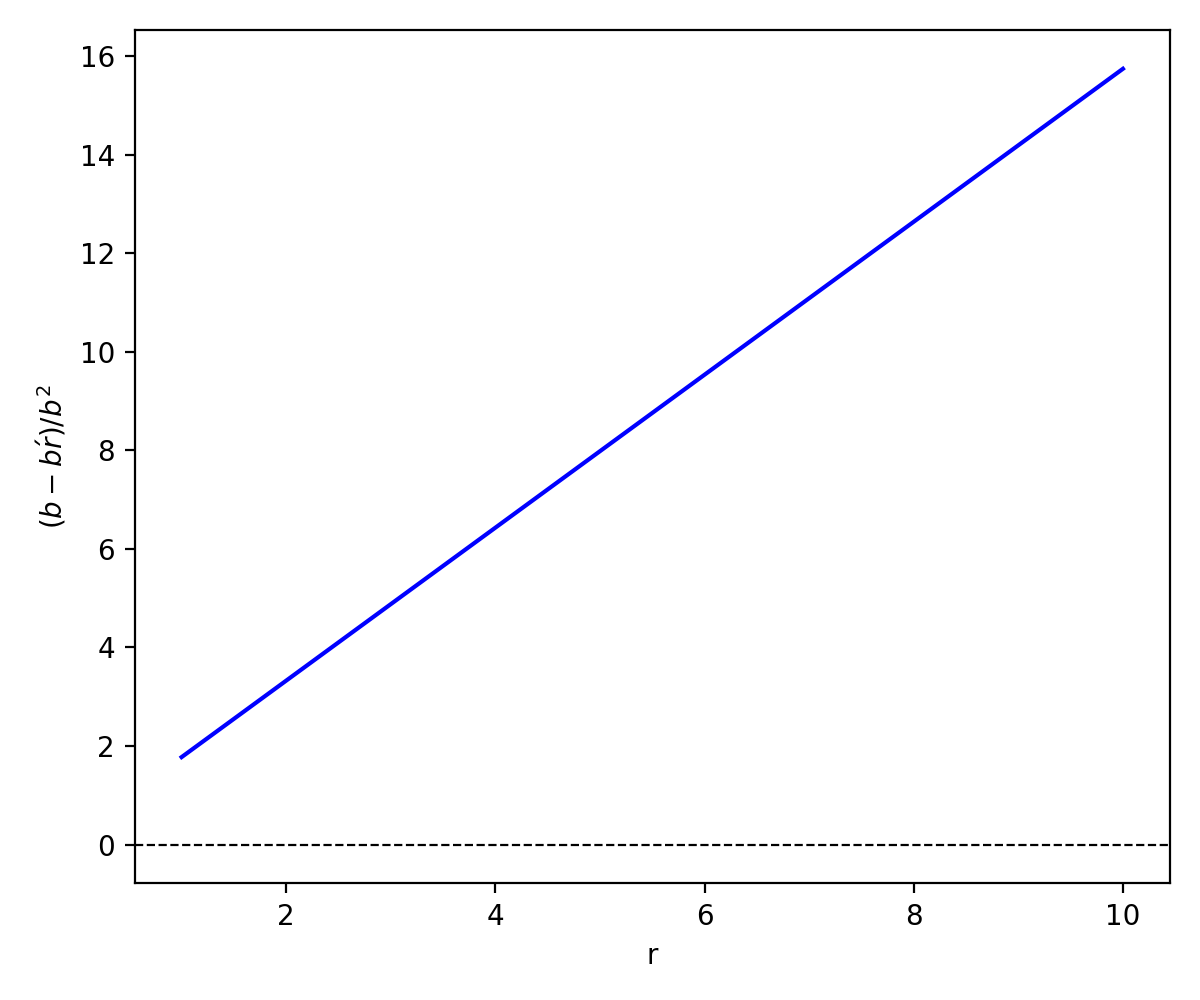}
    \caption{The flaring-out quantity $\bigl[b(r)-b'(r)\,r\bigr]/b^{2}$ for the
    $[0/1]$-order shape function.}
    \label{fig:flareOutPade01}
\end{figure}

\begin{figure}[!htbp]
    \centering
    \includegraphics[width=0.7\linewidth]{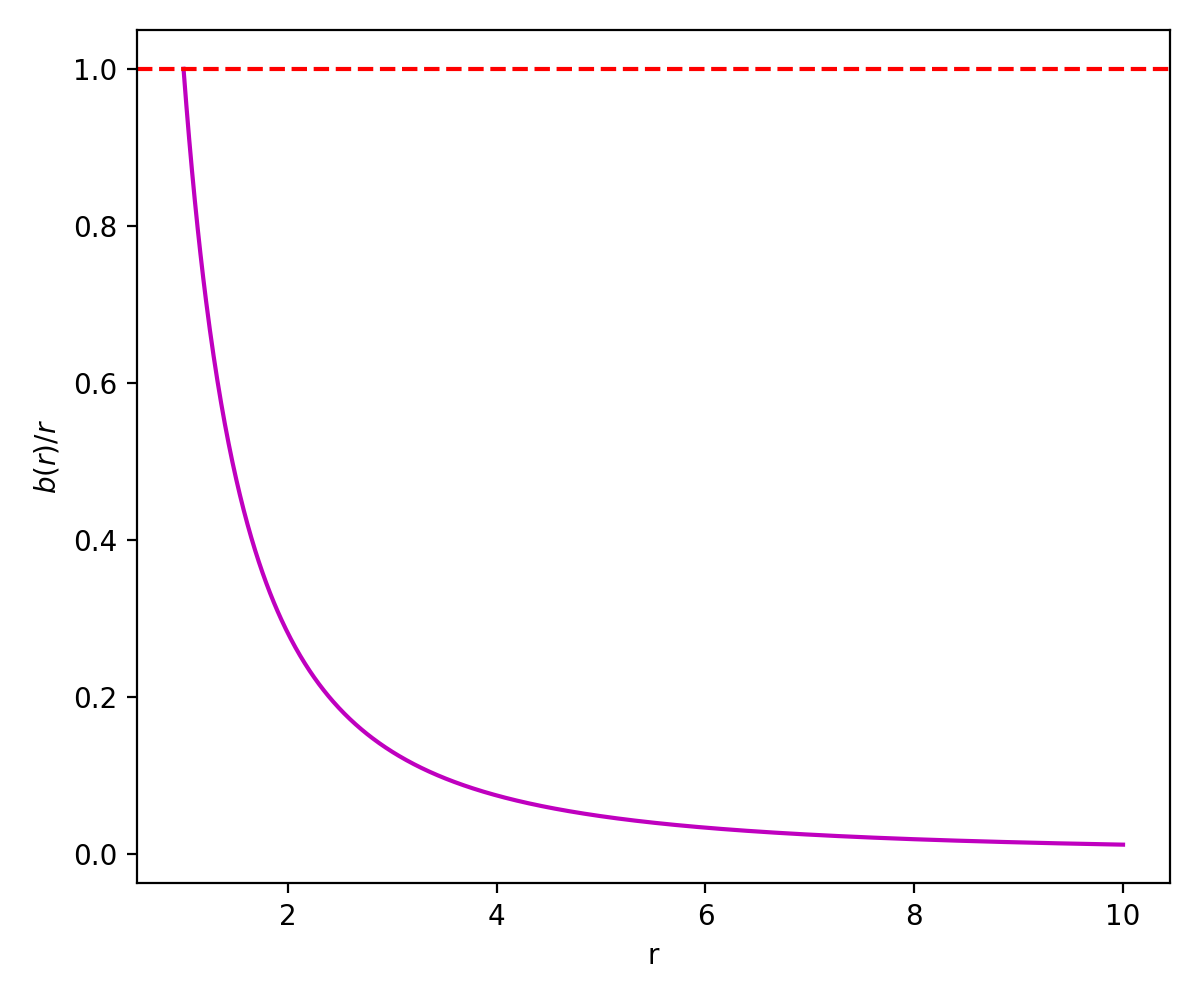}
    \caption{The ratio $b(r)/r$ for the $[0/1]$-order shape function.}
    \label{fig:bOverRPade01}
\end{figure}

We now examine the same geometric conditions with the parameters $r_0=1$, $a=-0.5$,
$M=1.5$, and $\alpha=\beta=2$. The flaring-out quantity
$\bigl[b(r)-b'(r)\,r\bigr]/b^{2}$ is shown in Fig.~\ref{fig:flareOutPade01},
where it remains strictly positive across the radial domain and grows steadily
with $r$, confirming that the flaring-out condition holds for this approximant.
The derivative of the shape function at the throat satisfies the bound
$b'(r_0)\leq 1$; here we obtain $b'(r_0)\simeq -0.776$, which is negative and
therefore comfortably below unity, reflecting the monotonic decay of the
$[0/1]$ shape function away from the throat. Lastly, the ratio $b(r)/r$ is
plotted in Fig.~\ref{fig:bOverRPade01}, where it begins at unity at the throat
and decreases monotonically, remaining strictly below $1$ for all $r>r_0$ and
approaching zero at large $r$.

As in the $[1/0]$ case, this expression satisfies the throat condition
$b(r_0)=r_0$, which is recovered immediately upon setting $r=r_0$ in the above
equation. Similarly, for the above $b(r)$ and the redshift function $\Phi(r)$ mentioned in section-\ref{sec:gravity}, the field Eqs. (\ref{eq:rho_gen}-\ref{eq:pt_gen}) reduces to,
\begin{eqnarray}
\rho &=& \frac{\beta}{2}
-\frac{a\alpha
\left(1+r_{0}\csch(r_{0})\sech(r_{0})\right)}
{r^{2}D^{2}},
\label{eq:rho01}
\\[2mm]
p_{r} &=& -\frac{\beta}{2}
+\frac{\alpha}{r^{3}}
\left[
\frac{r_{0}}{D}
-2M
+\frac{2Mr_{0}}{rD}
\right],
\label{eq:pr01}
\\[2mm]
p_{t} &=& -\frac{\beta}{2}
-\frac{\alpha}{2r^{3}}
\Biggl[
\frac{2M(M-r)}{r}
+\left(1+\frac{3M}{r}
-\frac{2M^{2}}{r^{2}}\right)\frac{r_{0}}{D}
\nonumber\\
&&\qquad
-\frac{a(M+r)
\left(1+r_{0}\csch(r_{0})\sech(r_{0})\right)}
{D^{2}}
\Biggr],
\label{eq:pt01}
\end{eqnarray}
where we have defined
\[
D\equiv 1-a\,(r-r_{0})\Big[\frac{1}{r_{0}}+\csch(r_{0})\sech(r_{0})\Big].
\]

\section{Physical Analysis}
\label{sec:ec}
\subsection{Energy Condition}
\begin{figure}[!htbp]
\centering
\includegraphics[width=0.95\columnwidth]{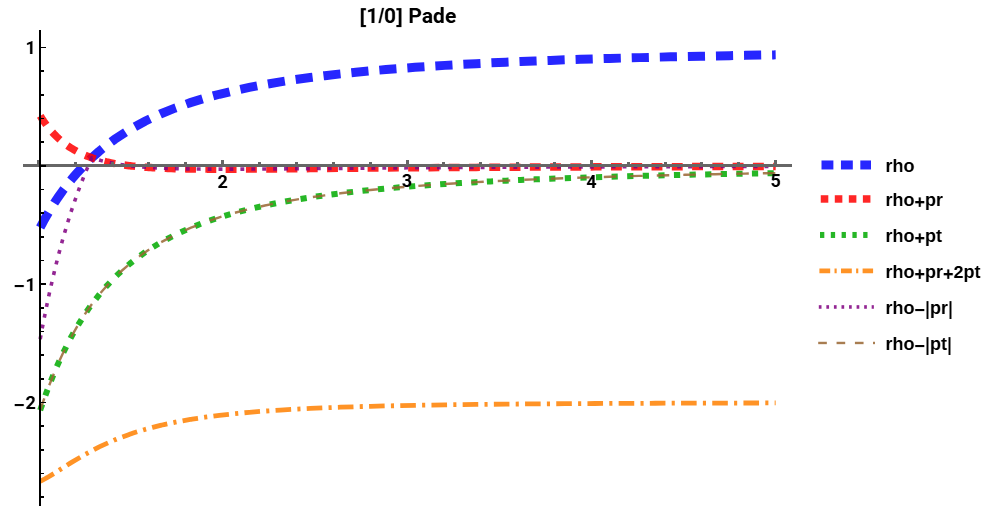}
\caption{Graphical analysis of the energy-conditions for the
$[1/0]$ Pad\'{e} shape function ($M=1.5$, $r_{0}=1$, $\alpha=2$, $\beta=2$,
$a=0.5$).}
\label{fig:ec10}
\end{figure}

\begin{figure}[!htbp]
\centering
\includegraphics[width=0.95\columnwidth]{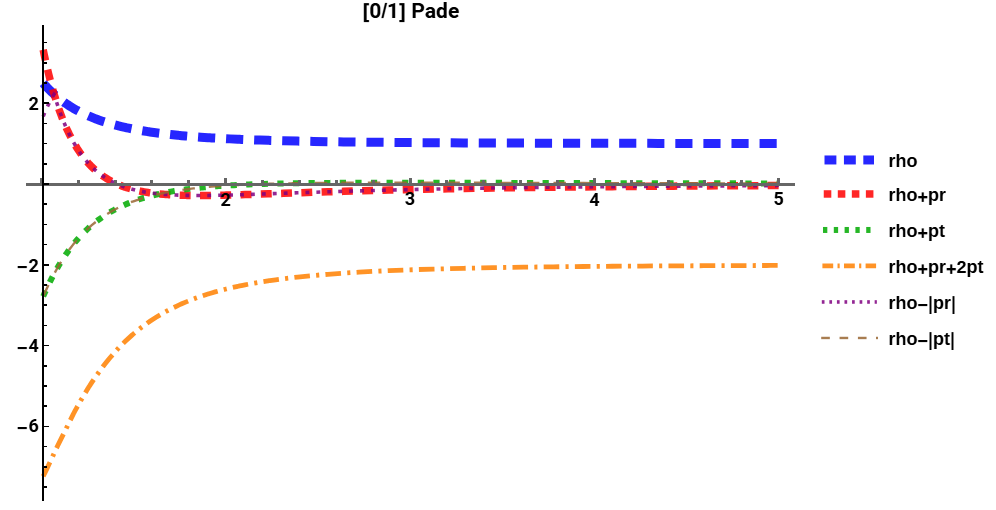}
\caption{Graphical analysis of the energy-condition combinations for the
$[0/1]$ Pad\'{e} shape function ($M=1.5$, $r_{0}=1$, $\alpha=2$, $\beta=2$,
$a=-0.5$).}
\label{fig:ec01}
\end{figure}

 The classical point wise energy
conditions for such a anisotropic matter source are expressed through the
combinations~\cite{Visser1995,Lobo2005}
\begin{align}
\text{NEC:}&\quad \rho+p_{r}\geq0,\qquad \rho+p_{t}\geq0,
\label{eq:NEC}\\[1mm]
\text{WEC:}&\quad \rho\geq0,\quad \rho+p_{r}\geq0,\quad \rho+p_{t}\geq0,
\label{eq:WEC}\\[1mm]
\text{SEC:}&\quad \rho+p_{r}\geq0,\quad \rho+p_{t}\geq0,\quad
            \rho+p_{r}+2p_{t}\geq0,
\label{eq:SEC}\\[1mm]
\text{DEC:}&\quad \rho\geq0,\quad \rho-|p_{r}|\geq0,\quad \rho-|p_{t}|\geq0.
\label{eq:DEC}
\end{align}
The null energy condition~\eqref{eq:NEC} is the weakest of the four and
controls the existence of the throat. For an anisotropic source the NEC
splits into two independent channels, the radial combination $\rho+p_{r}$
and the tangential combination $\rho+p_{t}$, and the NEC is violated whenever
either channel turns negative~\cite{Soni2026,Hassan2022,Lobo2005}. For the Morris--Thorne geometry
the flaring-out condition of Sec.~\ref{sec:gravity} forces at least one of
these null combinations negative in a neighbourhood of $r_{0}$, so a violation
of the NEC at the throat is the defining signature of the exotic matter
required for traversability. The energy conditions corresponding to the
$[1/0]$ and $[0/1]$ order approximants are illustrated in
Fig.~\ref{fig:ec10} and Fig.~\ref{fig:ec01}, respectively.

From Fig.~\ref{fig:ec10}, the energy density at the throat $r=r_{0}=1$ is
negative, $\rho=-0.5514$, and the tangential null combination is also negative,
$\rho+p_{t}=-2.1121$, while the radial null combination is positive,
$\rho+p_{r}=+0.4486$. Since one of the two null channels is negative, the
NEC~\eqref{eq:NEC} is violated at the throat through the tangential channel,
which suffices to identify the matter as exotic. The energy density $\rho$
increases monotonically away from the throat, changes sign in the near-throat
region, and approaches its asymptotic value $\rho\to\beta/2=1$ as $r\to\infty$.
The tangential combination $\rho+p_{t}$ remains negative across the entire
exterior and approaches zero from below, so the tangential NEC stays violated
everywhere and is only marginally restored at spatial infinity, whereas the
radial combination $\rho+p_{r}$ stays positive throughout. With $\rho<0$ at the
throat the WEC~\eqref{eq:WEC} fails as well, and the negative values
$\rho-|p_{r}|=-1.5514$ and $\rho-|p_{t}|=-2.1121$ show that the
DEC~\eqref{eq:DEC} also fails. The strong-energy combination
$\rho+p_{r}+2p_{t}=-2.6728$ at the throat and remains negative for all
$r\geq r_{0}$, so the SEC~\eqref{eq:SEC} is violated throughout.

For the $[0/1]$ order approximant, the energy density at the throat is large and
positive, $\rho=+2.5514$, the radial null combination is positive,
$\rho+p_{r}=+3.5514$, while the tangential null combination is negative,
$\rho+p_{t}=-2.8788$. The negative tangential channel again drives a violation
of the NEC~\eqref{eq:NEC} at the throat. Although $\rho>0$ here, the negative
$\rho+p_{t}$ means the WEC~\eqref{eq:WEC} still fails through its tangential
condition, and the values $\rho-|p_{r}|=+1.5514$ and $\rho-|p_{t}|=-2.8788$
show that the DEC~\eqref{eq:DEC} fails in the tangential channel. The
strong-energy combination is strongly negative at the throat,
$\rho+p_{r}+2p_{t}=-7.3272$, and remains negative across the exterior, so the
SEC~\eqref{eq:SEC} is violated as well.

For both Padé configurations, the radial null combination \(\rho+p_r\) remains positive throughout the exterior region \(r\geq r_0\), indicating that the radial NEC is satisfied. In contrast, the tangential combination \(\rho+p_t\) is negative at the throat for both branches and is responsible for the NEC violation. In the \([1/0]\) case, it remains negative throughout the exterior, approaching zero from below as \(r\to\infty\). The SEC combination, \(\rho+p_r+2p_t\), also remains negative throughout the exterior for both configurations. Thus, the violation of the energy conditions is primarily associated with the tangential pressure rather than the radial pressure. Although the energy density is positive at the throat for the \([0/1]\) branch, the negative tangential NEC still leads to WEC violation. These results highlight the role of pressure anisotropy in sustaining the wormhole geometries.
\subsection{Tolman--Oppenheimer--Volkoff Equilibrium}
\label{sec:tov}

The Tolman--Oppenheimer--Volkoff (TOV) equation~\cite{Oppenheimer1939}
governs the gravitational equilibrium of a static, spherically symmetric
matter distribution and provides a standard diagnostic of the stability of
wormhole solutions~\cite{Gorini2008}. For an anisotropic source, the
generalized TOV equation reads~\cite{Kuhfittig2013}
\begin{equation}
\frac{\nu'}{2}\bigl(\rho+p_r\bigr)
+ \frac{dp_r}{dr}
+ \frac{2}{r}\bigl(p_r-p_t\bigr) = 0,
\label{eq:tov}
\end{equation}
where $\nu(r)=2\Phi(r)$ is fixed by the redshift function. The three terms
represent the gravitational, hydrostatic, and anisotropic forces,
\begin{equation}
F_G = -\Phi'\bigl(\rho+p_r\bigr),
\quad
F_H = -\frac{dp_r}{dr},
\quad
F_A = \frac{2}{r}\bigl(p_t-p_r\bigr),
\label{eq:forces}
\end{equation}
and equilibrium requires their sum to vanish,
\begin{equation}
F_H + F_G + F_A = 0.
\label{eq:balance}
\end{equation}

Unlike wormhole studies adopting a constant redshift function, where
$\Phi'=0$ removes $F_G$ entirely, the present work employs the
asymptotically flat redshift $\Phi(r)=-M/r$ of Sec.~\ref{sec:gravity}, for
which $\Phi'(r)=M/r^{2}\neq0$. The gravitational force is therefore active
throughout, and all three contributions in Eq.~\eqref{eq:balance} must be
retained.

For the linear model $f(T)=\alpha T+\beta$ adopted here, $f_{TT}=0$ and the
theory is dynamically equivalent to the teleparallel equivalent of general
relativity supplemented by a constant. The effective source is then
identically conserved $\nabla_{\mu}\mathcal{T}^{\mu}{}_{\nu}=0$, by virtue
of the contracted Bianchi identity, so no additional modified-gravity force
$F_M$ arises and the standard anisotropic balance~\eqref{eq:balance} holds
exactly. Moreover, the constant $\beta$ enters $\rho$, $p_r$, and $p_t$ only
as an additive shift $\pm\beta/2$ that cancels in every combination in
Eq.~\eqref{eq:forces}; the equilibrium forces are thus independent of
$\beta$ and are controlled by $\alpha$, $M$, $r_0$, and the Pad\'{e}
parameter alone.

\begin{figure}[!htbp]
\centering
\includegraphics[width=0.8\columnwidth]{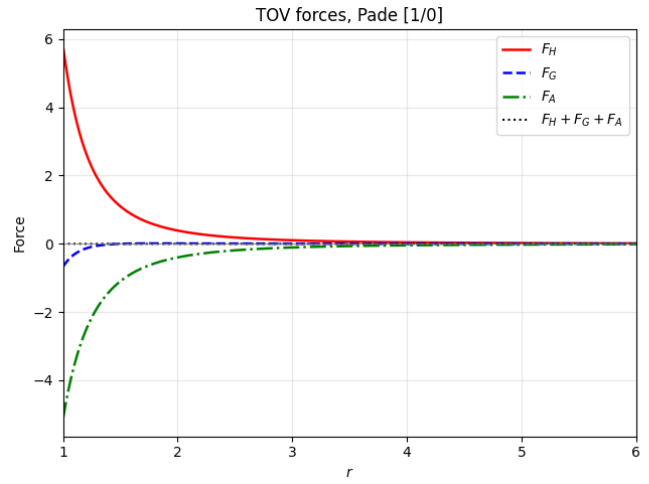}
\caption{Different kind of forces as functions of $r$ for the $[1/0]$ Pad\'{e}
wormhole, with $\alpha=2$, $\beta=2$, $M=1.5$, $r_0=1$, and $a=0.5$. The
dotted curve shows the sum $F_H+F_G+F_A$, which vanishes throughout,
confirming equilibrium.}
\label{fig:tov_left}
\end{figure}

\begin{figure}[!htbp]
\centering
\includegraphics[width=0.8\columnwidth]{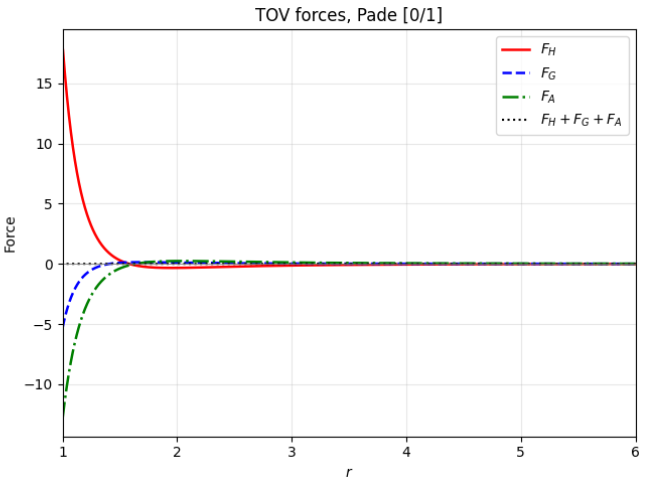}
\caption{Different kind of forces as functions of $r$ for the $[0/1]$ Pad\'{e}
wormhole, with $\alpha=2$, $\beta=2$, $M=1.5$, $r_0=1$, and $a=-0.5$.}
\label{fig:tov_right}
\end{figure}

The behaviour of the three forces is shown in Fig.~\ref{fig:tov_left} and
Fig.~\ref{fig:tov_right} for the $[1/0]$ and $[0/1]$ orders, with
$\alpha=2$, $\beta=2$, $M=1.5$, $r_0=1$, and the Pad\'{e} parameter fixed at
$a=0.5$ and $a=-0.5$ respectively. In both cases the hydrostatic force
$F_H$ is positive and directed outward, while the gravitational and
anisotropic forces $F_G$ and $F_A$ are negative and directed inward; their
sum vanishes identically throughout the spacetime, confirming that
condition~\eqref{eq:balance} is satisfied. The forces are largest near the
throat and decay rapidly at large $r$, where the geometry approaches flat
space. The two Pad\'{e} orders produce qualitatively identical equilibrium
structures, differing only quantitatively in the near-throat region: the
$[0/1]$ branch develops substantially larger force magnitudes at the throat
than the $[1/0]$ branch, consistent with its steeper shape function. This
exact cancellation demonstrates that the wormhole configurations obtained in
both Pad\'{e} orders are in stable hydrostatic equilibrium under the combined
action of the hydrostatic, gravitational, and anisotropic forces.

\subsection{Volume Integral Quantifier}
\label{subsec:VIQ}

The Volume Integral Quantifier (VIQ) serves as a crucial tool to measure the
total exotic matter required to maintain a traversable wormhole. For
spherically symmetric wormholes, it is defined as
\begin{equation} \label{eq:VIQ}
    I_V = \int_{r_0}^{R} (\rho + p_r)\, 8\pi r^2\, dr,
\end{equation}
where $\rho$ is the energy density, $p_r$ the radial pressure, $r_0$ the throat
radius, and $R$ a finite cutoff radius introduced to regularize the integral.
This quantity effectively captures the average violation of the NEC necessary
for wormhole stability~\cite{Balani2026}. While powerful, the VIQ may diverge
unless properly regularized, typically through junction conditions that impose a
cutoff radius; notably, wormholes with conformal symmetry or supported by
phantom energy can achieve arbitrarily small VIQ values, potentially connecting
wormhole physics to cosmic acceleration scenarios~\citep{Lobo2005}.

\begin{figure}[!htbp]
    \centering
    \includegraphics[width=1\linewidth]{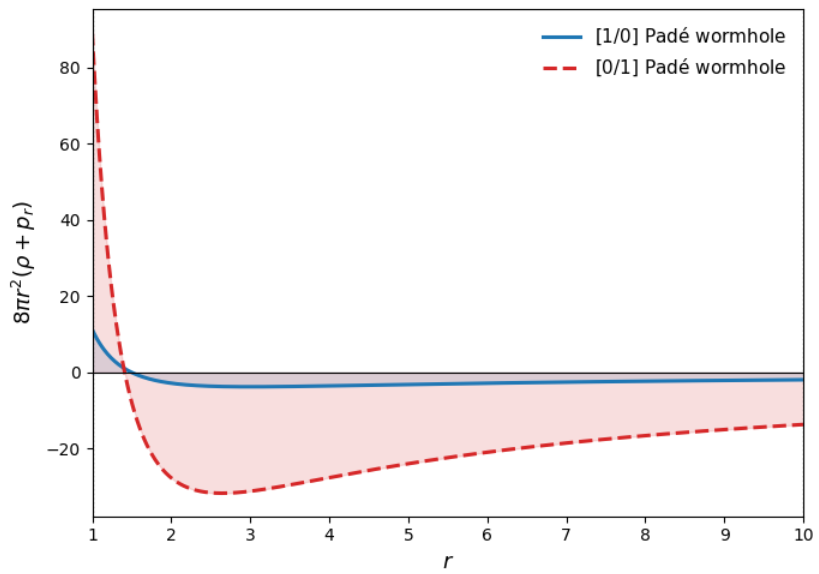}
    \caption{Radial profile of the VIQ integrand, $8\pi r^2(\rho+p_r)$,
    with $r_0=1$, $M=1.5$, $\alpha=\beta=2$, $a=0.5$, and $a=-0.5$. The
    shaded regions denote the integration domain from the throat to the cutoff
    radius $R=10r_0=10.0$.}
    \label{fig:VIQplot}
\end{figure}

We fix the throat radius at $r_0=1$ and plot the
integrand $8\pi r^2(\rho+p_r)$ as a function of $r$ for both wormhole solutions
in Fig.~\ref{fig:VIQplot}.

The integration domain runs from the throat $r_0=1$ to the cutoff radius
$R=10r_0=10.0$, chosen according to the junction-condition formalism, where the
interior geometry is matched to an exterior Schwarzschild geometry (Sec.~\ref{sec:junction}) at the boundary
where the influence of the exotic core has become negligible~\citep{Tayde2023}.
For both solutions, the integrand is largest in magnitude within a few throat radii and decays as $1/r$ at larger distances, as discussed below. The numerical integration of
Eq.~\eqref{eq:VIQ} yields
\begin{equation}
\label{eq:VIQvals}
I_V^{[1/0]} \approx -21.477963,
\qquad
I_V^{[0/1]} \approx -171.910047.
\end{equation}
Both results are finite and negative, confirming that NEC-violating (exotic)
matter is required to support each geometry within the linear
$f(T)=\alpha T+\beta$ model.

To examine the sensitivity of the VIQ to the integration boundary, we evaluated $I_V(R)$ for several cutoff radii with all model parameters fixed; the results are listed in Table~\ref{tab:VIQ_R}. The integrand is positive only in a thin layer next to the throat ($r\lesssim1.5\,r_0$ for $[1/0]$ and $r\lesssim1.41\,r_0$ for $[0/1]$), so $I_V(R)$ is small and positive for $R\simeq2r_0$ and becomes increasingly negative for larger $R$. At large $r$ the integrand decays only as $8\pi r^2(\rho+p_r)\simeq-C/r$, with $C=-8\pi\alpha\,[2(a-1)M+(1-a)r_0+a\,r_0(2M-r_0)\,\operatorname{cosech}(r_0)\,\operatorname{sech}(r_0)]\simeq22.5$ for $[1/0]$ and $C=16\pi\alpha M\simeq150.8$ for $[0/1]$; in the latter case the tail arises entirely from the redshift term $\Phi^{\prime}$ in $p_r$. Consequently $I_V(R)\simeq-C\ln R+\mathrm{const}$ grows logarithmically and does not converge as $R\to\infty$. The finite values in Eq.~\eqref{eq:VIQvals} therefore characterize the matched interior region $r_0\le r\le R$ only: the total amount of exotic matter is finite because the geometry is joined to the exterior Schwarzschild geometry at $r=R$ (Sec.~\ref{sec:junction}), beyond which $\rho+p_r=0$ and no further contribution arises, and it depends logarithmically on the choice of matching radius.

\begin{table}[!htbp]
\centering
\caption{Volume integral quantifier $I_V(R)$ for different cutoff radii $R$ ($r_0=1$, $M=1.5$, $\alpha=\beta=2$, $a=\pm0.5$).}
\label{tab:VIQ_R}
\begin{tabular}{ccc}
\hline
$R/r_0$ & $I_V^{[1/0]}$ & $I_V^{[0/1]}$ \\
\hline
2 & $1.28$ & $3.67$ \\
5 & $-9.23$ & $-82.34$ \\
10 & $-21.48$ & $-171.91$ \\
20 & $-35.42$ & $-269.45$ \\
50 & $-55.06$ & $-403.59$ \\
100 & $-70.35$ & $-506.80$ \\
\hline
\end{tabular}
\end{table}

As seen in Fig.~\ref{fig:VIQplot}, both
integrands are positive in a narrow region just outside the throat
($r_0\le r\lesssim 1.4$) before turning negative; the $[1/0]$ integrand then
saturates at a small, slowly decaying negative tail, so that $I_V^{[1/0]}$ is
dominated by this broad, shallow negative contribution. By contrast, the
$[0/1]$ integrand develops a deep negative minimum just outside the throat
($r\approx 2.7$) before relaxing slowly toward zero; the much larger magnitude
$|I_V^{[0/1]}|\gg|I_V^{[1/0]}|$ reflects this pronounced near-throat
concentration of exotic matter, consistent with the steeper fall-off of the
$[0/1]$ shape function. In both cases $I_V$ is finite over the matched interior region, so that, once the geometry is matched to the exterior vacuum at $r=R$, a bounded amount of NEC-violating matter suffices to sustain the wormhole throat for the chosen parameters; its value, however, depends on the matching radius (Table~\ref{tab:VIQ_R}).

\subsection{Junction Conditions and Thin Shell}
\label{sec:junction}

The interior geometry~\eqref{eq:MTmetric} is joined at $r=R$ to the exterior Schwarzschild geometry
\begin{equation}
ds^{2}_{\rm ext}=-\kappa^{2}\Big(1-\frac{2M_s}{r}\Big)dt^{2}+\Big(1-\frac{2M_s}{r}\Big)^{-1}dr^{2}+r^{2}d\Omega^{2},
\label{eq:ext}
\end{equation}
where $M_s$ is the exterior mass and the constant $\kappa$ rescales the exterior time. Although the linear form $f(T)=\alpha T+\beta$ may induce an
effective cosmological-constant contribution, we adopt the
Schwarzschild geometry as an approximate exterior spacetime for
the matching analysis. This approximation is valid when the
contribution of the effective cosmological constant is negligible
on the length scales considered, i.e.,
$|\Lambda_{\rm eff}|R^2/3\ll 1$, where $R$ is the matching
radius. A more general matching to a Schwarzschild--(anti-)de
Sitter exterior would be an interesting extension of the present
analysis. For the linear model, $G^{\mu}{}_{\nu}=0$ in Eq.~\eqref{eq:eff_eqs} leaves only the uniform term $\rho=-p_r=-p_t=\beta/2$, which acts as an effective cosmological-constant contribution with $\rho+p_r=\rho+p_t=0$, so the exterior does not contain no NEC-violating matter. Since $f_{TT}=0$, the junction conditions of $f(T)$ gravity reduce to the Darmois--Israel conditions of general relativity, with the coupling set by $\alpha$~\cite{Israel1966,delaCruzDombriz2014,VelayVitow2017}; a similar thin-shell matching for wormholes in modified gravity was carried out in Ref.~\cite{Balani2026}.

The first junction condition requires the induced metric on the hypersurface $\Sigma$ ($r=R$), $ds^{2}_{\Sigma}=-d\tau^{2}+R^{2}d\Omega^{2}$, to be continuous, which fixes $\kappa^{2}(1-2M_s/R)=e^{2\Phi(R)}$. The non-vanishing components of the extrinsic curvature on the interior ($-$) and exterior ($+$) sides are
\begin{align}
K^{\tau\,-}_{\ \tau}&=\Phi'(R)\sqrt{1-\tfrac{b(R)}{R}}, &
K^{\theta\,-}_{\ \theta}&=\frac{1}{R}\sqrt{1-\tfrac{b(R)}{R}},\nonumber\\
K^{\tau\,+}_{\ \tau}&=\frac{M_s/R^{2}}{\sqrt{1-2M_s/R}}, &
K^{\theta\,+}_{\ \theta}&=\frac{1}{R}\sqrt{1-\tfrac{2M_s}{R}}.
\label{eq:Kab}
\end{align}
The surface stress-energy tensor on the junction hypersurface
is determined by the Israel junction condition,
\begin{equation}
S_{ab}=-\frac{1}{8\pi}
\left([K_{ab}]-h_{ab}[K]\right),
\label{eq:israel}
\end{equation}
where $[K_{ab}]=K^+_{ab}-K^-_{ab}$ and
$[K]=h^{ab}[K_{ab}]$.
For a static, spherically symmetric shell, we write
$S^a{}_b=\mathrm{diag}(-\sigma,P,P)$.
The surface energy density and tangential pressure are then
\begin{align}
\sigma&=-\frac{1}{4\pi}[K^\theta{}_\theta],
\label{eq:sigma}\\
P&=\frac{1}{8\pi}
\left([K^\tau{}_\tau]+[K^\theta{}_\theta]\right).
\label{eq:pressure}
\end{align}
As an additional matching choice, we impose continuity of the radial metric component at the junction surface, which gives \(F(R)=1-b(R)/R\). For a Schwarzschild exterior this reduces to \(M_s=b(R)/2\).
 This condition ensures \([K^\theta{}_\theta]=0\) and consequently a vanishing surface energy density, \(\sigma=0\), within the adopted junction formalism.  
The matching therefore requires a thin shell with vanishing surface energy density and nonzero surface pressure. For $R=10r_0$ and the parameters of Sec.~\ref{sec:pade}, we find $b(R)\simeq7.98$, $M_s\simeq3.99$ and $P\simeq-0.164$ (a surface tension) for $[1/0]$, and $b(R)\simeq0.125$, $M_s\simeq0.063$ and $P\simeq0.029$ for $[0/1]$; in both cases $R>2M_s$, so no horizon forms. The VIQ defined in Sec.~\ref{subsec:VIQ} accounts only for the
bulk matter contribution in the interior region and does not
include the distributional stress-energy of the thin shell,
which is treated separately through the junction conditions.
Although the shell has vanishing surface energy density,
its nonzero tangential pressure contributes to its surface
stress tensor.

\subsection{Embedding Surface and Diagrams}
\label{subsec:embedding}
Following Morris and Thorne~\cite{Morris1988}, the spatial geometry of the
wormhole is visualized by embedding the equatorial slice $t=\text{const}$,
$\theta=\pi/2$ of the metric~\eqref{eq:MTmetric} into a three-dimensional
Euclidean space with cylindrical coordinates $(r,\phi,z)$, whose line element
is $ds^{2}=dz^{2}+dr^{2}+r^{2}d\phi^{2}$. Matching this to the induced
two-geometry of the slice yields the embedding function $z(r)$, satisfying
\begin{equation}
\frac{dz}{dr} = \pm\frac{1}{\sqrt{\dfrac{r}{b(r)}-1}},
\label{eq:dzdr}
\end{equation}
whose slope diverges at the throat, $dz/dr\to\infty$ as $r\to r_0$, so that the
embedded surface is vertical there---the defining geometric signature of a
wormhole throat. Integrating outward from the throat,
\begin{equation}
z(r) = \pm\int_{r_0^{+}}^{r}\frac{dr'}{\sqrt{\dfrac{r'}{b(r')}-1}},
\label{eq:zr}
\end{equation}
gives the two symmetric sheets $z(r)$ and $-z(r)$ that join smoothly at $r=r_0$
and connect the two asymptotic regions, which are asymptotically flat for the $[0/1]$ configuration and possess a solid-angle deficit for the $[1/0]$ configuration. The integral for both Pad\'{e} order and is evaluated numerically.

\begin{figure*}[tp]
\centering
\includegraphics[width=\textwidth]{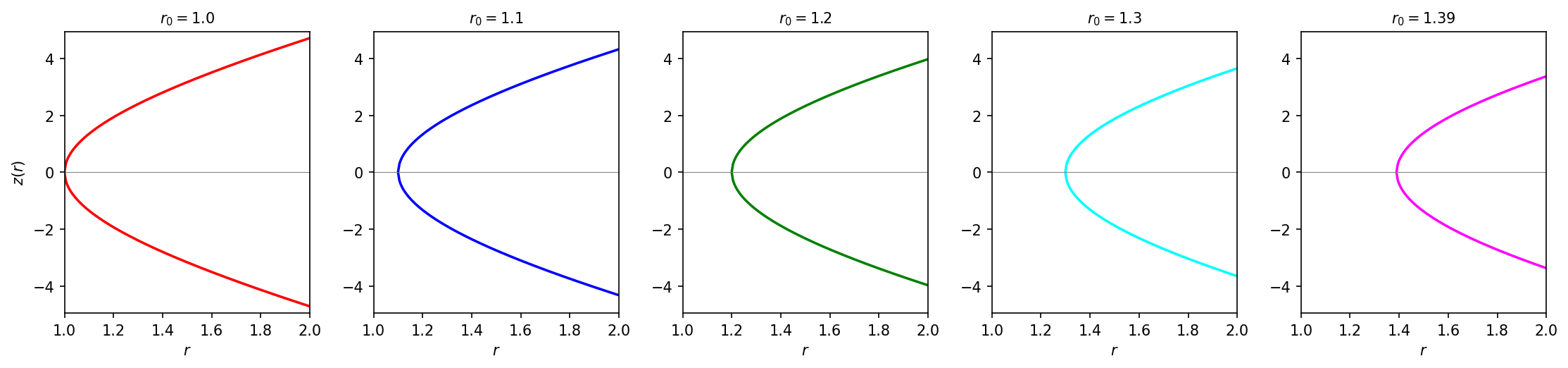}\\[4pt]
\includegraphics[width=\textwidth]{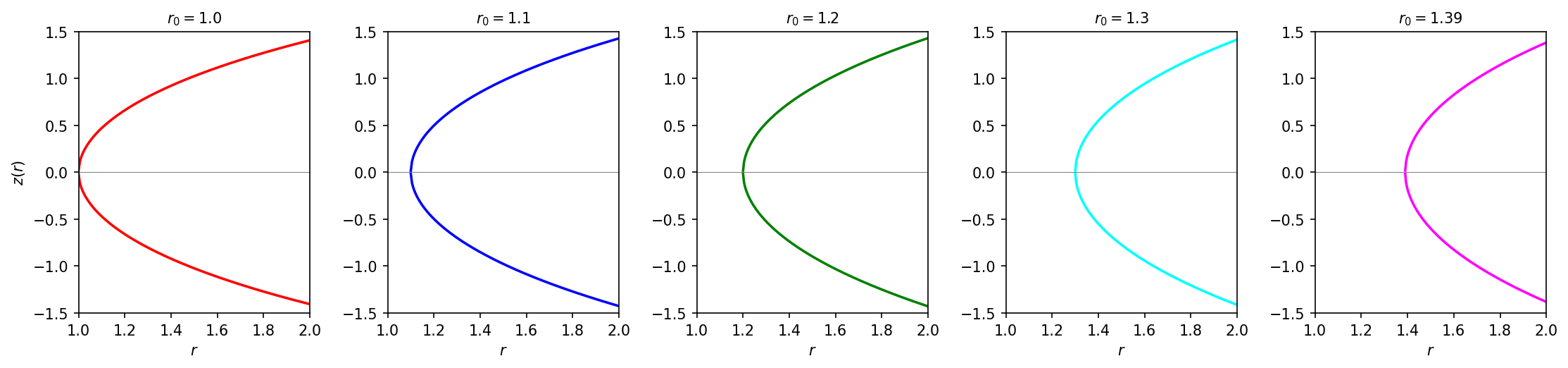}
\caption{Embedding diagrams $z(r)$ of the wormhole for several throat radii
$r_0$, shown for the $[1/0]$ (top row) and $[0/1]$ (bottom row) Pad\'{e} shape
functions.}
\label{fig:embed2D}
\end{figure*}

\begin{figure*}[tp]
\centering
\includegraphics[width=\textwidth]{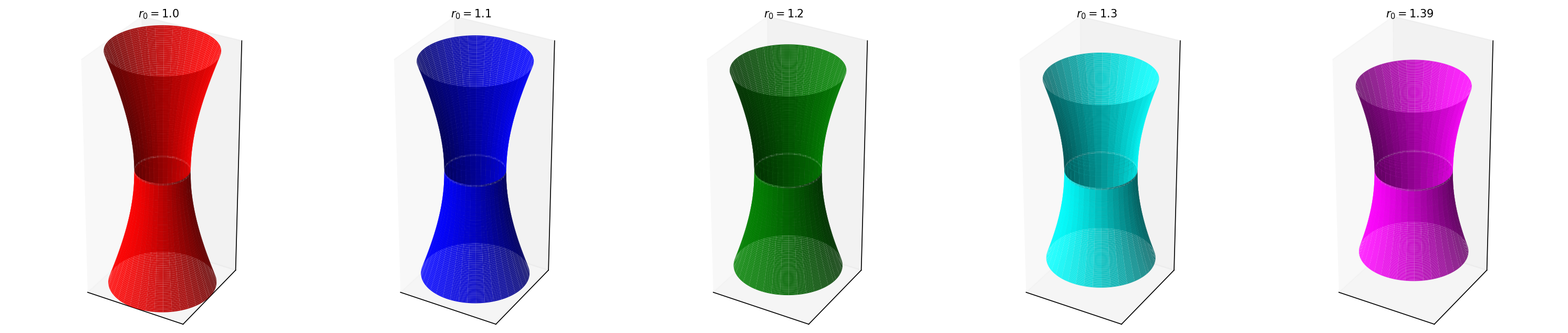}\\[4pt]
\includegraphics[width=\textwidth]{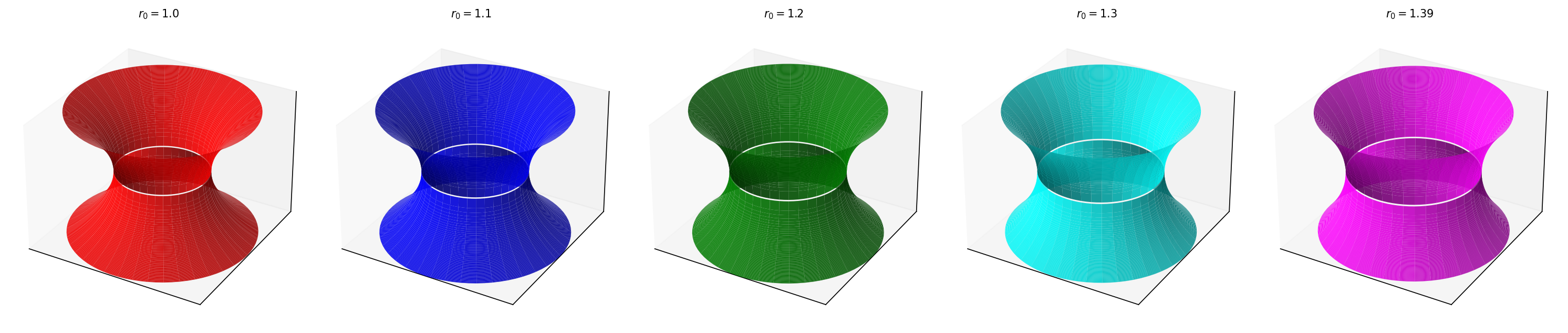}
\caption{Full three-dimensional visualization of the wormhole, obtained by
revolving the embedding profiles of Fig.~\ref{fig:embed2D} about the $z$-axis,
for the $[1/0]$ (top row) and $[0/1]$ (bottom row) Pad\'{e} shape functions and
several throat radii $r_0$.}
\label{fig:embed3D}
\end{figure*}

The resulting profiles for the $[1/0]$ and $[0/1]$ shape functions, computed for
several throat radii $r_0\in\{1.00,\allowbreak 1.10,\allowbreak 1.20,\allowbreak 1.30,\allowbreak 1.39\}$ with $a=0.5$ and
$a=-0.5$, are shown in Fig.~\ref{fig:embed2D} (with both the $+z$ and $-z$
sheets plotted) and tabulated in Table~\ref{tab:combined}. Each profile is
vertical at its own throat and flares outward monotonically, in agreement with
the flaring-out condition imposed in Sec.~\ref{sec:pade}. The full embedded
surfaces, obtained by revolving these profiles about the $z$-axis through
$\phi\in[0,2\pi)$, are displayed in Fig.~\ref{fig:embed3D}. They exhibit the
characteristic wormhole morphology: a narrow circular throat of radius $r_0$
that widens smoothly into two flared mouths. Both Pad\'{e} orders produce
qualitatively the same throat geometry, differing in the rate of flaring away
from the throat: the linear $[1/0]$ shape function yields a more steeply rising
profile reaching $|z|\simeq4.7$ at $r=2$, whereas the pole-free $[0/1]$ shape
function flares more gradually, reaching only $|z|\simeq1.4$ over the same
range, consistent with the slower growth of its shape function.

\subsection{Proper Radial Distance}
\label{subsec:proper_distance}
\begin{figure}[!htbp]
\centering
\includegraphics[width=0.95\columnwidth]{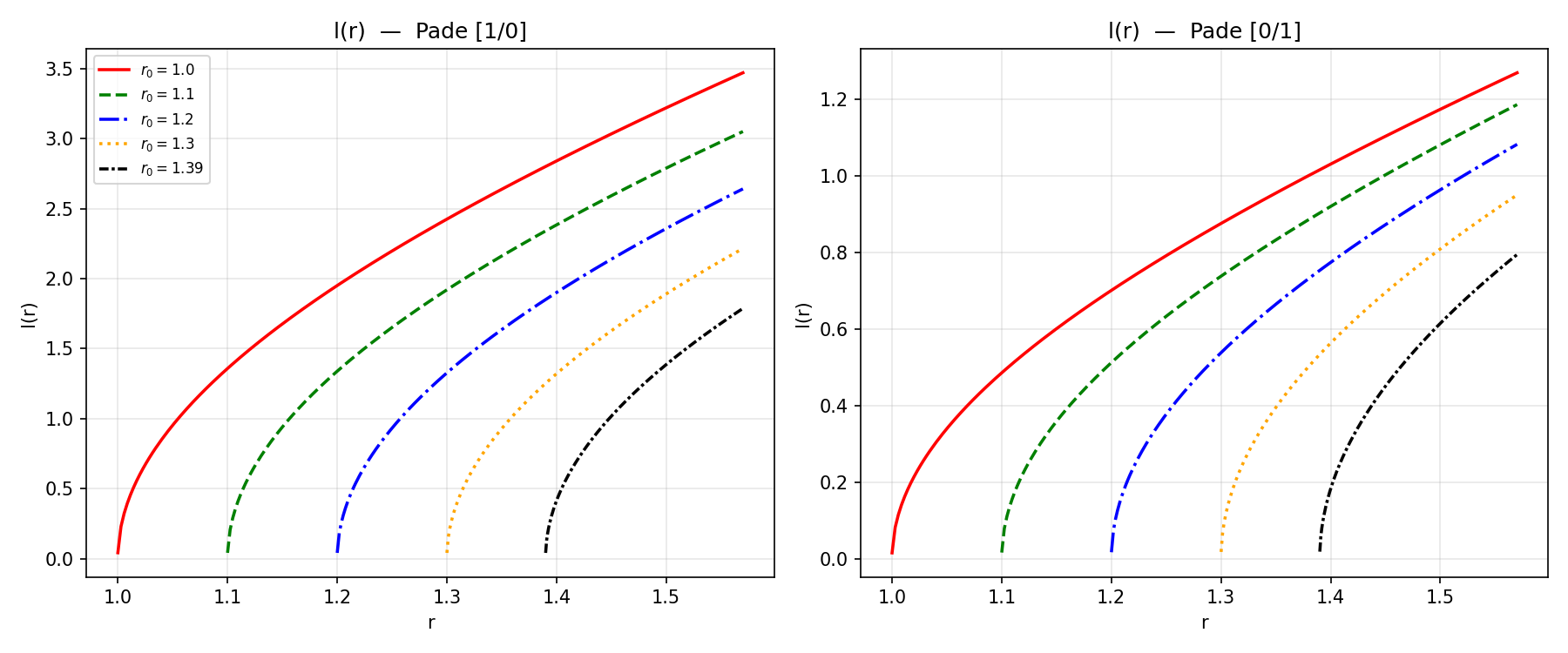}
\caption{Proper radial distance $l(r)$ against $r$ for the $[1/0]$ and
$[0/1]$ Pad\'{e} functions, for several throat radii $r_0$. Each
curve rises monotonically from $l=0$ at its own throat.}
\label{fig:lr}
\end{figure}

The proper radial distance measures the physical distance from the throat to a
point $r$ as measured by a static observer, and must remain finite and positive
everywhere outside the throat for the wormhole to be traversable. It is given by
\begin{equation}
l(r) = \pm\int_{r_0^{+}}^{r}\frac{dr'}{\sqrt{1-\dfrac{b(r')}{r'}}},
\label{eq:lr}
\end{equation}
with the plus (minus) sign corresponding to the upper (lower) universe connected
by the wormhole. As with $z(r)$, the integral cannot be solved analytically for
the Pad\'{e} shape functions of Sec.~\ref{sec:pade} and is evaluated
numerically. Since $b(r)/r<1$ holds throughout for $a=0.5$ and $a=-0.5$, the
integrand remains real and finite for all $r>r_0$, with an integrable inverse
square-root singularity at the throat itself. The resulting profiles of $l(r)$
are shown in Fig.~\ref{fig:lr} and tabulated alongside $z(r)$ and $E_g$ in
Table~\ref{tab:combined}. The $[1/0]$ profiles rise more steeply than the
$[0/1]$ profiles at fixed $r_0$, reaching $l\simeq3.5$ against $l\simeq1.3$ near
$r=1.57$ for $r_0=1$, reflecting the slower growth of the pole-free $[0/1]$
shape function.

\subsection{Gravitational Energy and Active Mass}
\label{subsec:Eg_mass}
Two integral diagnostics built from the energy density characterize the
mass-energy content of the wormhole. The first is the total gravitational
energy $E_g$, defined via the Lynden-Bell--Katz--Nandi
prescription~\cite{Nigmatzyanov2009,Katz2006,Nandi2009} as the difference
between the total energy $Mc^2$ and the total mechanical energy $E_M$,
$E_g = Mc^2 - E_M$, which reduces to
\begin{equation}
E_g = \frac{1}{2}\int_{r_0^{+}}^{r}\Bigl[1-\sqrt{g_{rr}}\Bigr]\rho\,r'^{2}\,dr'
+ \frac{r_0}{2},
\qquad g_{rr}=\Bigl(1-\tfrac{b(r)}{r}\Bigr)^{-1}.
\label{eq:Eg}
\end{equation}
The second is the active gravitational mass enclosed between the throat and a
radius $r$~\cite{Lobo2009,Cataldo2017},
\begin{equation}
M_{\rm active} = \int_{r_0^{+}}^{r} 4\pi\,\rho\,r'^{2}\,dr'.
\label{eq:Mactive_def}
\end{equation}
While Eq.~\eqref{eq:Eg} is evaluated numerically owing to the complexity of the
integrand, the active mass admits a closed form. For the linear $f(T)$ model
$\rho=\tfrac{\beta}{2}-\alpha\,b'/r^{2}$, so that
\begin{equation}
M_{\rm active}
= 4\pi\!\left[\frac{\beta}{6}\bigl(r^{3}-r_0^{3}\bigr)
- \alpha\bigl(b(r)-r_0\bigr)\right],
\label{eq:Mactive_closed}
\end{equation}
using the throat condition $b(r_0)=r_0$. This yields $M_{\rm active}(r_0)=0$, so
the active mass vanishes at the throat, and $dM_{\rm active}/dr=4\pi\rho r^{2}$,
so $M_{\rm active}$ increases where $\rho>0$ and decreases where $\rho<0$.

\begin{figure}[!htbp]
\centering
\includegraphics[width=0.95\columnwidth]{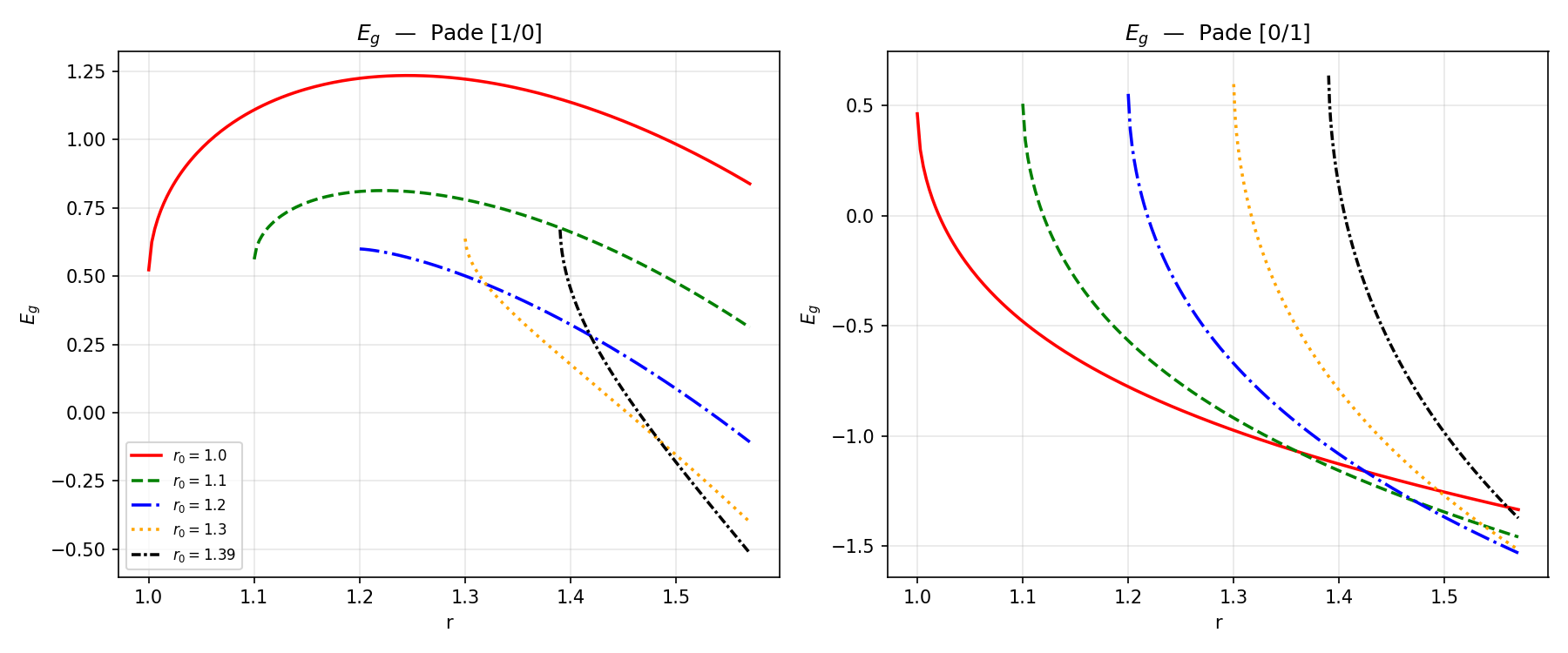}
\caption{Total gravitational energy $E_g$ against $r$ for the $[1/0]$ and
$[0/1]$ Pad\'{e} shape functions, for several throat radii $r_0$, with
$M=1.5$, $\alpha=2$, $\beta=2$. Each curve starts at $E_g=r_0/2$ at its own
throat.}
\label{fig:Eg}
\end{figure}

\begin{figure}[!htbp]
\centering
\includegraphics[width=0.95\columnwidth]{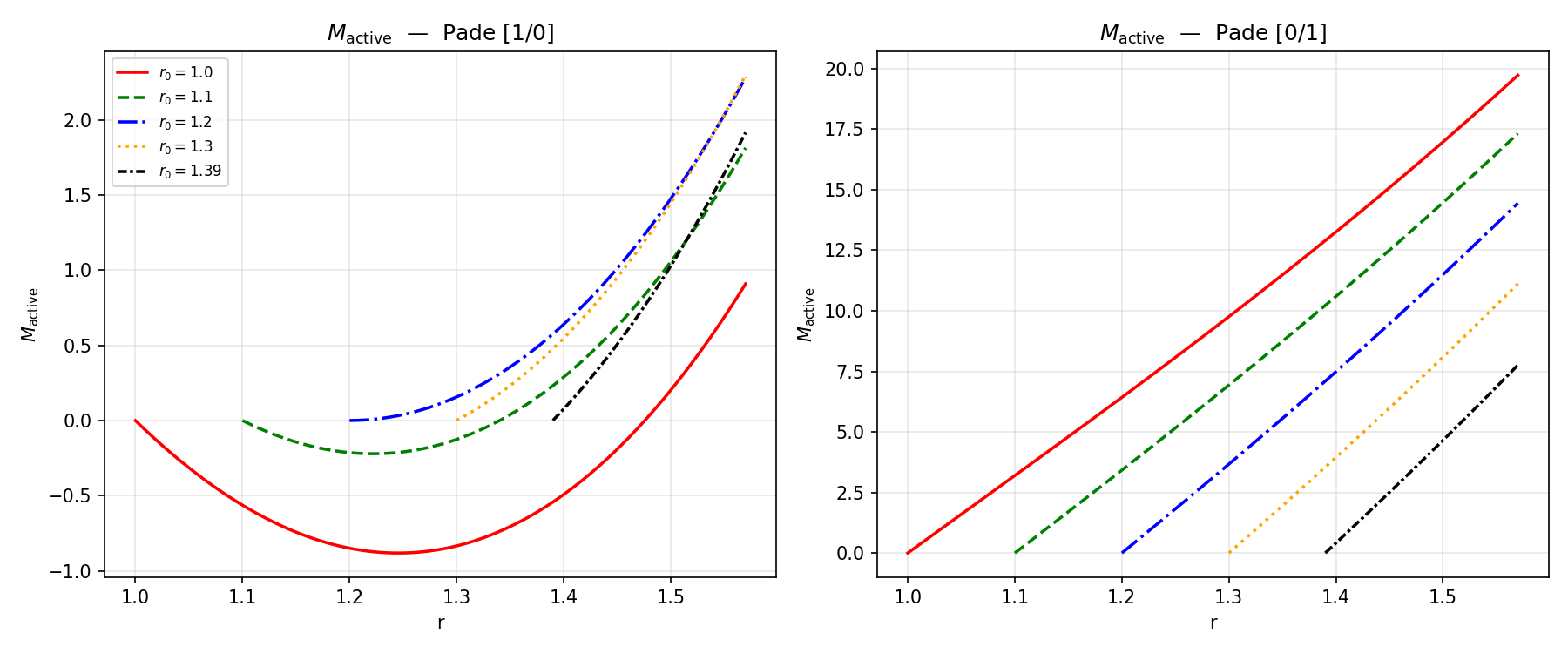}
\caption{Active mass function $M_{\rm active}$ against the radial coordinate $r$
for the $[1/0]$ and $[0/1]$ Pad\'{e} wormholes, for several throat
radii $r_0$, with $M=1.5$, $\alpha=2$, $\beta=2$.}
\label{fig:Mactive}
\end{figure}

The profiles of $E_g$ and $M_{\rm active}$ are shown in Fig.~\ref{fig:Eg} and
Fig.~\ref{fig:Mactive} respectively, for several throat radii $r_0$, and the
numerical values of $z(r)$, $l(r)$, and $E_g$ are collected in
Table~\ref{tab:combined} for three representative radii
$r_0=1.39,\,1.25,\,1.20$. The two diagnostics behave very differently between
the branches. For the $[1/0]$ order the gravitational energy starts positive at
the throat, $E_g=r_0/2$, rises to a maximum in the near-throat region, and then
decreases, turning negative at larger $r$ as the negative-energy-density region
begins to dominate the integral; the active mass likewise dips negative just
outside the throat before recovering and increasing. For the $[0/1]$ order the
gravitational energy decreases monotonically from $E_g=r_0/2$ and is negative
across most of the displayed range, whereas the active mass is large and
positive and grows steeply with $r$, since the energy density for this branch
is positive and rises rapidly away from the throat. We emphasize that these
sign changes are not pathological but are the expected quasilocal-energy
signature of the null-energy-condition-violating matter established in
Sec.~\ref{sec:ec}: a region of negative energy density generically drives the
gravitational energy and active mass negative in its vicinity, a feature
documented for exotic wormhole sources in general
relativity~\cite{Lobo2009,Nandi2009} and in modified
gravity~\cite{Cataldo2017}. The contrasting behaviour of the $[0/1]$ branch,
where $E_g<0$ even though $\rho>0$, arises from the geometric weighting factor
$1-\sqrt{g_{rr}}<0$ in Eq.~\eqref{eq:Eg}, which renders the gravitational energy
negative for a positive energy density and is therefore a property of the
strong-field geometry rather than of the sign of the source. The sign and slope
of both quantities are thus controlled by the sign of $\rho$ together with the
geometric factor in each branch, in accordance with
$dM_{\rm active}/dr=4\pi\rho r^{2}$. This behaviour is consistent with the
energy-condition analysis of Sec.~\ref{sec:ec}, where the $[1/0]$ branch has
negative throat energy density and the $[0/1]$ branch positive, and it confirms
a physically sensible distribution of gravitating mass-energy in both Pad\'{e}
orders.

\begin{table*}[tp]
\centering
\caption{Values of $z(r)$, $l(r)$, and $E_g$ for the $[1/0]$ and $[0/1]$
Pad\'{e} shape functions, for three throat radii $r_0$, with $M=1.5$,
$\alpha=2$, $\beta=2$, $a=0.5$, $a=-0.5$.}
\label{tab:combined}
\small
\begin{tabular}{c ccc ccc}
\hline\hline
& \multicolumn{3}{c}{$[1/0]$} & \multicolumn{3}{c}{$[0/1]$} \\
\cline{2-4}\cline{5-7}
$r$ & $z(r)$ & $l(r)$ & $E_g$ & $z(r)$ & $l(r)$ & $E_g$ \\
\hline
\multicolumn{7}{l}{$r_0=1.39$} \\
1.41 & 0.584207 & 0.584662 & 0.353857 & 0.257554 & 0.258587 & $-0.089466$ \\
1.43 & 0.827519 & 0.828803 & 0.204034 & 0.363888 & 0.366803 & $-0.382639$ \\
1.45 & 1.015119 & 1.017469 & 0.081529 & 0.445245 & 0.450590 & $-0.596728$ \\
1.49 & 1.314675 & 1.319697 & $-0.131414$ & 0.573719 & 0.585177 & $-0.918975$ \\
1.51 & 1.442425 & 1.449004 & $-0.230148$ & 0.627885 & 0.642920 & $-1.049686$ \\
\hline
\multicolumn{7}{l}{$r_0=1.25$} \\
1.30 & 0.927419 & 0.929206 & 0.457521 & 0.381691 & 0.386023 & $-0.379978$ \\
1.35 & 1.317641 & 1.322647 & 0.341257 & 0.538331 & 0.550521 & $-0.729689$ \\
1.40 & 1.621152 & 1.630261 & 0.218863 & 0.657554 & 0.679834 & $-0.977315$ \\
1.45 & 1.880397 & 1.894290 & 0.087068 & 0.757270 & 0.791400 & $-1.174313$ \\
1.50 & 2.117723 & 2.130962 & $-0.055018$ & 0.844440 & 0.891903 & $-1.340312$ \\
\hline
\multicolumn{7}{l}{$r_0=1.20$} \\
1.25 & 0.929645 & 0.931428 & 0.564282 & 0.372518 & 0.376954 & $-0.345118$ \\
1.30 & 1.321168 & 1.326158 & 0.501252 & 0.525319 & 0.537801 & $-0.671227$ \\
1.35 & 1.625926 & 1.635003 & 0.419881 & 0.641573 & 0.664382 & $-0.900962$ \\
1.40 & 1.886427 & 1.900265 & 0.323012 & 0.738766 & 0.773699 & $-1.082974$ \\
1.45 & 2.119034 & 2.138189 & 0.212068 & 0.823699 & 0.872268 & $-1.235819$ \\
\hline\hline
\end{tabular}
\end{table*}

\section{Observational Analysis}
\label{sec:obsanalysis}

\subsection{Null Geodesics and Photon Trajectories}
\label{sec:geodesics}

Light propagation is the most direct observational probe of the wormhole
geometry, since the location of the photon sphere and the associated critical
impact parameter fix the angular size of the shadow and govern strong-field
lensing~\cite{Virbhadra2000,Bozza2002,Tsukamoto2016,Shaikh2019jcap}. We
analyse equatorial null geodesics of the metric~\eqref{eq:MTmetric} for the
two Pad\'{e} shape functions.

By spherical symmetry, a null geodesic with initial data in the equatorial
plane remains in $\theta=\pi/2$. The Lagrangian
$2\mathcal{L}=g_{\mu\nu}\dot{x}^{\mu}\dot{x}^{\nu}$ is independent of $t$ and
$\phi$, so the energy and angular momentum
\begin{equation}
E = e^{2\Phi(r)}\dot{t},\qquad L = r^{2}\dot{\phi}
\label{eq:conserved}
\end{equation}
are conserved, with the overdot denoting $d/d\lambda$. Imposing the null
condition $g_{\mu\nu}\dot{x}^{\mu}\dot{x}^{\nu}=0$ and rescaling the affine
parameter so that $E=1$ yields
\begin{equation}
\left(\frac{dr}{d\lambda}\right)^{2}
= \left(1-\frac{b(r)}{r}\right)
\left(e^{-2\Phi(r)} - \frac{\tilde{b}^{2}}{r^{2}}\right),
\label{eq:radial}
\end{equation}
\begin{equation}
\frac{d\phi}{d\lambda} = \frac{\tilde{b}}{r^{2}},
\label{eq:angular}
\end{equation}
where $\tilde{b}\equiv L/E$ is the impact parameter. Equation~\eqref{eq:radial}
has the form $(dr/d\lambda)^{2}=(1-b/r)\,[e^{-2\Phi}-\tilde{b}^{2}V_{\rm
ph}(r)]$ with photon potential $V_{\rm ph}(r)=r^{-2}$, so the radial motion is
controlled by the function
\begin{equation}
h(r) \equiv r\,e^{-\Phi(r)},
\label{eq:hdef}
\end{equation}
through the turning-point condition $\tilde{b}=h(r)$.

A circular null orbit requires $(dr/d\lambda)^{2}=0$ together with its radial
derivative, which away from the throat reduces to $h'(r)=0$, i.e.
$\Phi'(r_{ph})=1/r_{ph}$. This condition involves only the redshift function,
so the photon-sphere radius is independent of the shape function and of the
Pad\'{e} order. For $\Phi(r)=-M/r$,
\begin{equation}
h(r)=r\,e^{M/r},\qquad
h'(r)=e^{M/r}\!\left(1-\frac{M}{r}\right),
\end{equation}
which has a single extremum, a minimum, at
\begin{equation}
r_{ph}=M,\qquad
\tilde{b}_{ph}=h(M)=e\,M .
\label{eq:photonsphere}
\end{equation}
The minimum of $h$ corresponds to an unstable circular photon
orbit~\cite{Tsukamoto2016,Nandi2017a}: rays with $\tilde{b}>\tilde{b}_{ph}$
reach an outer turning point and are deflected, rays with
$\tilde{b}<\tilde{b}_{ph}$ have no turning point and plunge to the throat, and
$\tilde{b}=\tilde{b}_{ph}=eM$ is the critical curve that bounds the shadow.
When $M>r_{0}$ the photon sphere lies in the accessible exterior; when
$M\leq r_{0}$ the extremum is hidden behind the throat, $h(r)$ is monotonic on
$r\geq r_{0}$, and the throat itself acts as the capture boundary with limiting
impact parameter $\tilde{b}_{ph}=r_{0}e^{M/r_{0}}$. The shape function does not
move the photon sphere; it controls the bending away from $r_{ph}$ through the
factor $(1-b/r)$ in~\eqref{eq:radial}, and hence the winding structure and the
asymptotic behaviour of the deflected rays.

The two branches are integrated with the coupling parameters fixed at the
values that keep each shape function well behaved on $[r_{0},\infty)$. For the
$[1/0]$ form we take $a=0.5>0$, for which $b(r)$ stays
positive; for the $[0/1]$ form, we take $a=-0.5<0$, which
moves the pole $r_{p}=r_{0}+[aG]^{-1}$ below the throat, leaving
$b(r)=r_{0}/[1+|a|G(r-r_{0})]$ pole-free and asymptotically flat, with
$G\equiv 1/r_{0}+\csch(r_{0})\sech(r_{0})$. The redshift mass is fixed in each
panel so that the photon sphere lies outside the throat: $M=\tfrac32\,r_{\star}$
with $r_{\star}=r_{0}+[aG]^{-1}$ for the $[1/0]$ branch, and
$M=\tfrac32\,r_{0}$ for the $[0/1]$ branch. In every panel $M>r_{0}$, so a
photon sphere of radius $r_{ph}=M$ and a shadow of radius
$\tilde{b}_{ph}=eM$ are present; the choice of $M$ rescales the figure without
altering these qualitative features.

\begin{figure*}[tp]
\centering
\begin{subfigure}[b]{0.7\textwidth}
\includegraphics[width=\textwidth]{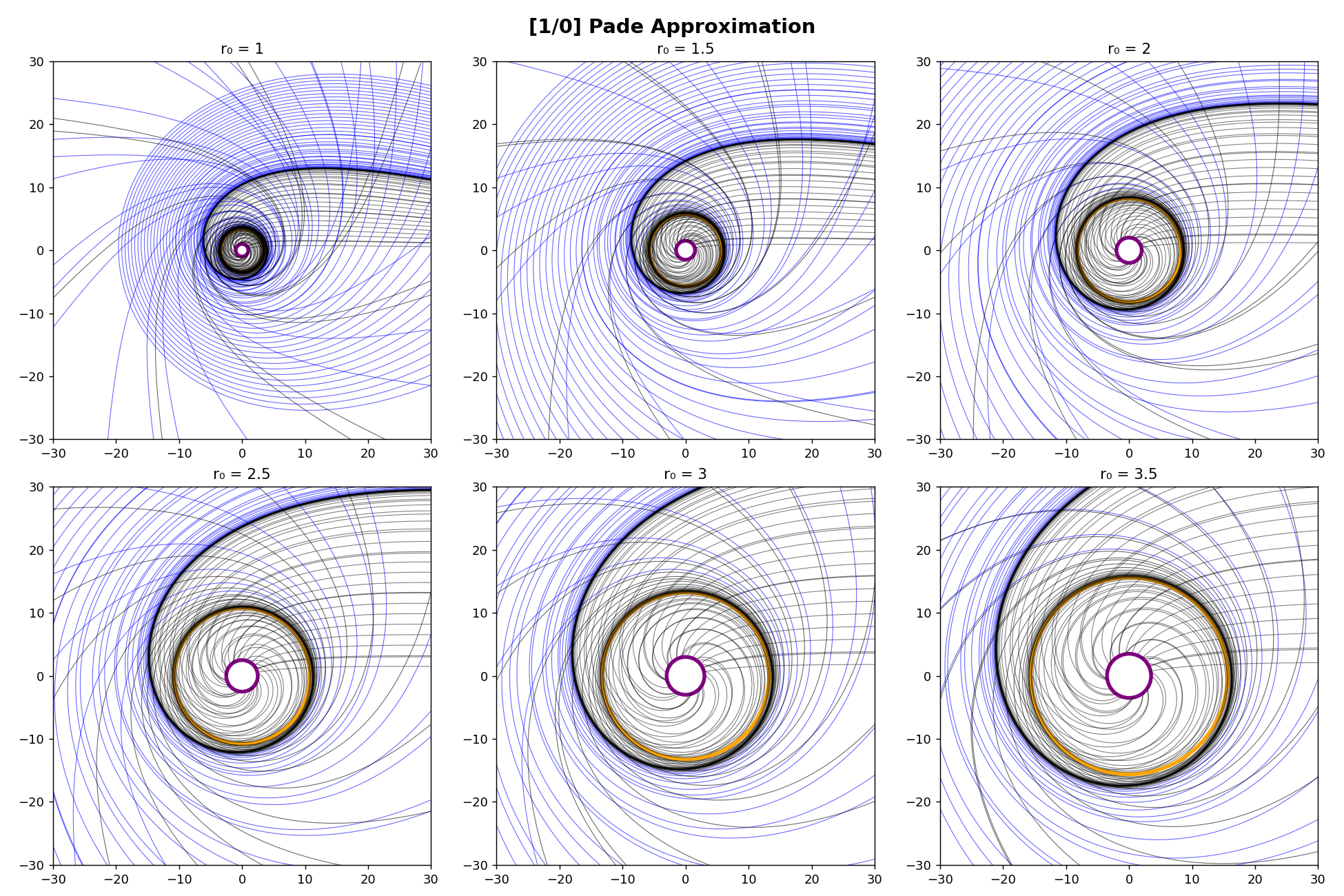}
\caption{$[1/0]$-order Pad\'{e} wormhole}
\label{fig:pade10}
\end{subfigure}
\hfill
\begin{subfigure}[b]{0.7\textwidth}
\includegraphics[width=\textwidth]{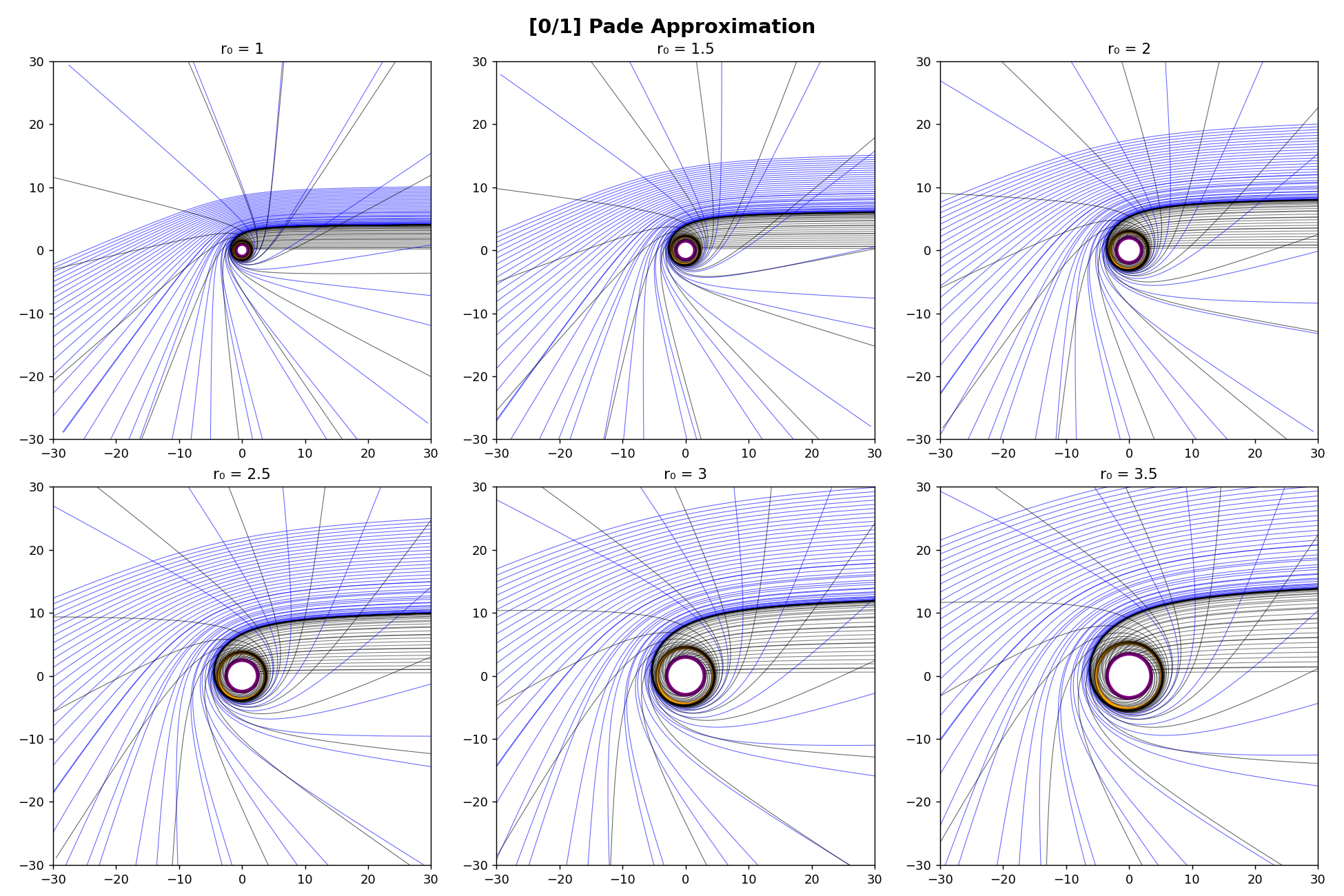}
\caption{$[0/1]$-order Pad\'{e} wormhole}
\label{fig:pade01}
\end{subfigure}
\caption{Equatorial photon trajectories for the $[1/0]$-order (left) and 
$[0/1]$-order (right) Pad\'{e} wormholes. The black disk marks the throat of 
radius $r_0$. Photon paths are shown as colored curves, illustrating the 
gravitational lensing effects in the wormhole spacetimes.}
\label{fig:pade}
\end{figure*}

Equations~\eqref{eq:radial}--\eqref{eq:angular} are integrated with an adaptive
Runge--Kutta scheme (relative tolerance $10^{-9}$) for six throat radii
$r_{0}\in\{1.0,1.5,2.0,2.5,3.0,3.5\}$. Rays are launched from $x=40$ with a
range of impact parameters densely sampled around $\tilde{b}_{ph}$, integration
terminating at the radial turning point (deflected rays) or at the throat
$r=r_{0}+10^{-4}$ (captured rays). The deflected branch is completed by the
reflection symmetry of the orbit about its turning point. Results are shown in
Fig.~\ref{fig:pade10} for the $[1/0]$ branch and Fig.~\ref{fig:pade01} for the
$[0/1]$ branch.

Both branches share the same photon-sphere radius $r_{ph}=M$ and critical
impact parameter $\tilde{b}_{ph}=eM$, because these depend on $\Phi$
alone~\eqref{eq:photonsphere}. The difference between Figs.~\ref{fig:pade10}
and~\ref{fig:pade01} is therefore entirely due to the shape function entering
$(1-b/r)$. In the $[1/0]$ case the linear shape function approaches a constant
fraction $b/r\to a[1+r_{0}\csch r_{0}\sech r_{0}]$, so the exterior carries
a residual deficit and even distant rays are curved, giving the broad winding
seen in Fig.~\ref{fig:pade10}; here $\tilde{b}$ is a coordinate quantity rather
than the true geometric impact parameter at infinity. In the $[0/1]$ case with
$a<0$ the shape function decays as $b\sim r_{0}/[\,|a|G\,r\,]$, so
$b/r\to0$, the exterior is asymptotically flat, $\tilde{b}$ coincides with the
asymptotic impact parameter, and the lensing is confined to a compact region
around $r_{ph}$, as in Fig.~\ref{fig:pade01}. The asymptotically flat $[0/1]$
geometry is the most suitable of the two for shadow and lensing observables,
while both reproduce the same shadow radius $eM$ set by the
redshift~\cite{Shaikh2019jcap,Shaikh2019plb,Jusufi2018}.





\subsection{Intensity Profiles and Shadow Maps}
\label{sec:shadow}

The optical appearance of a compact object illuminated by an optically thin emission region encodes the geometry of its photon region and serves as a direct observational probe of the underlying spacetime \cite{Vagnozzi2023,Shaikh2019jcap}. The Event Horizon Telescope images of M87$^*$ and Sgr~A$^*$ have established horizon-scale imaging as a quantitative test of gravity theories and exotic alternatives to black holes, including wormholes \cite{Vagnozzi2023}. Wormhole shadows and their gravitational-lensing signatures have consequently been studied across a wide range of geometries and matter models \cite{Shaikh2018,Nedkova2013,Shaikh2019jcap,Shaikh2019plb,Bambhaniya2021}, with strong-deflection lensing providing a complementary diagnostic of the throat region \cite{Tsukamoto2016,Tsukamoto2017,Bozza2002,Nandi2006}.

For the static spherically symmetric metric of Eq.~\eqref{eq:MTmetric} with redshift function $\Phi(r)=-M/r$, null geodesics are governed by the conserved energy and the angular momentum, whose ratio defines the impact parameter $b$. Photon trajectories follow from the radial equation derived from the null condition, and the effective potential governing the radial motion is set by the combination $e^{-2\Phi(r)}/r^2$, whose extremum fixes the photon sphere and the associated critical impact parameter $b_c$ \cite{Virbhadra2000,Bozza2002}. For the redshift profile adopted here, the critical impact parameter takes the compact form $b_c = eM$, in agreement with the value obtained for exponential-metric wormholes in the literature \cite{Boonserm2018,Tsukamoto2019}. This critical curve separates photons that are captured or transmitted through the throat from those that are deflected back to a distant observer, and its angular size on the observer's sky determines the boundary of the shadow \cite{Shaikh2018,Shaikh2019jcap}.

We modeled the observed specific intensity by integrating the emissivity of an optically thin accretion flow along each null geodesic, accounting for the gravitational redshift of the radiation through the factor $g = e^{\Phi(r)}$ \cite{Bambhaniya2021,Vagnozzi2023}. The radiative-transfer integral assigns to each impact parameter $b$ a total intensity built from the redshift-weighted emissivity and the path-length element along the trajectory, with photons possessing a radial turning point that contributes an additional passage through the emitting region. The numerical integration of null geodesics is carried out with an adaptive scheme, terminating each ray either at its radial turning point or upon reaching the throat, which acts as the effective capture surface for the present geometry \cite{Shaikh2018}.

We evaluated the intensity for both Pad\'{e}-approximant wormhole geometries constructed in Sec.~\ref{sec:pade}: the $[1/0]$-order branch, Eq.~\eqref{eq:b_pade10}, and the pole-free $[0/1]$-order branch, Eq.~\eqref{eq:Fprime_r0_again}. For each branch we consider a sequence of throat radii $r_0\in\{1.0,1.5,2.0,2.5,3.0,3.5\}$, with $M=1.5\,r_{\rm pole}(r_0)$ for the $[1/0]$ branch and $M=1.5\,r_0$ for the $[0/1]$ branch. To enable a faithful comparison across throat radii and between the two branches, all intensity profiles are normalized by a single global maximum, and all impact-parameter scans share a common range, so that both the relative brightness and the relative angular size of the features are preserved rather than rescaled panel by panel.

\begin{figure*}[tp]
    \centering
    \begin{subfigure}[b]{0.7\textwidth}
        \centering
        \includegraphics[width=\textwidth]{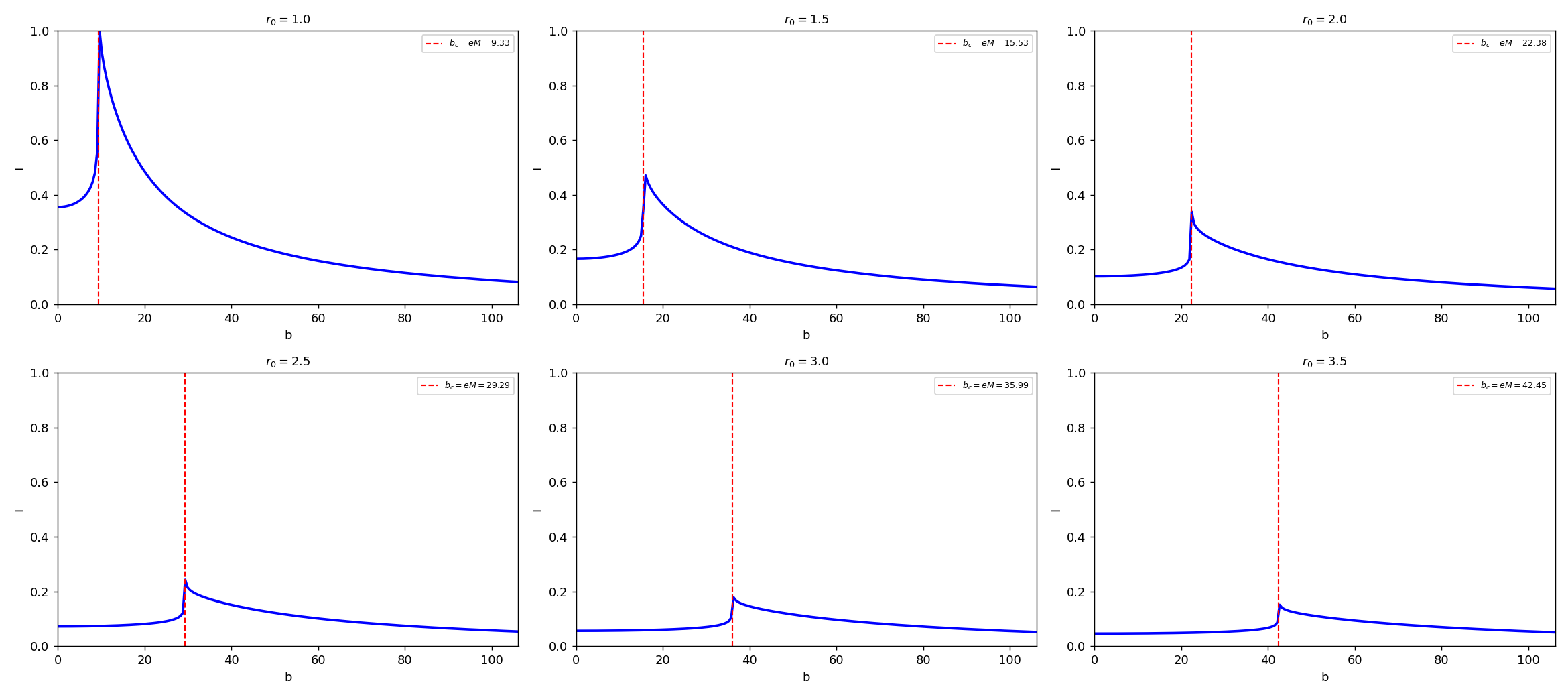}
        \caption{Normalized radial intensity profile $I(b)$ for the $[1/0]$-order Pad\'{e} wormhole, for throat radii $r_0\in\{1.0,1.5,2.0,2.5,3.0,3.5\}$.}
        \label{fig:intensity_10}
    \end{subfigure}
    \hfill
    \begin{subfigure}[b]{0.7\textwidth}
        \centering
        \includegraphics[width=\textwidth]{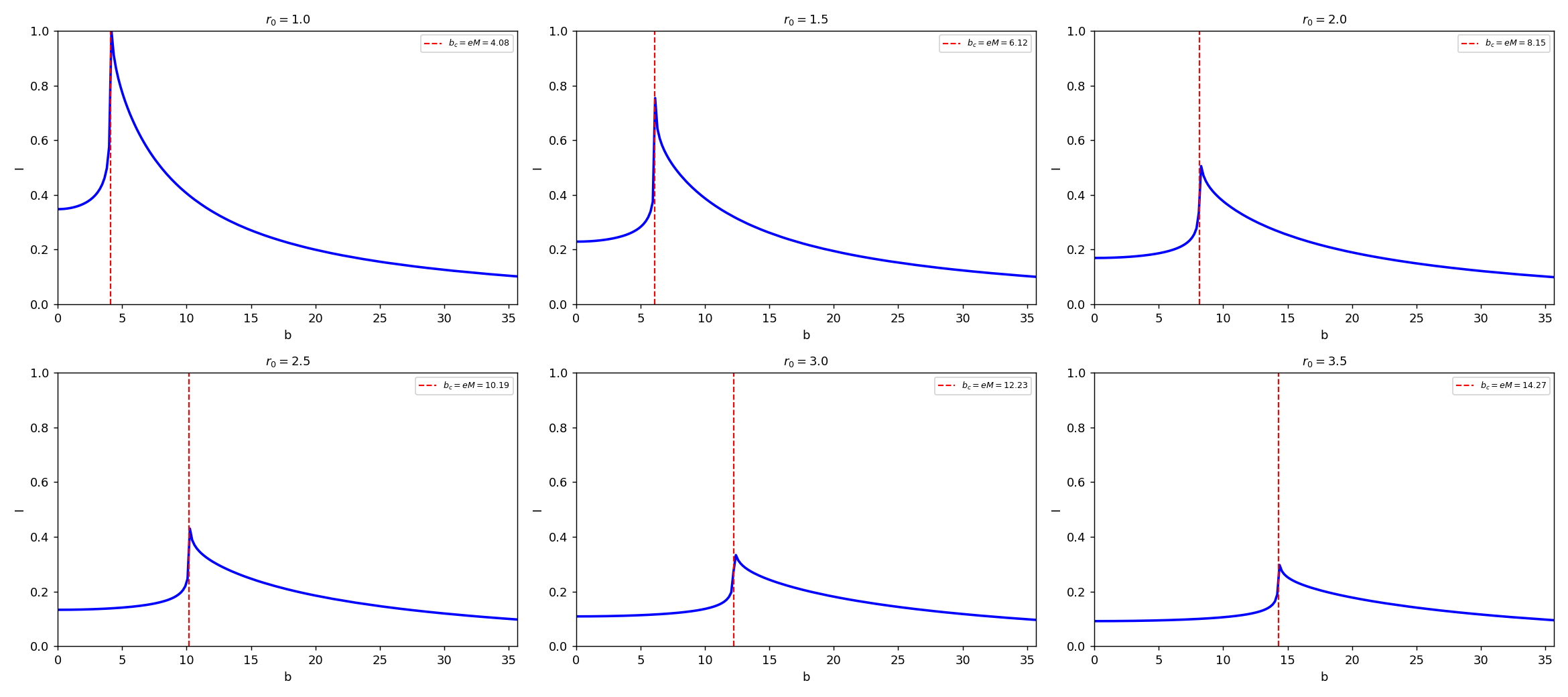}
        \caption{Normalized radial intensity profile $I(b)$ for the $[0/1]$-order Pad\'{e} wormhole, for throat radii $r_0\in\{1.0,1.5,2.0,2.5,3.0,3.5\}$.}
        \label{fig:intensity_01}
    \end{subfigure}
    \caption{Radial intensity profiles for the Pad\'{e} wormholes. Left: $[1/0]$-order branch. Right: $[0/1]$-order branch.}
    \label{fig:intensity_profiles}
\end{figure*}

The radial intensity profiles $I(b)$ are shown in Fig.~\ref{fig:intensity_10} for the $[1/0]$-order wormhole and in Fig.~\ref{fig:intensity_01} for the $[0/1]$-order wormhole. In every panel the intensity rises from a finite value near the optical axis, attains a sharp maximum at the critical impact parameter $b_c=eM$ (marked by the dashed line), and decays monotonically toward larger $b$. The peak at $b_c$ is the photon-ring enhancement: rays with $b$ close to $b_c$ wind multiple times near the photon sphere and accumulate a large path length through the emitting medium, producing the characteristic bright ring familiar from black-hole and wormhole imaging \cite{Shaikh2019jcap,Bambhaniya2021,Vagnozzi2023}. As $r_0$ increases, the associated mass $M$ and hence $b_c$ shift to larger values, so the peak migrates outward while its normalized height decreases, reflecting the reduced redshift enhancement for the more extended, lower-compactness configurations.

\begin{figure*}[tp]
    \centering
    \begin{subfigure}[b]{1\textwidth}
        \centering
        \includegraphics[width=1\textwidth]{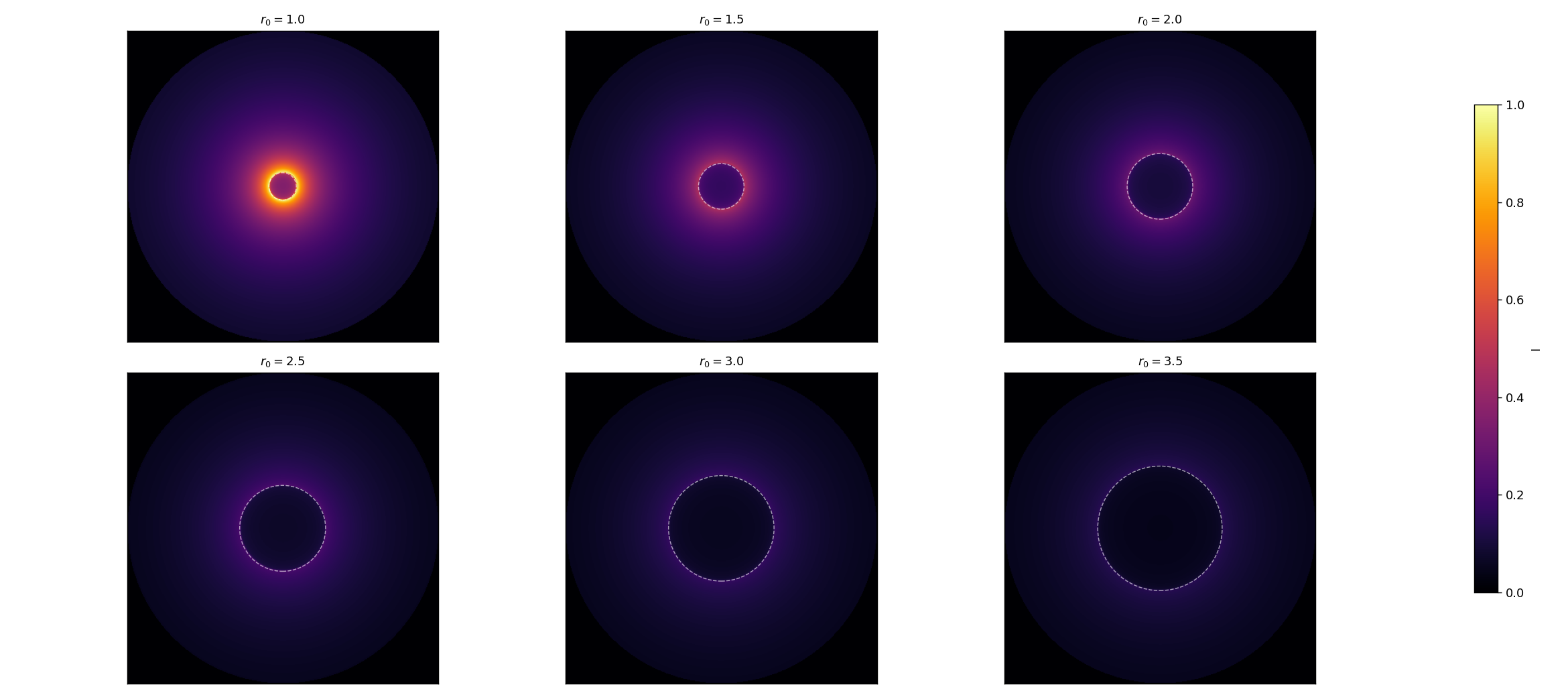}
        \caption{Shadow maps for the $[1/0]$-order Pad\'{e} wormhole for $r_0\in\{1.0,1.5,2.0,2.5,3.0,3.5\}$. The dashed white circle marks the critical curve at $b_c=eM$.}
        \label{fig:shadow_10}
    \end{subfigure}
    \hfill
    \begin{subfigure}[b]{1\textwidth}
        \centering
        \includegraphics[width=1\textwidth]{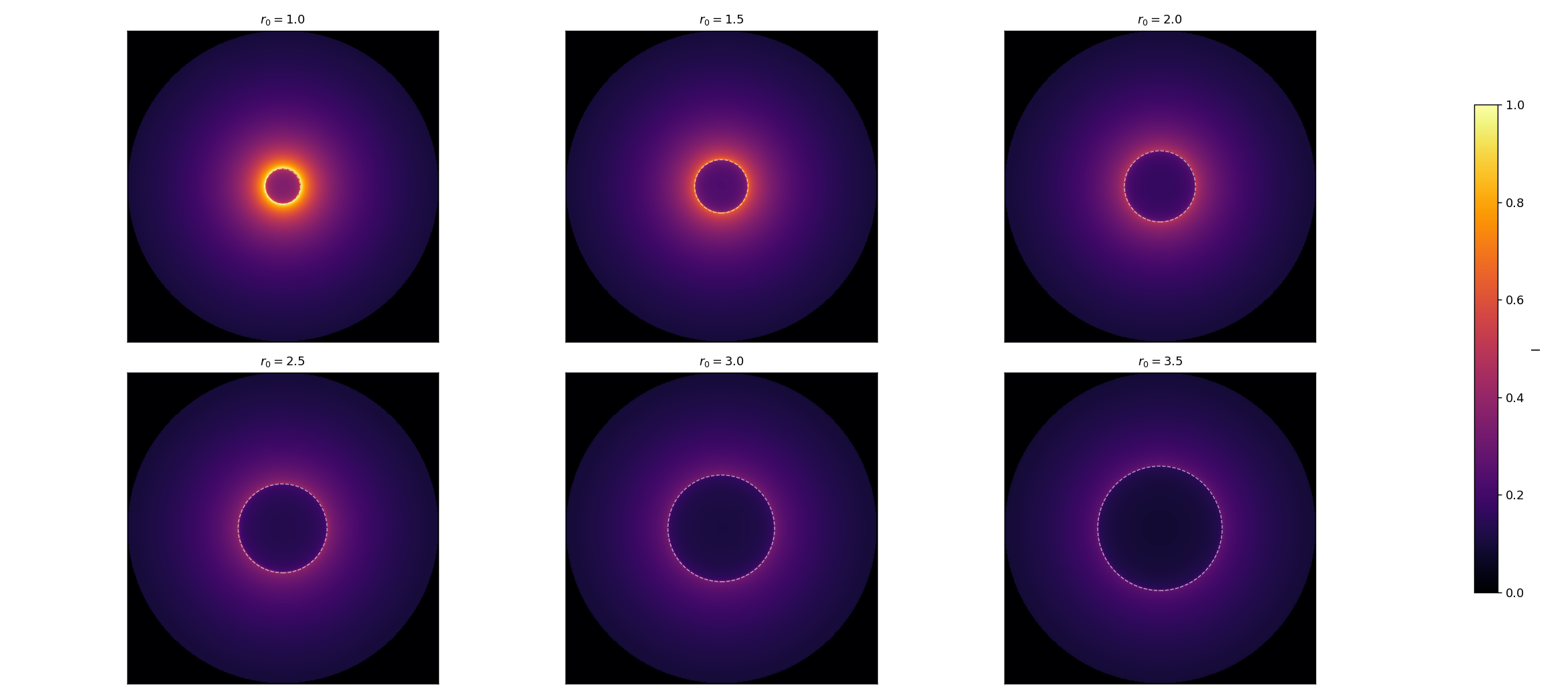}
        \caption{Shadow maps for the $[0/1]$-order Pad\'{e} wormhole for $r_0\in\{1.0,1.5,2.0,2.5,3.0,3.5\}$. The dashed white circle marks the critical curve at $b_c=eM$.}
        \label{fig:shadow_01}
    \end{subfigure}
    \caption{Shadow maps for the Pad\'{e} wormholes. Left: $[1/0]$-order branch. Right: $[0/1]$-order branch. The dashed white circle indicates the critical curve $b_c=eM$ in each panel.}
    \label{fig:shadow_maps}
\end{figure*}

The corresponding two-dimensional shadow maps, rendered as an axially symmetric radial colour gradient with the critical curve $b_c=eM$ overlaid, are displayed in Fig.~\ref{fig:shadow_10} for the $[1/0]$-order wormhole and in Fig.~\ref{fig:shadow_01} for the $[0/1]$-order wormhole. In each panel the dashed white circle marks the critical curve at the radius $b_c$, which sets the boundary of the shadow on the observer's sky. Because all panels are rendered on a common radial scale, the systematic growth of the critical curve with increasing $r_0$ is directly visible: larger throat radii yield larger $b_c$ and therefore larger shadows on the observer's sky. A comparison of the two branches at fixed $r_0$ reveals that the $[1/0]$-order geometry, for which $M=1.5\,r_{\rm pole}(r_0)$ exceeds the $[0/1]$ value $M=1.5\,r_0$ by a factor of a few, produces a substantially larger critical curve, 
while the pole-free $[0/1]$ geometry yields a more compact, centrally concentrated critical curve. This contrast underscores the sensitivity of the shadow size to the specific shape-function ansatz, in line with the broader finding that wormhole observables discriminate effectively between competing geometries and gravity models \cite{Shaikh2018,Shaikh2019jcap,Vagnozzi2023}.

Taken together, the intensity profiles and shadow maps demonstrate that both Pad\'{e}-approximant wormholes in the linear $f(T)$ model produce the photon-ring and shadow morphology characteristic of horizon-scale compact objects, with the shadow size scaling monotonically with the throat radius and depending sensitively on the shape-function branch \cite{Shaikh2018,Shaikh2019jcap,Vagnozzi2023}. These features place the present geometries within the class of wormhole spacetimes whose optical signatures could, in principle, be confronted with high-resolution very-long-baseline interferometric observations \cite{Vagnozzi2023,Shaikh2019jcap}.

\subsection{Accretion Disc Images}
\label{sec:disc}
 
The intensity profiles and shadow maps of Sec.~\ref{sec:shadow} describe the lensing of an idealized, axisymmetric emission ring. A complementary and more directly observable proxy is the relativistically imaged accretion disc, whose brightness distribution on the sky carries the combined imprint of gravitational lensing, frame-dependent Doppler boosting, and gravitational redshift in the strong-field region \cite{Luminet1979,Cunningham1975,Vagnozzi2023}. The horizon-scale images of M87$^*$ and Sgr~A$^*$ recorded by the Event Horizon Telescope have made such disc images a quantitative test of the near-horizon geometry \cite{EHTM87I,EHTSgrAI,Falcke2000}, and horizonless compact objects, including wormholes, are now routinely examined as black-hole mimickers whose disc morphology can be confronted with very-long-baseline interferometry \cite{Shaikh2017,Nandi2017a,Nandi2017b,Bambhaniya2021,Shaikh2019jcap}. Accretion onto wormhole geometries and the associated optical signatures have been studied for a range of matter models and gravity theories \cite{DeFalco2020,DeFalco2021,Harko2011}.
 
We construct the images by backward ray tracing. For each pixel of the observer's image plane, parametrized by the celestial coordinates $(\alpha,\beta)$ at inclination $i=80^{\circ}$, a null geodesic is launched toward the source and integrated with a fixed-step fourth-order Runge--Kutta scheme. The trajectory is governed by the geodesic equations of the metric in Eq.~\eqref{eq:MTmetric} with redshift function $\Phi(r)=-M/r$ and the Pad\'{e}-approximant shape functions of Sec.~\ref{sec:pade}, the $[1/0]$-order branch of Eq.~\eqref{eq:b_pade10} and the pole-free $[0/1]$-order branch of Eq.~\eqref{eq:Fprime_r0_again}. The conserved energy and axial angular momentum $L$ fix the initial momenta on the image plane, with the radial momentum set by the null condition \cite{Bozza2002,Virbhadra2000}. Each ray is terminated either upon escaping to large radius or upon reaching the throat at $r_{\rm stop}=1.001\,r_0$, which acts as the effective capture surface for the present horizonless geometry \cite{Shaikh2018,Tsukamoto2016,Tsukamoto2017}.
 
The emitting matter is modeled as a geometrically thin optically thin disc lying in the equatorial plane, extending from an inner edge $r_{\rm in}=2M$ to an outer edge $r_{\rm out}=11M$, with a Gaussian vertical profile of half-thickness $h=0.5M$ and a phenomenological radial emissivity $\propto r^{-q}$ with $q=2.5$ \cite{NovikovThorne1973,PageThorne1974,ShakuraSunyaev1973}. The disc material follows circular geodesics with orbital angular velocity
\begin{equation}
    \Omega(r) = e^{\Phi(r)}\sqrt{\frac{\Phi'(r)}{r}} ,
    \label{eq:Omega_disc}
\end{equation}
and the radiation is transported along the null geodesics with the frequency ratio of observed to emitted photons,
\begin{equation}
    g = \frac{e^{\Phi}\sqrt{1-r\,\Phi'}}{1-\Omega\, L} ,
    \label{eq:gfactor_disc}
\end{equation}
which combines the gravitational redshift with the Doppler shift of the orbiting source \cite{Cunningham1975,Luminet1979,Bambhaniya2021}. By Liouville's theorem the observed specific intensity accumulated along each ray scales as
\begin{equation}
    I_{\rm obs} \propto \int g^{4}\, r^{-q}\, e^{-(z/h)^2}\, d\lambda ,
    \label{eq:Iobs_disc}
\end{equation}
with $z=r\cos\theta$ the height above the equatorial plane, so that the steep $g^4$ dependence produces the strong brightness asymmetry between the approaching and receding sides of the disc characteristic of near-edge-on viewing \cite{Luminet1979,JohannsenPsaltis2010,Vagnozzi2023}.
 
\begin{figure*}[tp]
    \centering
    \begin{subfigure}[b]{0.7\textwidth}
        \centering
        \includegraphics[width=\textwidth]{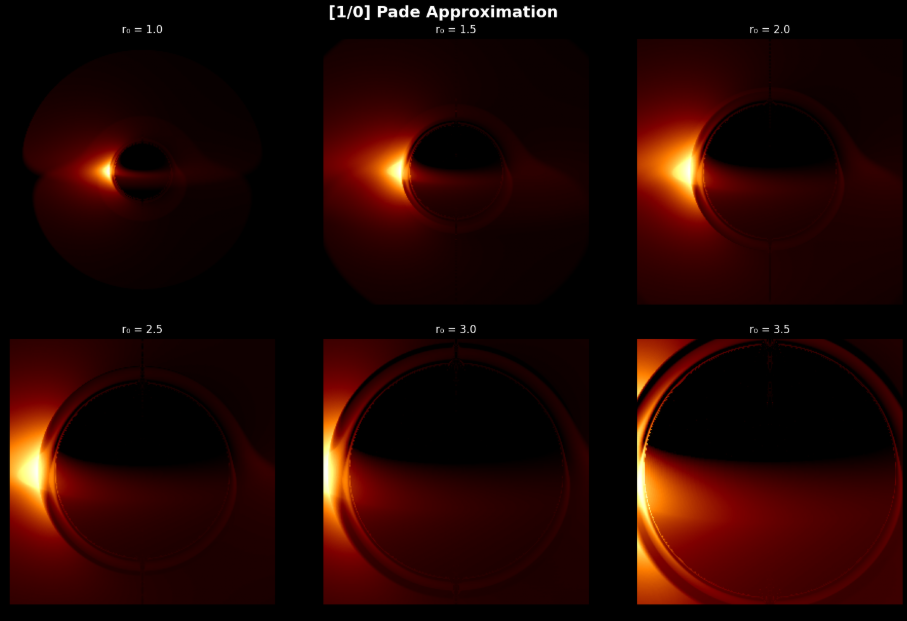}
        \caption{Geometrically thin, optically thin accretion disc around the $[1/0]$-order Pad\'{e} wormhole,for $r_0\in\{1.0,1.5,2.0,2.5,3.0,3.5\}$ }
        \label{fig:disc_10}
    \end{subfigure}
    \hfill
    \begin{subfigure}[b]{0.7\textwidth}
        \centering
        \includegraphics[width=\textwidth]{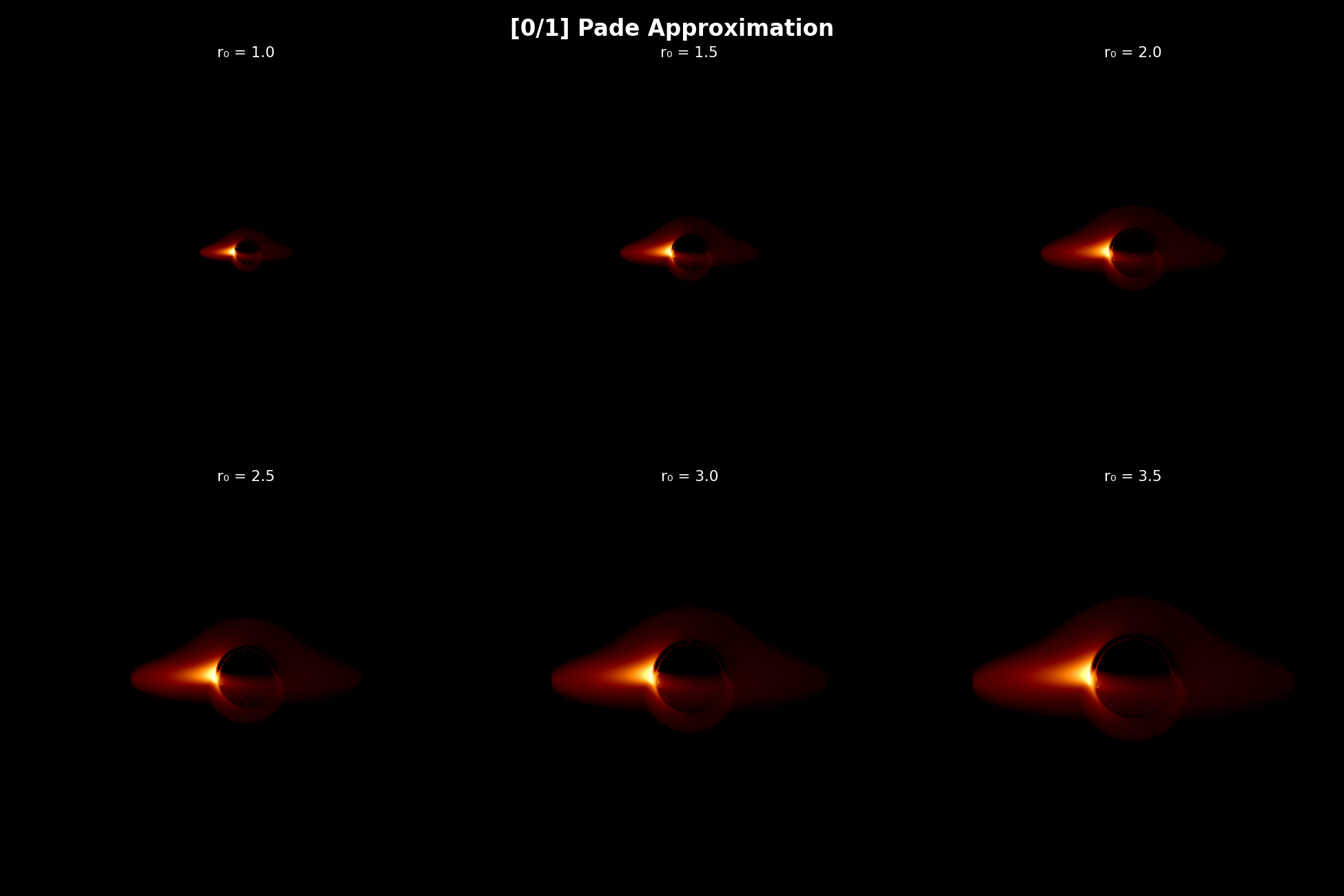}
        \caption{Geometrically thin, optically thin accretion disc around the $[0/1]$-order Pad\'{e} wormhole,for $r_0\in\{1.0,1.5,2.0,2.5,3.0,3.5\}$ }
        \label{fig:disc_01}
    \end{subfigure}
    
    \caption{Comparison of relativistically ray-traced accretion disc images for Pad\'{e} wormholes. Left: $[1/0]$-order wormhole with $\Phi(r)=-M/r$ and $M=1.5\,r_{\rm pole}(r_0)$. Right: pole-free $[0/1]$-order wormhole with $M=1.5\,r_0$ and $a<0$. Both sets show images for different throat radii viewed at inclination $i=80^{\circ}$.}
    \label{fig:disc_comparison}
\end{figure*}

Figure~\ref{fig:disc_10} shows the resulting images for the $[1/0]$-order wormhole with $M=1.5\,r_{\rm pole}(r_0)$, for throat radii $r_0\in\{1.0,1.5,2.0,2.5,3.0,3.5\}$. Each panel displays the bright, Doppler-boosted crescent on the approaching side of the disc, a central brightness depression bounded by the photon ring, and a lensed secondary image of the far side of the disc arching above and below the shadow, the hallmark morphology of strongly lensed accretion flows around compact objects \cite{Luminet1979,Falcke2000,Gralla2019}. Because the field of view and the observer distance are held fixed in physical units across the sequence, the steady growth of the shadow and photon ring with $r_0$ directly reflects the increase of the associated mass $M$ with throat radius \cite{Boonserm2018,Tsukamoto2019}.
 
Figure~\ref{fig:disc_01} shows the corresponding images for the pole-free $[0/1]$-order wormhole with $M=1.5\,r_0$. Here, the field of view and the observer distance are scaled in proportion to $M$, so the rendered images are approximately self-similar across $r_0$ and isolate the morphology intrinsic to the shape-function branch. The $[0/1]$ geometry yields a more compact, centrally concentrated crescent with a rounder shadow and a thinner lensed ring than the $[1/0]$ case at fixed $r_0$, consistent with its smaller mass and smaller critical impact parameter $b_c=eM$ \cite{Boonserm2018,Tsukamoto2019,Shaikh2019jcap}. The asymmetric ring with a bright approaching side and a faint receding side closely resembles the observed structure of relativistically imaged accretion flows \cite{EHTM87I,EHTSgrAI,Vagnozzi2023}.
 
To assess the dependence on the adopted emission model, we repeated the ray tracing for $q\in\{2,2.5,3\}$, $r_{\rm in}\in\{1.2M,2M,3M\}$, $r_{\rm out}\in\{8M,11M,15M\}$ and $i\in\{17^\circ,45^\circ,80^\circ\}$. In all cases the dark central region remains bounded by the critical curve $b_c=eM$ to within the image resolution ($\sim4\%$), and the peak of the intensity profiles of Sec.~\ref{sec:shadow} stays at $b=b_c$ to within $1\%$ even for emissivities $j\propto r^{-3}$. The emissivity index and the outer boundary only redistribute the brightness, while the inner boundary sets the location of the brightest emission and, for $r_{\rm in}=3M$, opens a dark gap outside the critical curve. The inclination mainly controls the Doppler asymmetry, with the left/right flux ratio decreasing from $\sim15$ at $i=80^\circ$ to $\sim2$ at $i=17^\circ$; for the $[1/0]$ branch at low inclination, foreground disc emission partially fills the region inside $b_c$. Thus the shadow size is fixed by the wormhole geometry, whereas the intensities and image asymmetry depend on the assumed disc model.

Taken with the intensity profiles and shadow maps of Sec.~\ref{sec:shadow}, the disc images show that both Pad\'{e}-approximant wormholes in the linear $f(T)$ model reproduce the Doppler-boosted, lensed-crescent morphology expected of accretion flows around horizonless compact objects, with the size of the shadow and photon ring scaling monotonically with the throat radius and the detailed brightness distribution depending sensitively on the shape-function branch \cite{Shaikh2018,Shaikh2019jcap,Nandi2017b}. These optical signatures place the present geometries within the class of wormhole spacetimes whose appearance could, in principle, be tested with current and next-generation horizon-scale imaging \cite{Vagnozzi2023,Falcke2000,JohannsenPsaltis2010}.


\subsection{Rosette Orbits}
\label{subsec:timelike}
The null geodesics of Sec.~\ref{sec:geodesics} probe the photon region; the
motion of massive test particles provides a complementary diagnostic of the
strong-field geometry, since bound timelike orbits and their periastron
precession are sensitive to both the redshift and the shape function
\cite{Nandi2017a,Bambhaniya2021,Cataldo2017}. We integrate timelike geodesics
of the metric~\eqref{eq:MTmetric} with $\Phi(r)=-M/r$ for the two Pad\'{e}
shape functions of Sec.~\ref{sec:pade}.

A timelike geodesic $x^{\mu}(\tau)$ obeys
\begin{equation}
\frac{d^{2}x^{\mu}}{d\tau^{2}}
+\Gamma^{\mu}{}_{\alpha\beta}
\frac{dx^{\alpha}}{d\tau}\frac{dx^{\beta}}{d\tau}=0,
\qquad
g_{\mu\nu}\dot{x}^{\mu}\dot{x}^{\nu}=-1,
\label{eq:geodesic_timelike}
\end{equation}
where $\tau$ is proper time and the overdot denotes $d/d\tau$. As in the null
case, stationarity and axial symmetry give the conserved energy and angular
momentum
\begin{equation}
E=e^{2\Phi(r)}\dot{t},
\qquad
L=r^{2}\dot{\phi},
\label{eq:conserved_timelike}
\end{equation}
the orbit remains planar by spherical symmetry. Substituting
Eq.~\eqref{eq:conserved_timelike} into the normalization condition yields the
radial equation
\begin{equation}
\left(\frac{dr}{d\tau}\right)^{2}
=\left(1-\frac{b(r)}{r}\right)
\left[E^{2}e^{-2\Phi(r)}-\frac{L^{2}}{r^{2}}-1\right],
\label{eq:radial_timelike}
\end{equation}
so that the radial motion is governed by the effective potential
\begin{equation}
V_{\rm eff}(r)=\left(1-\frac{b(r)}{r}\right)
\left(1+\frac{L^{2}}{r^{2}}\right),
\label{eq:Veff_timelike}
\end{equation}
with $e^{-2\Phi}=e^{2M/r}$ for the adopted redshift function. The shape
function enters both Eqs.~\eqref{eq:radial_timelike}
and~\eqref{eq:Veff_timelike} through the factor $(1-b/r)$, which controls the
turning-point structure and hence the radial range of the bound orbits
\cite{Bambhaniya2021,Cataldo2017}.

\begin{figure*}[tp]
\centering
\begin{subfigure}[b]{0.7\textwidth}
\includegraphics[width=\textwidth]{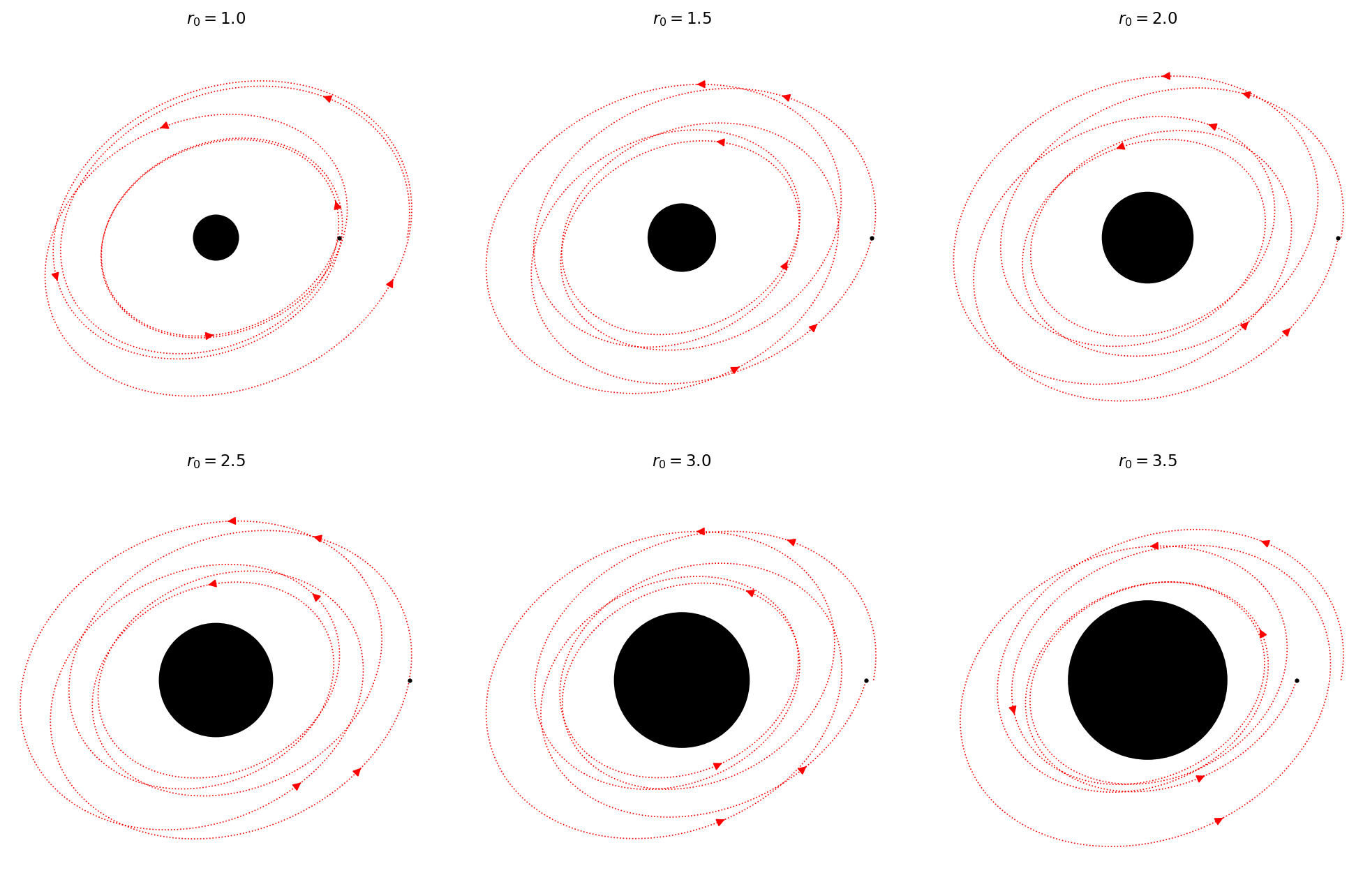}
\caption{$[1/0]$-order Pad\'{e} wormhole}
\label{fig:timelike10}
\end{subfigure}
\hfill
\begin{subfigure}[b]{0.7\textwidth}
\includegraphics[width=\textwidth]{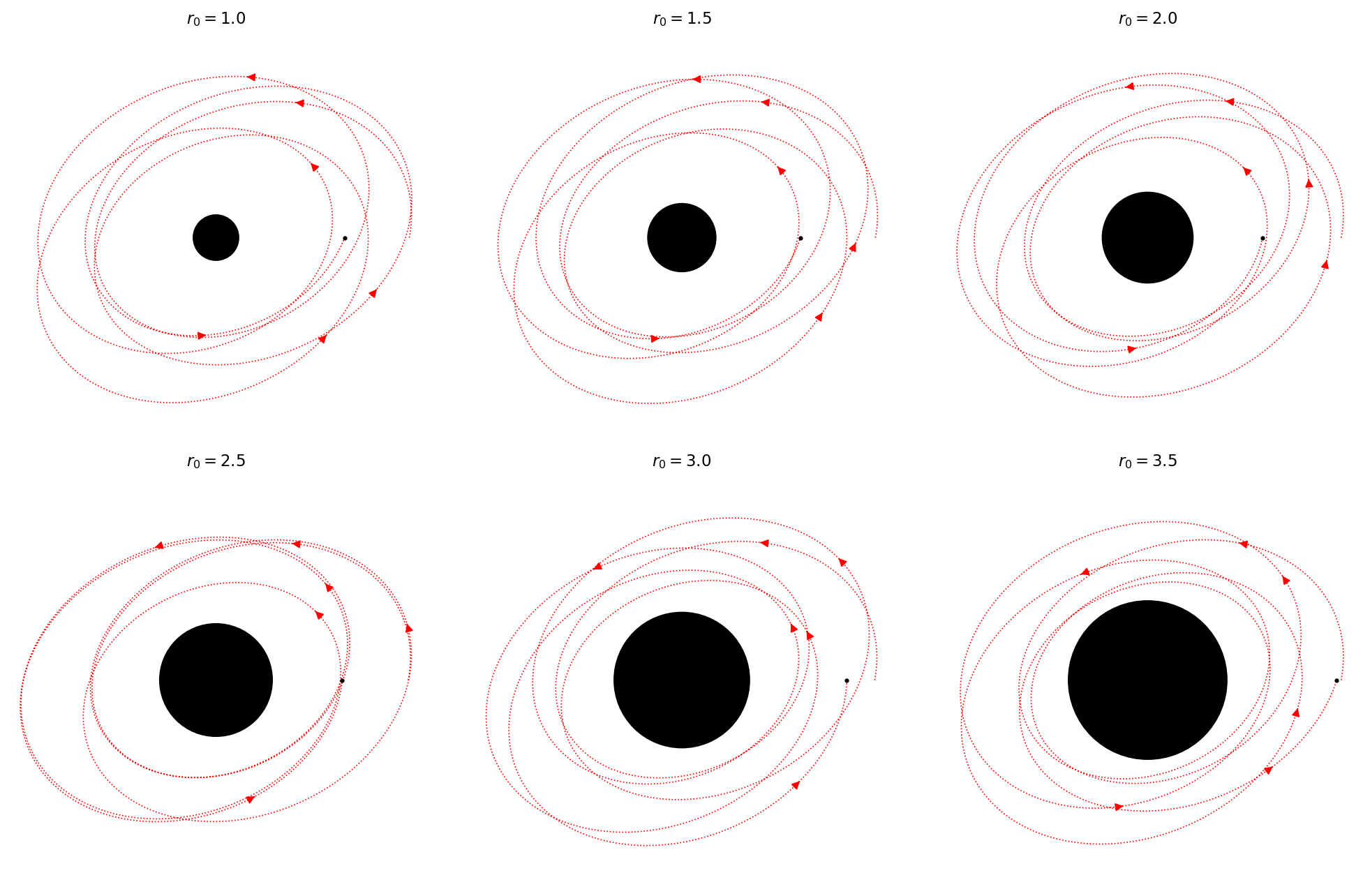}
\caption{Pole-free $[0/1]$-order Pad\'{e} wormhole}
\label{fig:timelike01}
\end{subfigure}
\caption{Timelike bound orbits for Pad\'{e} wormholes with $M=1.5$, for the throat
radii $r_0\in\{1.0,1.5,2.0,2.5,3.0,3.5\}$. The black disk marks the throat of
radius $r_{0}$. Each red dotted curve is a single bound trajectory integrated
over five azimuthal windings; the arrows indicate the sense of motion and the
black dot marks the launch point.}
\label{fig:timelike}
\end{figure*}

We integrate the full system~\eqref{eq:geodesic_timelike} with an explicit
eighth-order Runge--Kutta scheme (relative tolerance $10^{-9}$), with the
energy and angular momentum fixed by the initial four-velocity at the launch
radius $r=9$ and $M=1.5$. The Christoffel symbols are evaluated numerically from
the diagonal metric. Each trajectory is terminated either after five azimuthal
windings or upon reaching the capture radius $r=r_{0}+0.05$ at the throat,
which acts as the effective absorbing surface for the present horizonless
geometry \cite{Shaikh2018,Tsukamoto2016}. The coupling parameters are fixed at
$a=0.5$ for the $[1/0]$ branch and $a=-0.5$ for the pole-free $[0/1]$
branch, so that $b(r)$ stays positive and well behaved on $[r_{0},\infty)$, in
line with Sec.~\ref{sec:geodesics}. Results are shown in
Fig.~\ref{fig:timelike10} for the $[1/0]$ branch and
Fig.~\ref{fig:timelike01} for the $[0/1]$ branch, for the six throat radii
$r_{0}\in\{1.0,1.5,2.0,2.5,3.0,3.5\}$.

In both branches the orbits are non-closed rosettes whose periastra advance
from one radial cycle to the next, the strong-field analogue of relativistic
perihelion precession \cite{Nandi2017a,Bambhaniya2021}. As $r_{0}$ increases,
the throat disk grows and the inner turning point migrates outward, so the
accessible region of the orbit is progressively truncated and the innermost
windings are absorbed at the throat. The two Pad\'{e} orders share the same
redshift function and therefore the same near-throat precession, but differ in
the radial extent of the bound region through the factor $(1-b/r)$ in
Eq.~\eqref{eq:Veff_timelike}: the linear $[1/0]$ shape function retains a
residual deficit at large $r$, broadening the outer loops, whereas the
asymptotically flat $[0/1]$ shape function confines the precessing orbit to a
more compact region around the throat, consistent with the null-geodesic
behaviour of Sec.~\ref{sec:geodesics} \cite{Cataldo2017,Shaikh2019jcap}.

\section{Stability and Quasinormal Mode Analysis}
\label{sec:qnm}
The response of a compact object to small perturbations is encoded in its
quasinormal modes (QNMs), the complex frequencies at which the system rings
down under purely outgoing boundary conditions \cite{Regge1957,Zerilli1970}.
The real part of a QNM frequency fixes the oscillation frequency of the
emitted radiation and the imaginary part fixes its damping rate, so that the
mode spectrum constitutes a characteristic fingerprint of the underlying
geometry that is in principle accessible to gravitational-wave observation
\cite{Kokkotas1999,Berti2009,Konoplya2011}. For horizonless objects such as
wormholes the QNM problem differs qualitatively from the black-hole case: in
the absence of an event horizon the effective potential is typically symmetric
about the throat, the boundary conditions are outgoing on both sides, and the
late-time signal can develop features, including echoes, that distinguish the
wormhole from a black hole of comparable mass
\cite{Cardoso2016,Bueno2018,Bronnikov2021}. In this section we carry out a
detailed stability and QNM analysis of the two Pad\'{e}-approximant wormholes
constructed in Sec.~\ref{sec:pade}, using the global parameter set $M=1.5$,
$r_0=1$, $l=2$, $a=0.5$ for the $[1/0]$ branch, and $a=-0.5$ for the
pole-free $[0/1]$ branch unless stated otherwise.

\subsection{Master Equation and Effective Potential}
\label{subsec:master}
A massless test field propagating on the static, spherically symmetric
background of Eq.~\eqref{eq:MTmetric} separates, after a multipole and
frequency decomposition, into a one-dimensional wave equation of
Regge--Wheeler form \cite{Regge1957,Zerilli1970,Konoplya2011},
\begin{equation}
\frac{d^{2}\Psi}{dr_*^{2}} + \bigl[\omega^{2}-V_s(r)\bigr]\Psi = 0,
\label{eq:master}
\end{equation}
where the tortoise coordinate is defined through
\begin{equation}
\frac{dr_*}{dr} = \frac{e^{-\Phi(r)}}{\sqrt{1-b(r)/r}},
\label{eq:tortoise}
\end{equation}
so that $r_*\to\pm\infty$ on the two asymptotic sheets connected by the throat (asymptotically flat for the $[0/1]$ configuration and with a solid-angle deficit for the $[1/0]$ configuration). For a field of spin $s$ the effective potential takes the form
\begin{equation}
V_s(r) = e^{2\Phi(r)}\,\frac{l(l+1)}{r^{2}}
+ \frac{1-s^{2}}{2r}\,\frac{d}{dr}\!\left[e^{2\Phi(r)}
\left(1-\frac{b(r)}{r}\right)\right],
\label{eq:Vs}
\end{equation}
with $s=0,1,2$ corresponding to test scalar, electromagnetic, and axial
gravitational perturbations respectively. $l$ is the angular momentum quantum number (also called the multipole number or Legendre index) associated with the spherical harmonic decomposition of the perturbing field. $l = 0$ corresponds to the monopole mode (spherically symmetric), $l = 1$ corresponds to the dipole mode, and $l = 2$ corresponds to the quadrupole mode (this is typically the dominant mode for gravitational wave ringdown). The first term is the centrifugal
barrier weighted by the redshift factor, while the second, spin-dependent term
encodes the curvature coupling and vanishes identically for $s=1$. We adopt
$\Phi(r)=-M/r$ throughout, so that $e^{2\Phi}=e^{-2M/r}$ and the potential is
controlled by the shape function $b(r)$ through the factor $1-b(r)/r$.

\begin{figure}[!htbp]
\centering
\includegraphics[width=0.95\columnwidth]{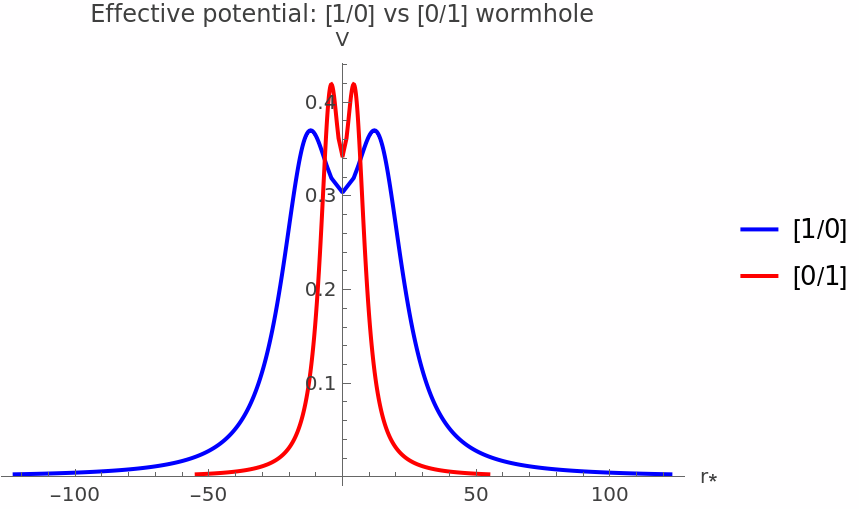}
\caption{Scalar ($s=0$) effective potential $V(r_*)$ as a function of the
tortoise coordinate for the $[1/0]$ (blue) and $[0/1]$ (red) Pad\'{e}
wormholes, with $M=1.5$, $r_0=1$, $l=2$, $a=0.5$, and $a=-0.5$ respectively.}
\label{fig:Veff}
\end{figure}

The scalar effective potentials of the two branches are shown in
Fig.~\ref{fig:Veff} as functions of the tortoise coordinate, obtained by
integrating Eq.~\eqref{eq:tortoise} outward from the throat on each sheet and
reflecting the result to the second sheet. Both potentials are positive
definite and vanish at spatial infinity on either side, which already
indicates that neither configuration supports an unstable bound state: a
nodeless negative-energy mode cannot exist for a strictly positive potential,
so both wormholes are dynamically stable against linear perturbations
\cite{Konoplya2011,Cardoso2009}. The two branches differ markedly in the
near-throat region. The $[1/0]$ potential is broad and develops a shallow local
minimum at the throat flanked by two maxima, the signature of a wide effective
cavity, whereas the $[0/1]$ potential is more sharply peaked and spatially
compact, with a narrower central well. This difference in barrier width is the
origin of the contrasting damping behaviour found below, since a broader, more
slowly varying barrier supports longer-lived, less-damped oscillations
\cite{SchutzWill1985,IyerWill1987}.

\subsection{WKB Determination of the Fundamental Mode}
\label{subsec:wkb}
The QNM frequencies of Eq.~\eqref{eq:master} are computed using the
semianalytic Wentzel--Kramers--Brillouin (WKB) method, which matches
asymptotic outgoing solutions across the potential barrier through a
Taylor expansion about its peak \cite{SchutzWill1985,IyerWill1987}. In the
third-order Iyer--Will formulation the frequency satisfies
\begin{equation}
\frac{i\bigl(\omega^{2}-V_0\bigr)}{\sqrt{-2V_0''}}
- \Lambda - \Lambda_3 = n + \frac{1}{2},
\label{eq:wkb}
\end{equation}
where $V_0$ and $V_0''$ are the value and second tortoise derivative of the
potential at its peak, $n$ is the overtone number, and $\Lambda,\Lambda_3$
are the second- and third-order correction terms built from the higher
derivatives of the potential at the peak \cite{IyerWill1987,Konoplya2003}. For
the symmetric wormhole barrier the peak is determined separately on each sheet
and the corresponding derivatives are evaluated with respect to $r_*$.

\begin{figure}[!htbp]
\centering
\includegraphics[width=0.95\columnwidth]{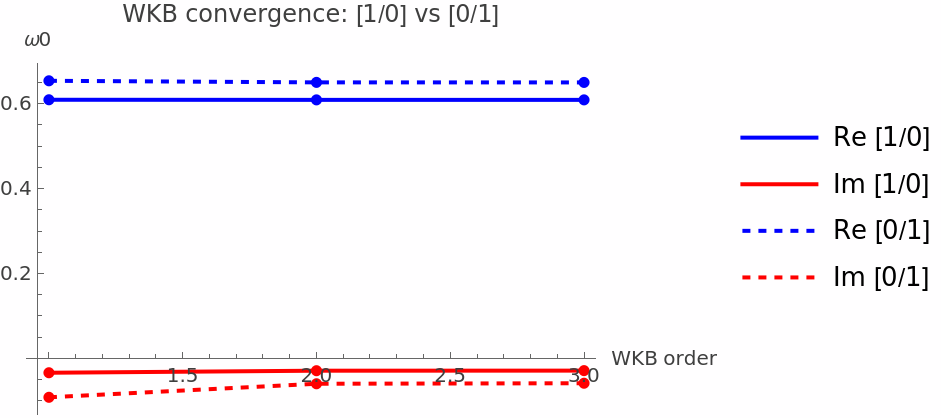}
\caption{Convergence of the fundamental scalar QNM frequency
($n=0$, $l=2$, $s=0$) with WKB order for the $[1/0]$ and $[0/1]$ Pad\'{e}
wormholes.}
\label{fig:wkbconv}
\end{figure}

Figure~\ref{fig:wkbconv} shows the convergence of the fundamental
($n=0$, $l=2$, $s=0$) frequency with increasing WKB order for both branches.
The real and imaginary parts stabilize already at second order and change
negligibly between second and third order, confirming that the third-order
truncation is well converged for the fundamental mode
\cite{IyerWill1987,Konoplya2011}. The resulting fundamental scalar frequencies
are
\begin{align}
\omega_{[1/0]} &= 0.60887 - 0.02886\,i, \\
\omega_{[0/1]} &= 0.64998 - 0.05857\,i,
\end{align}
collected together with the higher multipoles and spin channels in
Table~\ref{tab:qnm_main}. The $[0/1]$ branch has both a higher oscillation
frequency and a substantially larger damping rate than the $[1/0]$ branch,
consistent with its narrower, more sharply peaked potential.

\begin{table}[!htbp]
\centering
\caption{Fundamental and low-overtone WKB ($n=0$) QNM frequencies for the
$[1/0]$ and $[0/1]$ Pad\'{e} wormholes, with $M=1.5$, $r_0=1$, $a=0.5$,
$a=-0.5$, for multipoles $l=2,3,4$ and spin channels $s=0,1,2$.}
\label{tab:qnm_main}
\small
\begin{tabular}{cc cc}
\hline\hline
$l$ & channel & $\omega_{[1/0]}$ & $\omega_{[0/1]}$ \\
\hline
2 & $s=0$ & $0.60887-0.02886\,i$ & $0.64998-0.05857\,i$ \\
2 & $s=1$ & $0.60142-0.02841\,i$ & $0.60296-0.05161\,i$ \\
2 & $s=2$ & $0.57848-0.02691\,i$ & $0.45142-0.05811\,i$ \\
3 & $s=0$ & $0.85539-0.03018\,i$ & $0.88553-0.06610\,i$ \\
3 & $s=1$ & $0.85011-0.02996\,i$ & $0.85191-0.06290\,i$ \\
3 & $s=2$ & $0.83406-0.02925\,i$ & $0.74424-0.04693\,i$ \\
4 & $s=0$ & $1.10133-0.03091\,i$ & $1.12513-0.07067\,i$ \\
4 & $s=1$ & $1.09724-0.03078\,i$ & $1.09897-0.06888\,i$ \\
4 & $s=2$ & $1.08486-0.03037\,i$ & $1.01733-0.06194\,i$ \\
\hline\hline
\end{tabular}
\end{table}

\subsection{Parameter Dependence of the Spectrum}
\label{subsec:sweeps}
\begin{figure}[!htbp]
\centering
\includegraphics[width=0.95\columnwidth]{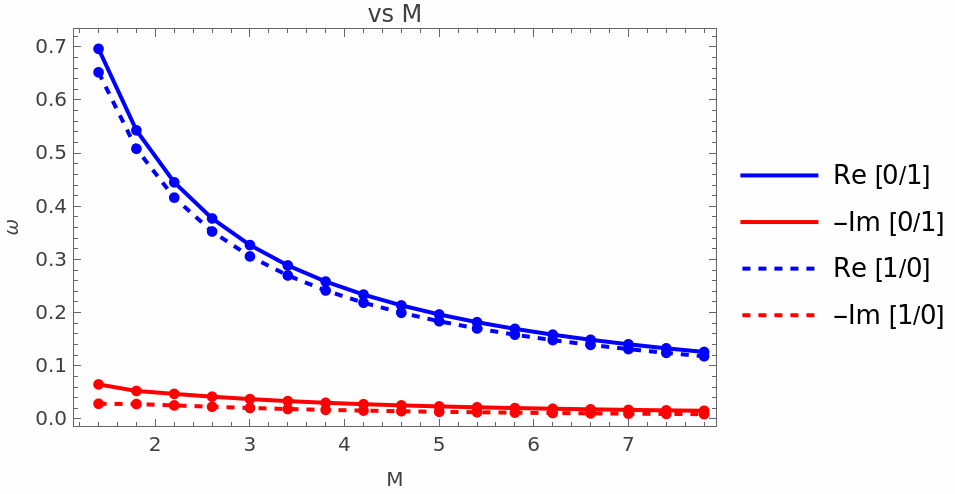}
\caption{Fundamental $l=2$ scalar QNM frequency as a function of the mass
parameter $M$ for the $[0/1]$ and $[1/0]$ Pad\'{e} wormholes.}
\label{fig:sweepM}
\end{figure}

\begin{figure}[!htbp]
\centering
\includegraphics[width=0.95\columnwidth]{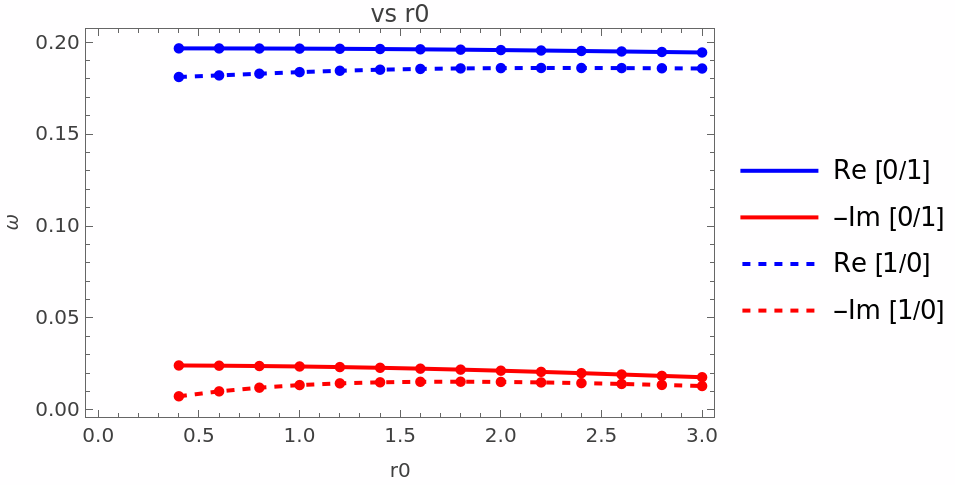}
\caption{Fundamental $l=2$ scalar QNM frequency as a function of the throat
radius $r_0$ for the $[0/1]$ and $[1/0]$ order Pad\'{e} wormholes.}
\label{fig:sweepr0}
\end{figure}

\begin{figure}[!htbp]
\centering
\includegraphics[width=0.95\columnwidth]{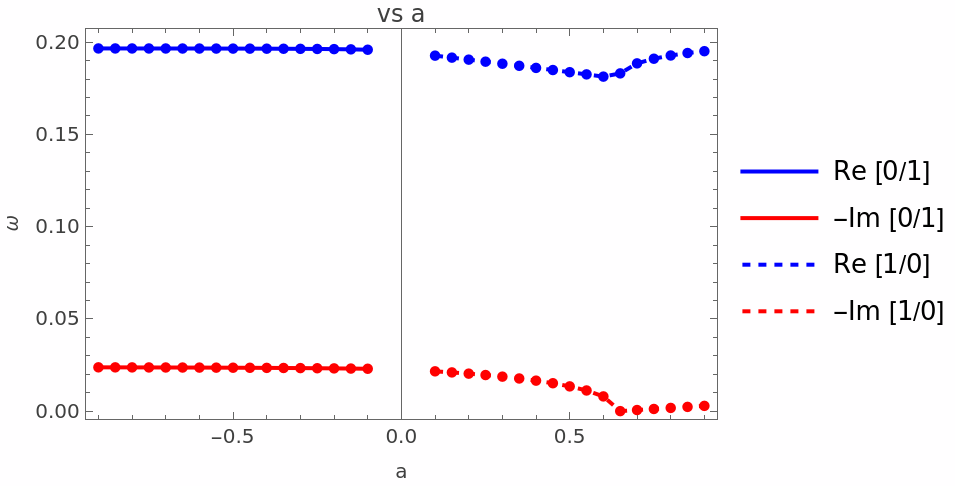}
\caption{Fundamental $l=2$ scalar QNM frequency as a function of the Pad\'{e}
coefficient $a$, with $a<0$ for the $[0/1]$ branch and $a>0$ for the $[1/0]$
branch.}
\label{fig:sweepa}
\end{figure}

To map the dependence of the fundamental mode on the model parameters we vary
each of $M$, $r_0$, and the Pad\'{e} coefficient $a$ in turn, holding the
others fixed, and recompute the $l=2$ scalar frequency at third WKB order. The
results are shown in Fig.~\ref{fig:sweepM}, Fig.~\ref{fig:sweepr0}, and
Fig.~\ref{fig:sweepa}.

The dependence on the mass parameter, Fig.~\ref{fig:sweepM}, is the strongest
of the three. Both the real part and the damping rate decrease monotonically
with increasing $M$, with the real part falling from roughly $0.69$ at $M=1.4$
to about $0.12$ at $M=7.8$ for the $[0/1]$ branch, and the two branches
tracking one another closely over the entire range. This inverse scaling
reflects the fact that increasing $M$ deepens the redshift factor $e^{-2M/r}$
and broadens the potential, lowering both the oscillation frequency and the
decay rate, in the manner familiar from black-hole and exponential-metric
geometries \cite{Boonserm2018,Tsukamoto2019,Konoplya2011}.

The dependence on the throat radius, Fig.~\ref{fig:sweepr0}, is comparatively
weak. The $[0/1]$ frequencies are almost independent of $r_0$, while the
$[1/0]$ frequencies vary only mildly, with the real part nearly flat and the
damping rate increasing slowly with $r_0$. The two branches converge as $r_0$
grows, which is expected since both shape functions approach the same throat
behaviour as the Pad\'{e} parameter contribution becomes subdominant
\cite{Baker1961,Stahl1998}.

The dependence on the Pad\'{e} coefficient, Fig.~\ref{fig:sweepa}, displays the
clearest distinction between the branches. For the $[0/1]$ branch, accessible
for $a<0$, both the real part and the damping rate are essentially constant
across $-0.9\le a\le-0.1$. For the $[1/0]$ branch, accessible for $a>0$, the
real part dips and then recovers as $a$ increases toward unity, while the
damping rate decreases steadily and approaches very small values near
$a\simeq0.6$, signalling the onset of long-lived, weakly damped modes as the
barrier flattens. The discontinuity at $a=0$ in the figure simply reflects the
fact that the two branches occupy opposite signs of $a$ by construction, the
$[1/0]$ form requiring $a>0$ and the pole-free $[0/1]$ form requiring $a<0$, as
established in Sec.~\ref{sec:pade}.

\subsection{Eikonal Limit and the Photon-Sphere Correspondence}
\label{subsec:eikonal}
In the eikonal limit of large multipole number the QNM frequencies are governed
by the properties of the unstable circular null geodesic at the photon sphere:
the real part approaches the orbital angular frequency $\Omega_{\rm ph}$ and
the damping rate approaches the Lyapunov exponent $\lambda_L$ of the orbit
\cite{Cardoso2009,Konoplya2011},
\begin{equation}
\omega_{l\gg1} \simeq \Omega_{\rm ph}\Bigl(l+\tfrac12\Bigr)
- i\Bigl(n+\tfrac12\Bigr)\lambda_L.
\label{eq:eikonal}
\end{equation}
Since the photon sphere of the present geometry lies at $r_{\rm ph}=M$ with
critical impact parameter $\tilde b_{\rm ph}=eM$ independent of the shape
function, as shown in Sec.~\ref{sec:geodesics}, this correspondence provides a
stringent internal consistency check on the WKB frequencies.

\begin{figure}[!htbp]
\centering
\includegraphics[width=0.95\columnwidth]{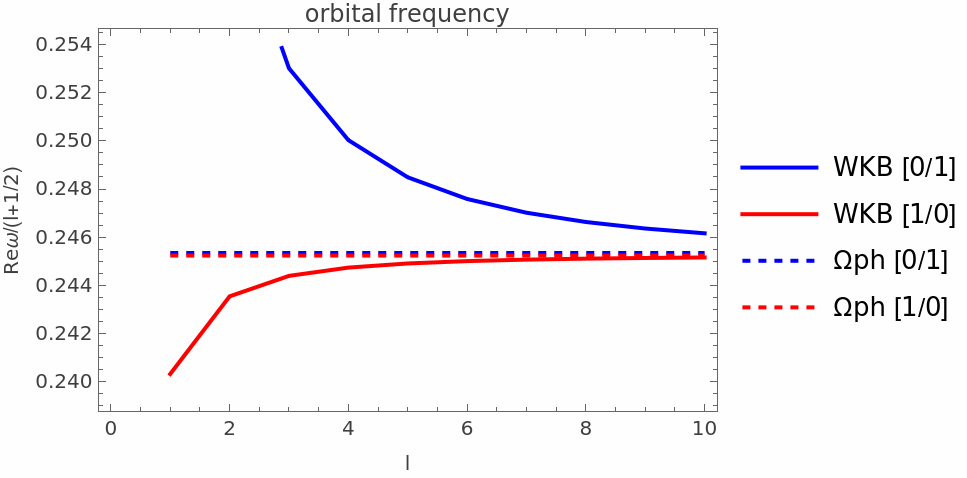}
\caption{Scaled real part $\mathrm{Re}\,\omega/(l+\tfrac12)$ of the QNM
frequency as a function of $l$ for the $[0/1]$ and $[1/0]$ branches, with the photon-sphere orbital frequencies $\Omega_{\rm ph}$.}
\label{fig:orbital}
\end{figure}

Figure~\ref{fig:orbital} shows the scaled real part
$\mathrm{Re}\,\omega/(l+\tfrac12)$ as a function of $l$ for both branches,
together with the orbital frequency $\Omega_{\rm ph}$ computed directly from
the null geodesics. The WKB ratio converges smoothly to $\Omega_{\rm ph}$ from
below for the $[1/0]$ branch and from above for the $[0/1]$ branch as $l$
increases, with both branches approaching the common limiting value to better
than one percent by $l=10$. The corresponding limiting orbital frequencies are
\begin{equation}
\Omega_{\rm ph}^{[1/0]} = 0.24524, \qquad
\Omega_{\rm ph}^{[0/1]} = 0.24535,
\end{equation}
which agree to four significant figures, as expected from the
shape-function-independence of the photon sphere.

\begin{figure}[!htbp]
\centering
\includegraphics[width=0.95\columnwidth]{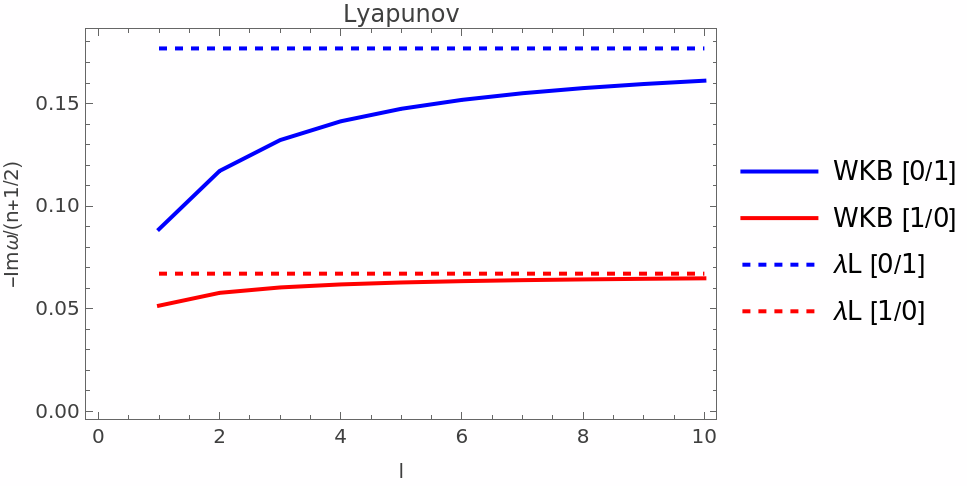}
\caption{Scaled damping rate $-\mathrm{Im}\,\omega/(n+\tfrac12)$ as a function
of $l$ for the $[0/1]$ and $[1/0]$ branches, with the
Lyapunov exponents $\lambda_L$ shown as dashed horizontal lines.}
\label{fig:lyapunov}
\end{figure}

Figure~\ref{fig:lyapunov} shows the analogous comparison for the scaled damping
rate against the Lyapunov exponent. Here, the two branches differ
substantially, with limiting values
\begin{equation}
\lambda_L^{[1/0]} = 0.06706, \qquad
\lambda_L^{[0/1]} = 0.17685.
\end{equation}
The much larger Lyapunov exponent of the $[0/1]$ branch reflects the steeper
curvature of its more sharply peaked potential at the photon sphere, and is the
eikonal counterpart of the larger fundamental damping rate found in
Sec.~\ref{subsec:wkb}. The orbital frequencies and Lyapunov exponents are
collected in Table~\ref{tab:eikonal}, and the convergence of the scaled real
part toward $\Omega_{\rm ph}$ with increasing $l$ is tabulated in
Table~\ref{tab:orbital_conv}.

\begin{table}[!htbp]
\centering
\caption{Photon-sphere orbital frequency $\Omega_{\rm ph}$ and Lyapunov
exponent $\lambda_L$ for the two branches, with $M=1.5$.}
\label{tab:eikonal}
\begin{tabular}{ccc}
\hline\hline
Branch & $\Omega_{\rm ph}$ & $\lambda_L$ \\
\hline
$[1/0]$ & $0.24524$ & $0.06706$ \\
$[0/1]$ & $0.24535$ & $0.17685$ \\
\hline\hline
\end{tabular}
\end{table}

\begin{table}[!htbp]
\centering
\caption{Convergence of the scaled real part $\mathrm{Re}\,\omega/(l+\tfrac12)$
toward the photon-sphere orbital frequency with increasing $l$, for the
fundamental scalar mode at $M=1.5$. The last row gives the eikonal limit.}
\label{tab:orbital_conv}
\begin{tabular}{ccc}
\hline\hline
$l$ & $[1/0]$ & $[0/1]$ \\
\hline
2        & $0.24355$ & $0.25999$ \\
3        & $0.24440$ & $0.25301$ \\
4        & $0.24474$ & $0.25003$ \\
5        & $0.24491$ & $0.24849$ \\
6        & $0.24501$ & $0.24759$ \\
8        & $0.24511$ & $0.24663$ \\
10       & $0.24516$ & $0.24616$ \\
$\infty$ & $0.24524$ & $0.24535$ \\
\hline\hline
\end{tabular}
\end{table}

\subsection{Overtone Spectrum}
\label{subsec:overtones}
\begin{figure}[!htbp]
\centering
\includegraphics[width=0.95\columnwidth]{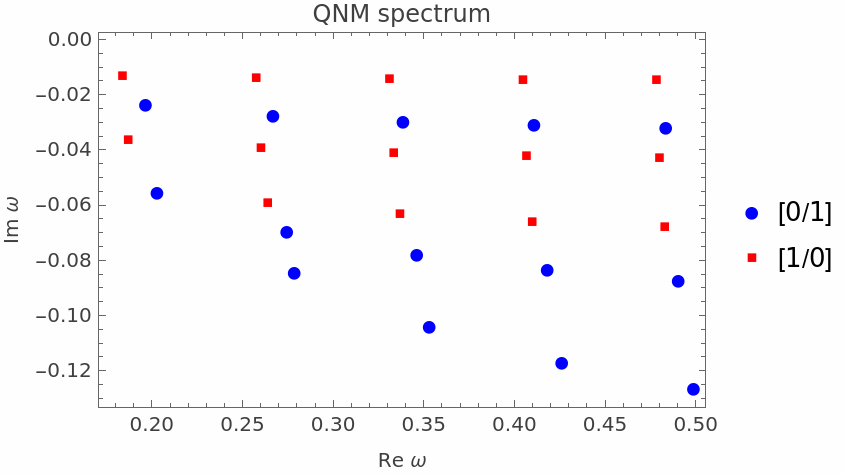}
\caption{Scalar QNM spectrum in the complex frequency plane for the $[0/1]$ and $[1/0]$ Pad\'{e} wormholes, for multipoles
$l=2$ to $6$ and overtones up to $n=\min(l-1,2)$. The $[0/1]$ modes are more
strongly damped.}
\label{fig:spectrum}
\end{figure}

The distribution of QNM frequencies in the complex plane characterizes the full
ringdown response. Figure~\ref{fig:spectrum} shows the scalar QNM spectrum for
both branches, with multipoles $l=2$ through $6$ and overtones up to
$n=\min(l-1,2)$. The modes organize into the characteristic pattern of
increasing the real part with $l$ and increasing the damping with the overtone number $n$,
with the $[0/1]$ modes systematically deeper in the lower half-plane than
the $[1/0]$ modes, in agreement with the higher damping rates found above
\cite{Berti2009,Konoplya2011}.

The overtone structure of the fundamental multipole sequence is detailed in
Table~\ref{tab:overtones}. For most multipoles, the damping rate increases
monotonically with overtone number, as expected. We note that for the $[0/1]$
branch the deep overtones near $n\simeq l$, specifically $l=4,n=3$ and
$l=5,n=4$, show a non-monotonic damping rate. This is a known limitation of the
WKB method, whose accuracy degrades for overtones with $n\gtrsim l$ where the
expansion about the potential peak is no longer well controlled
\cite{IyerWill1987,Konoplya2003,Konoplya2011}. These entries should therefore
be regarded as indicative rather than quantitatively reliable, and we retain
them only for completeness.

\begin{table}[!htbp]
\centering
\caption{Scalar ($s=0$) QNM frequencies for overtones $n$ at multipoles
$l=2$ to $5$, with $M=1.5$. The deepest $[0/1]$ overtones at $l=4,n=3$ and
$l=5,n=4$ lie outside the reliable regime of the WKB expansion.}
\label{tab:overtones}
\small
\begin{tabular}{cc cc}
\hline\hline
$l$ & $n$ & $\omega_{[1/0]}$ & $\omega_{[0/1]}$ \\
\hline
2 & 0 & $0.60887-0.02886\,i$ & $0.64998-0.05857\,i$ \\
2 & 1 & $0.61304-0.07698\,i$ & $0.66118-0.13459\,i$ \\
3 & 0 & $0.85539-0.03018\,i$ & $0.88553-0.06610\,i$ \\
3 & 1 & $0.85891-0.08329\,i$ & $0.89783-0.16222\,i$ \\
3 & 2 & $0.86351-0.12195\,i$ & $0.90072-0.17750\,i$ \\
4 & 0 & $1.10133-0.03091\,i$ & $1.12513-0.07067\,i$ \\
4 & 1 & $1.10432-0.08695\,i$ & $1.13736-0.18078\,i$ \\
4 & 2 & $1.10873-0.13156\,i$ & $1.14500-0.22383\,i$ \\
4 & 3 & $1.11318-0.16492\,i$ & $1.14109-0.20292\,i$ \\
5 & 0 & $1.34701-0.03138\,i$ & $1.36668-0.07372\,i$ \\
5 & 1 & $1.34961-0.08933\,i$ & $1.37840-0.19393\,i$ \\
5 & 2 & $1.35369-0.13786\,i$ & $1.38865-0.25688\,i$ \\
5 & 3 & $1.35823-0.17702\,i$ & $1.39025-0.26539\,i$ \\
5 & 4 & $1.36246-0.20699\,i$ & $1.38230-0.21997\,i$ \\
\hline\hline
\end{tabular}
\end{table}

\subsection{Time-Domain Evolution and Late-Time Behaviour}
\label{subsec:timedomain}
\begin{figure*}[tp]
\centering
\includegraphics[width=0.9\textwidth]{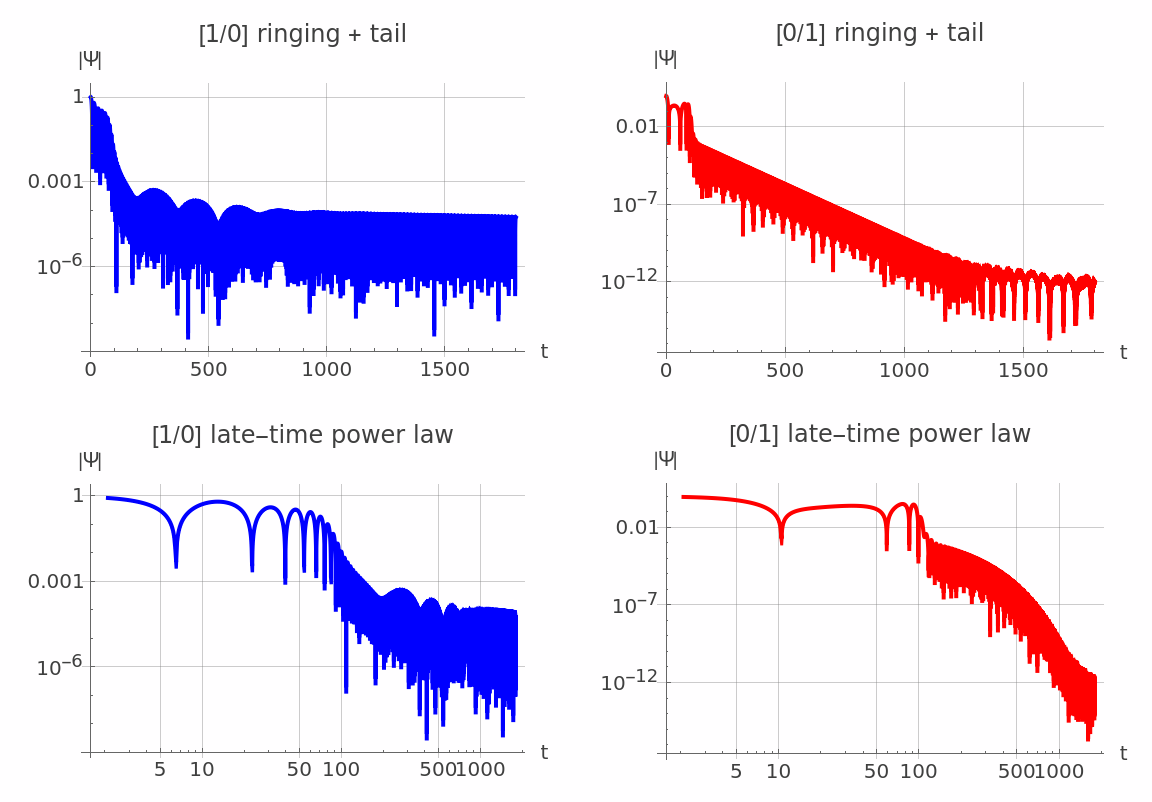}
\caption{Time-domain evolution of a scalar perturbation. Upper panels: the
ringdown and tail of the $[1/0]$ (left, blue) and $[0/1]$ (right, red)
wormholes on a semi-logarithmic scale. Lower panels: the same signals on a
log-log scale, revealing the inverse power-law late-time tail. The intermediate
stage is dominated by the fundamental quasinormal mode.}
\label{fig:timedomain}
\end{figure*}

To corroborate the frequency-domain results and to probe the full signal beyond
the dominant mode, we evolve Eq.~\eqref{eq:master} directly in the time domain.
Recasting the wave equation in null coordinates and integrating with a
characteristic finite-difference scheme \cite{Gundlach1994,Price1972}, we launch
a Gaussian wave packet on one sheet and record the field at a fixed observation
point. The resulting waveforms are shown in Fig.~\ref{fig:timedomain}.

Both branches exhibit the expected three-stage evolution: an initial transient
that depends on the precise form of the initial data, an intermediate
quasinormal ringing stage, and a late-time tail
\cite{Price1972,Ching1995,Gundlach1994}. During the early part of the ringing
stage the signal decays steeply, since the pulse excites a superposition of
overtones that decay faster than the fundamental. As the overtones die out the
slowest-damped fundamental mode emerges and dominates the signal just before the
onset of the tail. Fitting the envelope of the $[1/0]$ waveform over this
late-ringdown window, where the fundamental dominates, yields a damping rate
\begin{equation}
-\mathrm{Im}\,\omega_{[1/0]}^{\rm TD} = 0.02966,
\end{equation}
in agreement with the WKB value $0.02886$ of Table~\ref{tab:qnm_main} to better
than three percent. This independent time-domain confirmation validates the WKB
determination of the fundamental mode for the $[1/0]$ branch.



\subsection{Effective Cavity and Long-Time Signal}
\label{subsec:echo}
A distinctive feature of horizonless compact objects is the possibility of
gravitational-wave echoes, repeated pulses produced when radiation is trapped
between the photon-sphere barrier and an inner reflecting feature
\cite{Cardoso2016,Bueno2018,Bronnikov2021,Damour2007}. To assess this
possibility we examine the structure of the effective potential and evolve the
field over a long time window. Inspecting the potential on each branch we find
\begin{align}
[1/0]:\quad & V_{\rm throat}=0.30434, \;\; V_{\rm peak}=0.36989, \\
[0/1]:\quad & V_{\rm throat}=0.34297, \;\; V_{\rm peak}=0.41871,
\end{align}
so that in both cases the throat value lies below the peak, $V_{\rm
throat}<V_{\rm peak}$, and the potential takes the form of a double barrier
enclosing a shallow central well. This is the geometric prerequisite for an
effective cavity that can in principle support trapped, repeatedly reflected
modes \cite{Cardoso2016,Bueno2018}.

\begin{figure*}[tp]
\centering
\includegraphics[width=0.9\textwidth]{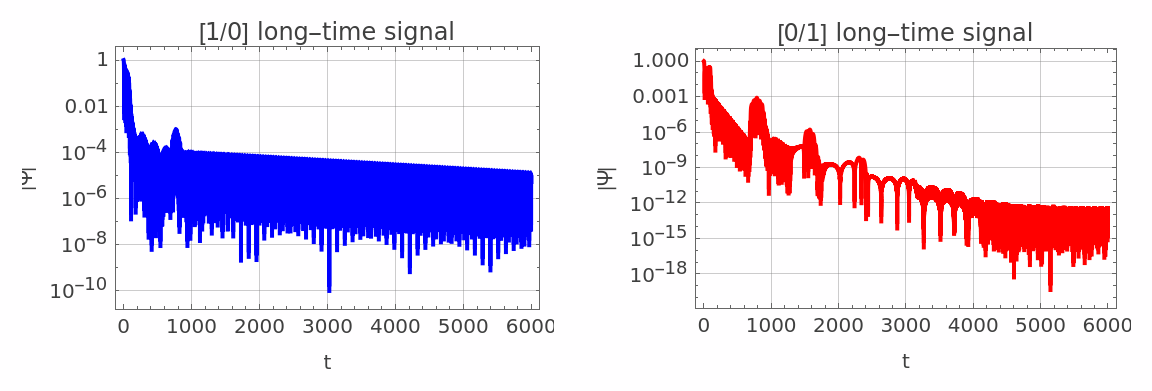}
\caption{Long-time evolution of a scalar perturbation for the $[1/0]$ (left,
blue) and $[0/1]$ (right, red) wormholes, integrated to $t=6000$. The shallow
central well produces a modulated decaying signal rather than a train of
discrete echoes.}
\label{fig:longtime}
\end{figure*}

The long-time signals are shown in Fig.~\ref{fig:longtime}. Both branches show
the ringdown followed by a slowly decaying tail, but the well in the present
geometries is shallow, with $V_{\rm throat}/V_{\rm peak}\simeq0.82$ for both
branches, so the cavity is leaky and no sharp, well-separated echo train
develops within the integration window. Instead, the trapped component
manifests itself as a modulation of the decaying signal rather than as discrete echo
pulses, consistent with the expectation that pronounced echoes require a deep,
high-contrast cavity \cite{Bueno2018,Bronnikov2021}. The absence of strong
echoes is therefore itself a diagnostic feature of these Pad\'{e} wormholes,
distinguishing them from ultra compact configurations with near-total inner
reflection.

\subsection{Spin Dependence of the Spectrum}
\label{subsec:spins}
\begin{figure}[!htbp]
\centering
\includegraphics[width=0.95\columnwidth]{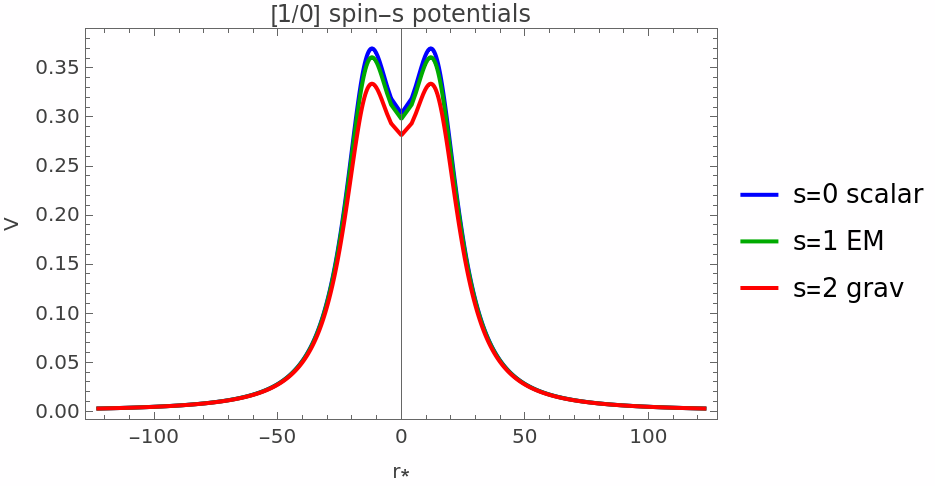}\\[4pt]
\includegraphics[width=0.95\columnwidth]{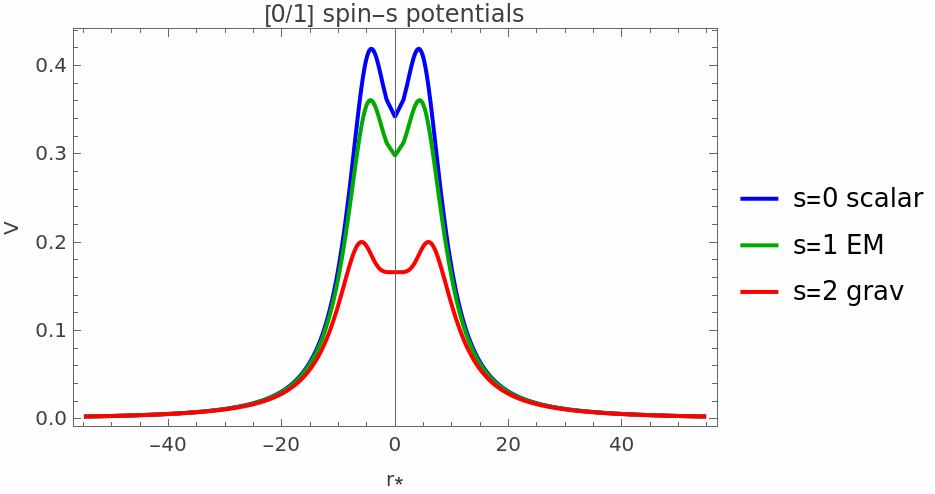}
\caption{Spin-$s$ effective potentials for the $[1/0]$ and $[0/1]$ Pad\'{e} wormholes, for scalar ($s=0$, blue), electromagnetic
($s=1$, green), and axial gravitational ($s=2$, red) perturbations, with
$M=1.5$, $r_0=1$, $l=2$.}
\label{fig:spinpot}
\end{figure}

Finally, we examine how the QNM spectrum depends on the spin of the perturbing
field, comparing test scalar ($s=0$), electromagnetic ($s=1$), and axial
gravitational ($s=2$) perturbations through the potential of Eq.~\eqref{eq:Vs}.
The spin-$s$ potentials are shown in Fig.~\ref{fig:spinpot} for both branches.
For the $[1/0]$ branch the three potentials are nearly degenerate, since the
spin-dependent term is small relative to the centrifugal barrier for this
broad, slowly varying potential. For the $[0/1]$ branch the spin dependence is
much more pronounced: the scalar barrier is highest and the gravitational
barrier is significantly lower and flatter, reflecting the stronger curvature
coupling of the more compact $[0/1]$ geometry.

The corresponding fundamental ($n=0$, $l=2$) frequencies are listed by channel
in Table~\ref{tab:spin}. For the $[1/0]$ branch the frequency decreases only
slightly from the scalar to the gravitational channel, by about five percent in
the real part. For the $[0/1]$ branch the variation is far larger: the
gravitational real part drops to $0.45142$, roughly thirty percent below the
scalar value of $0.64998$, while the damping rates remain comparable across
channels. This strong spin dependence of the $[0/1]$ branch is a direct
consequence of the sharper, more spin-sensitive potential, and provides a
further observable that distinguishes the two Pad\'{e} orders.

\begin{table}[!htbp]
\centering
\caption{Fundamental ($n=0$, $l=2$) QNM frequencies by spin channel for the two
branches, with $M=1.5$, $r_0=1$, $a=0.5$, $a=-0.5$.}
\label{tab:spin}
\begin{tabular}{c cc}
\hline\hline
Channel & $\omega_{[1/0]}$ & $\omega_{[0/1]}$ \\
\hline
$s=0$ scalar & $0.60887-0.02886\,i$ & $0.64998-0.05857\,i$ \\
$s=1$ EM     & $0.60142-0.02841\,i$ & $0.60296-0.05161\,i$ \\
$s=2$ grav   & $0.57848-0.02691\,i$ & $0.45142-0.05811\,i$ \\
\hline\hline
\end{tabular}
\end{table}


\section{Concluding Remarks}
\label{sec:conclusion}
In this work, we constructed traversable wormhole solutions in the linear
teleparallel model $f(T)=\alpha T+\beta$, with the asymptotically flat,
horizon-free redshift function $\Phi(r)=-M/r$ and two Pad\'{e}-approximated
shape functions for the $[1,0]$ and $[0,1]$ order. Both reproduce the same first-order Taylor expansion about
the throat, satisfy $b(r_0)=r_0$, and meet the flaring-out and admissibility
constraints of Sec.~\ref{sec:pade}. Because $f_{TT}=0$, the field equations
reduce to the teleparallel equivalent of general relativity shifted by the
constant $\beta$, and the effective source
$\mathcal{T}^{\mu}{}_{\nu}=-\alpha G^{\mu}{}_{\nu}-\tfrac{\beta}{2}\delta^{\mu}{}_{\nu}$
is identically conserved. Throughout the analysis we adopted the unified
parameter set $M=1.5$, $r_0=1$, $\alpha=2$, $\beta=2$, with $a=0.5$ for the
$[1/0]$ branch and $a=-0.5$ for the $[0/1]$ branch.

The energy-condition analysis shows that the tangential null combination is
negative at the throat for both orders, so the NEC is violated and the matter
is exotic, as required by the flaring-out condition. For the $[1/0]$ order the
throat values are $\rho=-0.5514$, $\rho+p_r=+0.4486$, and $\rho+p_t=-2.1121$,
with the strong-energy combination $\rho+p_r+2p_t=-2.6728$, so that the radial
null channel is satisfied while the tangential channel drives the violation;
the NEC, WEC, SEC, and DEC all fail. For the $[0/1]$ order the energy density at
the throat is large and positive, $\rho=+2.5514$, with $\rho+p_r=+3.5514$ and
$\rho+p_t=-2.8788$, and the strong-energy combination $\rho+p_r+2p_t=-7.3272$.
Although the energy density is positive here, the negative tangential null
combination leaves the NEC violated through the tangential channel, and the WEC,
SEC, and DEC fail as well. In both orders the violation is therefore carried
entirely by the tangential channel while the radial channel remains satisfied, a
feature shared with other anisotropic wormhole constructions in modified
gravity. The two orders differ in the radial spread of the violation, with the
$[0/1]$ combinations exhibiting a more pronounced near-throat concentration of
exotic matter.

The physical diagnostics confirm viability for both orders. The volume integral
quantifier, evaluated over the matched interior region from the throat $r_0=1$
to the cutoff $R=10r_0=10$, is finite and negative, $I_V^{[1/0]}\approx-21.4780$ and $I_V^{[0/1]}\approx-171.9100$, so a bounded amount of NEC-violating matter sustains each throat for a given matching radius, with $|I_V|$ growing logarithmically with $R$, with the much larger $|I_V^{[0/1]}|$ reflecting the deep
near-throat concentration of exotic matter in that branch. The TOV analysis
gives an exact cancellation $F_H+F_G+F_A=0$, with the hydrostatic force $F_H$
directed outward and the gravitational and anisotropic forces $F_G$ and $F_A$
directed inward, the equilibrium forces being independent of $\beta$ and the
$[0/1]$ branch developing the larger near-throat force magnitudes. The embedding
surface and proper radial distance are finite and monotonic in both orders. The
gravitational energy and active mass reflect the underlying exotic-matter
distribution rather than ordinary positive mass: for the $[1/0]$ branch the
gravitational energy starts at $E_g=r_0/2$, rises, and then turns negative as
the negative-energy-density region dominates, while the active mass dips
negative just outside the throat before recovering; for the $[0/1]$ branch the
gravitational energy is negative across most of the range through the geometric
factor $1-\sqrt{g_{rr}}<0$, even though the energy density is positive, while the
active mass grows large and positive. These sign changes are the expected
quasilocal-energy signature of NEC-violating matter and are fully consistent
with the energy-condition results.

The optical analysis fixes the photon sphere at $r_{ph}=M$ and the critical
impact parameter at $\tilde{b}_{ph}=eM$. Both quantities depend on the redshift
function alone and are therefore identical for the two Pad\'{e} orders. The
shape function controls the bending away from the photon sphere through the
factor $(1-b/r)$. The $[1/0]$ geometry retains a constant deficit
$b/r\to a[1+r_0\csch r_0\sech r_0]$, producing broad winding and a non-flat
exterior in which $\tilde{b}$ is a coordinate quantity. The $[0/1]$ geometry
decays as $b/r\to0$, giving asymptotic flatness, lensing confined to a compact
region, and $\tilde{b}$ equal to the true asymptotic impact parameter. The
intensity profiles peak at $b_c=eM$ through the photon-ring enhancement, and
both the shadow maps and the $i=80^{\circ}$ accretion-disc images show shadow
and photon-ring sizes growing monotonically with $r_0$, with the $[0/1]$ branch
giving a more compact, centrally concentrated crescent at fixed $r_0$. The
embedding diagrams are vertical at the throat in both orders, with the $[1/0]$
profile flaring more steeply than the pole-free $[0/1]$ profile, and the
timelike rosette orbits precess with the same near-throat rate set by $\Phi$,
the $[1/0]$ branch spreading over larger radii.

The quasinormal-mode analysis gives a positive-definite, symmetric
double-barrier potential about the throat for both orders, the $[0/1]$ barrier
taller and sharper and the $[1/0]$ barrier lower and wider, with peak values
$V_{\rm peak}^{[1/0]}=0.36989$ and $V_{\rm peak}^{[0/1]}=0.41871$ respectively. The peak
height decreases with spin following the $(1-s^2)$ weighting, the effect being
small for the broad $[1/0]$ barrier and pronounced for the compact $[0/1]$
barrier. The third-order WKB fundamental scalar frequencies ($l=2$, $n=0$) are
$\omega=0.60887-0.02886\,i$ for $[1/0]$ and $\omega=0.64998-0.05857\,i$ for
$[0/1]$ respectively. The strictly negative imaginary parts establish linear stability, which
is confirmed by the direct time-domain evolution: the $[1/0]$ ringdown yields a
damping rate of $0.02966$ from the late-ringdown envelope, in agreement with the
WKB value to better than three percent, while the more strongly damped $[0/1]$
mode rings down within a few cycles so that its fundamental is best quoted from
WKB. Both signals cross over at late times to a universal inverse power-law
tail. The characteristic frequencies scale approximately as $\omega\propto
M^{-1}$, depend only weakly on $r_0$, and approach the photon-sphere values in
the eikonal limit, with $\Omega_{ph}^{[1/0]}=0.24524$,
$\Omega_{ph}^{[0/1]}=0.24535$, $\lambda_L^{[1/0]}=0.06706$, and
$\lambda_L^{[0/1]}=0.17685$. The $[1/0]$ damping rate drops toward zero near
$a\simeq0.6$, signaling the onset of long-lived, weakly damped oscillations,
and the shallow central well, with $V_{\rm throat}/V_{\rm peak}\simeq0.82$ for
both branches, produces a modulated decaying signal rather than a train of
discrete echoes.

These results establish the Pad\'{e} approximation as a practical tool for 
constructing analytically tractable wormhole geometries in torsion-based 
gravity. Both Pad\'{e}-order wormholes in the linear $f(T)$ model are stable, 
traversable configurations sustained by a bounded amount of NEC-violating 
matter, whose optical and dynamical signatures are testable by horizon-scale 
very-long-baseline interferometry and gravitational-wave ringdown observations. The asymptotically flat $[0/1]$
geometry is the more suitable of the two for shadow and lensing observables,
since its coordinate impact parameter coincides with the asymptotic one, while
the two branches remain clearly distinguishable through their shadow
compactness, ringdown damping, Lyapunov exponents, and spin-dependent spectra.
Natural extensions include rotating generalizations through a Lense--Thirring 
or Newman--Janis construction, as well as higher-order Pad\'{e} approximants. 
A direct confrontation of the predicted shadow radii and ringdown frequencies 
with current Event Horizon Telescope and future gravitational-wave detector 
data remains a compelling observational avenue.

\section*{Data availability} No new data are associated with this article.

\section*{Acknowledgments}
SP acknowledges the support by Chennai Mathematical Institute (CMI). PKS thanks IUCAA, Pune, India, for its assistance through the visiting associateship program.

\bibliographystyle{apsrev4-2}
\bibliography{references}
\end{document}